\documentclass[preprint,authoryear,10pt,letter]{elsarticle}
\usepackage[top=0.75in, bottom=0.75in, left=0.75in, right=0.75in]{geometry}

\usepackage{graphicx}
\usepackage[labelsep=endash, figurename=Figure, tablename=Table, textfont=it]{caption}
\usepackage{float}
\usepackage[hyphens]{url}
\usepackage[colorlinks=true,linkcolor=black,citecolor=black,urlcolor=blue]{hyperref}
\usepackage{multicol}

\usepackage{fancyhdr}
\usepackage{array}
\usepackage{textcomp}
\usepackage{verbatim}
\usepackage{cite}
\usepackage[dvipsnames]{xcolor}
\usepackage{microtype}
\usepackage{graphicx}
\usepackage{subfigure}
\usepackage{booktabs} % for professional tables
\usepackage{bbm}

\usepackage{pgfplots}
\pgfplotsset{compat=1.7}

\usepackage{xurl}
\usepackage[linesnumbered,ruled,vlined,commentsnumbered]{algorithm2e}
\usepackage{amsmath,amssymb,amsfonts}
\usepackage{tikz}
\usetikzlibrary{decorations.pathreplacing}
\usetikzlibrary{positioning,arrows.meta,quotes}
\usetikzlibrary{shapes}
\usetikzlibrary{decorations.pathmorphing}
\usetikzlibrary{bayesnet}
\tikzset{>=latex}
\tikzstyle{plate caption} = [caption, node distance=0, inner sep=0pt, below left=5pt and 0pt of #1.south]

\usepackage[normalem]{ulem}
\usepackage{multirow}

\usepackage{pifont}
\newcommand{\cmark}{\ding{51}}%
\newcommand{\xmark}{\ding{55}}%

\newcommand{\setR}{\mathbb{R}}

\journal{arXiv}

\begin{document}

\begin{frontmatter}

% \title {Driving Regime Identification in Car-Following Behaviors}

\title{Markov Regime-Switching Intelligent Driver Model for \\Interpretable Car-Following Behavior}

\author[label1]{Chengyuan Zhang}
% \ead{enzozcy@gmail.com}
\author[label2]{Cathy Wu}
\author[label1]{Lijun Sun\corref{cor1}}
\ead{lijun.sun@mcgill.ca}

\address[label1]{Department of Civil Engineering, McGill University, Montreal, QC H3A 0C3, Canada}
\address[label2]{Laboratory for Information \& Decision Systems (LIDS), Massachusetts Institute of Technology, Cambridge, MA 02139, USA}

\cortext[cor1]{Corresponding author.}

\begin{abstract}

Accurate and interpretable car-following models are essential for traffic simulation and autonomous vehicle development. However, classical models like the Intelligent Driver Model (IDM) are fundamentally limited by their parsimonious and single-regime structure. They fail to capture the multi-modal nature of human driving, where a single driving state (e.g., speed, relative speed, and gap) can elicit many different driver actions. This forces the model to average across distinct behaviors, reducing its fidelity and making its parameters difficult to interpret. To overcome this, we introduce a regime-switching framework that allows driving behavior to be governed by different IDM parameter sets, each corresponding to an interpretable behavioral mode. This design enables the model to dynamically switch between interpretable behavioral modes, rather than averaging across diverse driving contexts. We instantiate the framework using a Factorial Hidden Markov Model with IDM dynamics (FHMM-IDM), which explicitly separates intrinsic driving regimes (e.g., aggressive acceleration, steady-state following) from external traffic scenarios (e.g., free-flow, congestion, stop-and-go) through two independent latent Markov processes. Bayesian inference via Markov chain Monte Carlo (MCMC) is used to jointly estimate the regime-specific parameters, transition dynamics, and latent state trajectories. Experiments on the HighD dataset demonstrate that FHMM-IDM uncovers interpretable structure in human driving, effectively disentangling internal driver actions from contextual traffic conditions and revealing dynamic regime-switching patterns. This framework provides a tractable and principled solution to modeling context-dependent driving behavior under uncertainty, offering potential improvements in the fidelity of traffic simulations, the efficacy of safety analyses, and the development of more human-centric Advanced Driver-Assistance Systems (ADAS). The code will be released at \url{https://github.com/Chengyuan-Zhang/Markov_Switching_IDM} upon acceptance of the paper.
\end{abstract}

\end{frontmatter}

%\newpage

%\tableofcontents

%\newpage

\section{INTRODUCTION}\label{introduction}
Understanding and modeling human driving behavior is fundamental to the development of intelligent transportation systems, ADAS technologies, and autonomous vehicles \citep{li2021car}. Driving actions arise from a complex interplay between internal decision-making processes and external traffic environments \citep{wang2022social}, often exhibiting diverse, context-dependent patterns. However, most existing car-following models adopt simplified or fixed behavioral assumptions, limiting their ability to capture the stochasticity and adaptability observed in naturalistic driving. To address this gap, we advocate for a regime-switching framework that models driving behavior as a sequence of latent behavioral modes and contextual scenarios, each governed by interpretable dynamics.

Traditional car-following models, such as the IDM \citep{treiber2000congested}, typically assume a deterministic formulation in which a fixed set of parameters maps directly to the driver's longitudinal control behavior. The IDM computes acceleration as a function of speed, relative speed, and spacing:
\begin{subequations}
\begin{align}
    \mathrm{IDM}(\boldsymbol{x}_t;\boldsymbol{\theta}) &= a_{\mathrm{max}}\Bigg(1-\left(\frac{v_t}{v_f}\right)^\delta -\left(\frac{s^*}{s_t}\right)^2\Bigg),\label{IDM_formulation}\\
    s^* &= s_0 +v_tT+\frac{v_t\Delta v_t}{2\sqrt{a_{\mathrm{max}}b}},
\end{align}
\end{subequations}
where $\boldsymbol{x}_t = [v_t, \Delta v_t, s_t]^\top$ represents the state variables (speed, relative speed, and gap), and $\boldsymbol{\theta} = \{v_f, s_0, T, a_{\mathrm{max}}, b\} \in \mathbb{R}^5$ denotes the model parameters governing driver behavior.

While effective in controlled or homogeneous traffic conditions, classical IDM-type models are built on a deterministic assumption: they posit a one-to-one mapping between the state of the driver-vehicle system (e.g., speed, spacing, and relative speed) and the driver's acceleration response. However, real-world driving is inherently stochastic and context-sensitive, often exhibiting one-to-many mappings. That is, the same traffic state may correspond to multiple plausible acceleration responses, depending on the driver’s latent intention or situational interpretation. Conversely, the same observed action may result from distinct underlying causes. For instance, a driver approaching a slower vehicle may choose to decelerate gently, maintain speed momentarily, or accelerate to change lanes, each being a valid response under the same traffic conditions but reflecting different behavioral modes. Similarly, a small acceleration might reflect a relaxed adjustment in free-flow, a hesitant reaction in uncertain conditions, or a defensive maneuver in dense congestion. This behavioral ambiguity is especially prominent in naturalistic data, where only a small fraction of actions are clearly purposeful or reactive; the majority occur in ambiguous or transitional states \citep{zhang2023interactive}. When such data are used to calibrate deterministic models via root mean squared error (RMSE) or Gaussian likelihoods---metrics that treat all data points as equally informative, the resulting model tends to regress toward the mean. This leads to an ``averaged'' behavior that fails to reproduce the variability and sharp transitions observed in real driving. Consequently, such models suffer from non-identifiability: multiple parameter settings may explain the data equally well, yet lack meaningful behavioral interpretation \citep{zhang2024bayesian}. This compromises both the interpretability and fidelity of driver modeling, especially in downstream applications such as behavior prediction, risk estimation, and simulation-based safety evaluation.

To address this challenge, it is essential to adopt a regime-switching scheme that recognizes driving as a composition of context-dependent behavioral modes. By segmenting the driving process into discrete regimes, each governed by its own interpretable set of behavioral parameters, such a framework allows for a one-to-many mapping from observed data to latent driving intentions and contexts. This structure enables the model to assign ambiguous observations to the most plausible regime given the surrounding traffic conditions, rather than fitting a single, fixed response. As a result, regime-switching models mitigate the tendency toward averaged behavior, preserve sharp behavioral transitions, and enhance both interpretability and predictive consistency.

To operationalize this regime-switching perspective, we first develop a hybrid probabilistic model, HMM-IDM, by integrating the classical Intelligent Driver Model with a Hidden Markov Model (HMM) \citep{rabiner1986introduction}. In this formulation, each latent state corresponds to a distinct driving regime, characterized by its own set of IDM parameters. The model captures how drivers dynamically transition among regimes such as aggressive acceleration, cruising, and deceleration, thereby accommodating temporal variability and regime-dependent responses. To further disentangle the influence of intrinsic behavioral modes from external driving contexts, we extend this model to a Factorial Hidden Markov Model \citep{ghahramani1995factorial} with IDM dynamics, namely, FHMM-IDM. FHMM-IDM introduces a structured latent state space with two independent components: the driving regime process, which encodes internal behavioral intent, and the traffic scenario process, which represents surrounding contextual conditions such as free-flow, congestion, or stop-and-go dynamics. Each joint latent state governs a separate set of IDM parameters and gives rise to distinct acceleration behaviors depending on both driver Regime and environmental context. This factorization not only improves behavioral interpretability but also enhances the model’s capacity to reflect real-world variability in a principled and data-driven manner. 
We validate the proposed framework using the HighD naturalistic driving dataset, demonstrating that FHMM-IDM effectively uncovers interpretable behavioral structures and captures realistic regime-switching patterns. Detailed case studies show how the model successfully disentangles intrinsic driver behaviors from contextual traffic variations, providing a richer and more faithful representation of human driving for simulation, prediction, and behavior analysis tasks.

\begin{figure*}[!t]
    \centering
    \subfigure{\centering
    \resizebox{0.9\textwidth}{!}{\begin{tikzpicture}[scale=1]

\draw[fill=green!10,draw=black!0] (0,0) rectangle (2.4,4);
\draw[fill=yellow!10,draw=black!0] (2.4,0) rectangle (5.5,4);
\draw[fill=blue!7,draw=black!0] (5.5,0) rectangle (9.7,4);
\draw[fill=black!10,draw=black!0] (9.7,0) rectangle (12,4);

\draw[fill=orange!30,draw=black!0] (0,-1.5) rectangle (4,0);
\draw[fill=Melon!60,draw=black!0] (4,-1.5) rectangle (8,0);
\draw[fill=ForestGreen!30,draw=black!0] (8,-1.5) rectangle (12,0);

% Draw axes
\draw[thick,->] (0,0) -- (12.5,0) node[right] {Time};
\draw[thick,->] (0,0) -- (0,4.5) node[above] {Driver Response};
\draw[thick,->] (0,0) -- (0,-2.) node[below] {Traffic Condition};

% Draw global nonlinear function
\draw[thick, blue!60, line width=.6mm, domain=0.2:12, samples=100] plot (\x,{sin(\x*0.8 r)+0.15*\x+0.37}) 
node[right] {}; % Real driver behavior

% Local nonlinear pieces (driving primitives)
\draw[red, thick, line width=.4mm, domain=0.2:2.4] plot (\x,{0.643*\x+0.371}) node[above] {};
\draw[green, thick, dotted, line width=.4mm,domain=2.4:4] plot (\x,{0.643*\x+0.371}) node[above] {};

\draw[red, thick, line width=.4mm,  domain=2.4:5.5] plot (\x,{-.574*\x +3.275}) node[above] {};
\draw[yellow, thick, dotted, line width=.4mm,domain=.5:2.4] plot (\x,{-.574*\x +3.275}) node[above] {};
% \draw[yellow, thick, dotted, line width=.4mm,domain=5.5:6] plot (\x,{-.574*\x +3.275}) node[above] {};

\draw[red, thick, line width=.4mm, domain=5.5:9.7] plot (\x,{0.65*\x-3.45}) node[above] {};
% \draw[blue!70, thick, dotted, line width=.4mm,domain=1.:5] plot (\x,{0.65*\x-3.45}) node[above] {};
\draw[blue!70, thick, dotted, line width=.4mm,domain=9.7:10.37] plot (\x,{0.65*\x-3.45}) node[above] {};

\draw[red, thick, line width=.4mm, domain=9.7:12] plot (\x,{-.3*\x+5.78}) node[above] {};
\draw[black!50, thick, dotted, line width=.4mm,domain=8.7:9.7] plot (\x,{-.3*\x+5.78}) node[above] {};

% Dotted vertical lines to show segmentation
\foreach \x in {2.4,5.5,9.7}{
    \draw[dashed,gray] (\x,0) -- (\x,4);
}
\foreach \x in {4,8}{
    \draw[dashed,gray] (\x,0) -- (\x,-1.5);
}

% Nodes indicating driving primitives
% Define nodes for primitives
\node[align=center] (P1) at (1.2,3.5) {\texttt{Regime A}};
\node[align=center] (P2) at (3.9,3.5) {\texttt{Regime B}};
\node[align=center] (P3) at (7.6,3.5) {\texttt{Regime C}};
\node[align=center] (P4) at (11,3.5) {\texttt{Regime D}};

\node[align=center] (S1) at (2.2,-.75) {\texttt{Scenario A}};
\node[align=center] (S2) at (6.2,-.75) {\texttt{Scenario B}};
\node[align=center] (S3) at (10.2,-.75) {\texttt{Scenario C}};

% Draw Markov transition arrows between states
% \path[->, thick, bend right=30, color=gray] (P1.north) edge [bend left] node {} (P2.north);
% \path[->, thick, bend right=30, color=gray] (P2.north) edge [bend left] node {} (P3.north);
% \path[->, thick, bend right=30, color=gray] (P3.north) edge [bend left] node {} (P4.north);

% \path[->, thick, bend left=30, color=gray] (S1.south) edge [bend right] node {} (S2.south);
% \path[->, thick, bend left=30, color=gray] (S2.south) edge [bend right] node {} (S3.south);
% --- Uniformly distributed Markov transitions (Regimes) ---
\foreach \x in {0,1,2,3,4,5,6,7,8,9,10,11} {
    \path[->, thick, bend right=30, color=gray!70]
        (\x,4.) edge[bend left] (\x+1,4.);
}

% --- Uniformly distributed Markov transitions (Scenarios) ---
\foreach \x in {0,1,2,3,4,5,6,7,8,9,10,11} {
    \path[->, thick, bend left=30, color=gray!70]
        (\x,-1.5) edge[bend right] (\x+1,-1.5);
}

\foreach \x in {1,2,3,4,5,6,7,8,9,10,11,12} {
    \draw[thick] (\x,-0.037) -- (\x,0.037);
}

% % Optional: Add transition probabilities (example values)
% \node [color=gray] at (2.5,4.4) {$p_{12}$};
% \node [color=gray] at (5.2,4.4) {$p_{23}$};
% \node [color=gray] at (8.3,4.5) {$p_{34}$};
\node [color=gray] at (5.9,4.4) {Markov Switching};
\node [color=gray] at (5.9,-2.) {Markov Switching};

% Legend
\draw[blue,thick, line width=.6mm] (13.,1.1)--(12.4,1.1);
\node[right] at (13.2,1.1) {Driving behaviors};
\draw[red,thick, line width=.4mm] (13.,0.6)--(12.4,0.6);
\node[right] at (13.2,0.6) {Regime-specific models};

\end{tikzpicture}}
    }
    \caption{Conceptual illustration of the FHMM-IDM model. The top panel depicts the evolution of driver response (e.g., acceleration), segmented into discrete latent driving regimes, while the bottom panel shows the corresponding traffic scenarios. Both latent processes evolve via Markov switching. The blue curve represents the observed behavioral trajectory, and the red straight lines show the model’s output (linear models for illustration purposes) within each driving regime using regime-specific models. Dotted lines indicate latent behavioral trends or variability beyond what each individual regime can capture, motivating the need for switching between multiple regimes. Note that \texttt{Regime A}–\texttt{D} and \texttt{Scenario A}–\texttt{C} are illustrative placeholders only; the data-driven regimes and scenarios inferred by the model appear in Figs.~\ref{fig:case2}-\ref{fig:case9}.}
    \label{fig:markov_process}
\end{figure*}
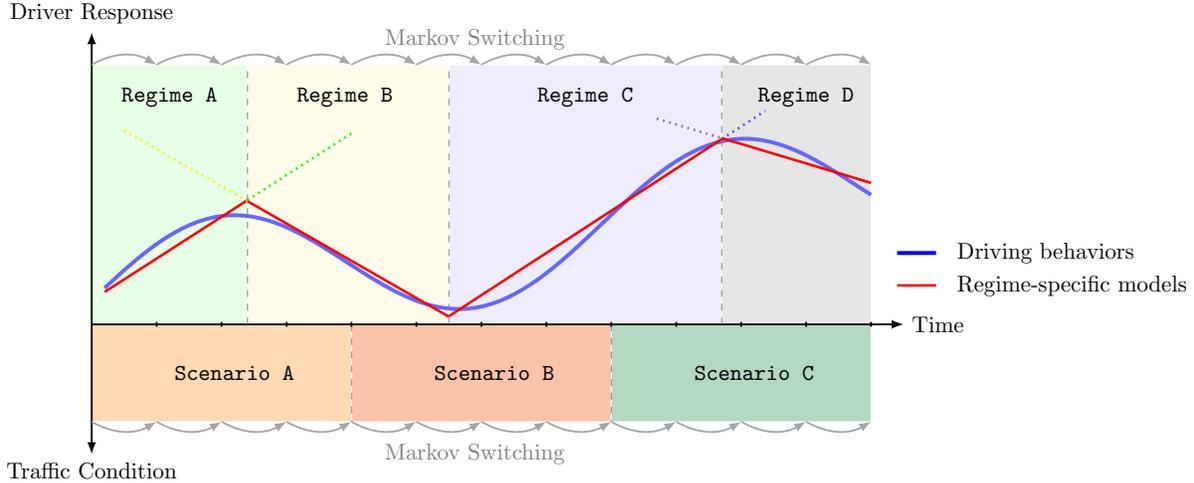

Conceptually, both the HMM-IDM and FHMM-IDM hybrid approaches embody a strategy of assembling multiple driving primitives\footnote{In the literature, driving primitives refer to fundamental, granular units of driving activity that serve as the building blocks of more complex behavior \citep{zhang2021spatiotemporal}. In other words, they are the basic maneuver elements or patterns that drivers execute, which can be sequenced and combined to produce higher-level driving strategies.} to approximate globally nonlinear driving behavior. Rather than relying on a single, overly complex model to capture all behavioral variability, these models decompose the driving process into a collection of simpler and granular components, i.e., driving regimes. This is analogous to approximating a nonlinear function using multiple local linear segments. Each regime-specific IDM instance corresponds to a driving primitive that governs the vehicle’s response under a particular regime, such as aggressive car-following, relaxed cruising, or defensive braking, conditioned on specific traffic scenarios.
The probabilistic structure of the HMM or FHMM governs transitions among these primitives, enabling the model to respond to evolving conditions by switching between regimes. The driver’s overall behavior is thus modeled as a piecewise sequence, where each segment reflects the output of a distinct IDM parameterization determined by the current latent state. This structure is illustrated in Fig.~\ref{fig:markov_process}, where the driver’s observed response trajectory (top) is interpreted as the outcome of latent driving regimes and traffic scenarios, each evolving over time via independent Markov chains. The blue curve represents the ground-truth driving action (e.g., acceleration), while the red curve shows the regime-specific model output (linear models for illustration purposes). Dashed lines indicate latent behavioral trends not captured by any single primitive, further motivating the need for switching among specialized regimes. Shaded regions delineate the segmentation imposed by the latent states, revealing how the model adaptively partitions the trajectory into interpretable behavioral modes and contextual scenarios.
This modular scheme captures both the stochastic and adaptive nature of real-world driving. As traffic conditions evolve, the model dynamically adjusts its active primitive, for example, transitioning from free-flow to stop-and-go conditions. By combining interpretable latent states with data-driven transitions, the HMM-IDM and FHMM-IDM frameworks provide a flexible yet structured approach to modeling human driving behavior with both realism and transparency.

This work makes the following key contributions and advantages:
\begin{enumerate}
    \item \textbf{A novel modeling framework:} We introduce a Markov regime-switching framework for car-following behavior that explicitly separates intrinsic driving regimes from external traffic scenarios. This addresses the long-standing challenge of one-to-many mappings in naturalistic data, providing a principled solution to the problem of behavioral non-identifiability in deterministic models.

    \item \textbf{A hybrid probabilistic model with rigorous inference procedure:} We instantiate the framework through FHMM-IDM, a novel integration of the FHMM with the IDM. In FHMM-IDM, each latent driving regime corresponds to a unique set of IDM parameters, while the factorial structure captures the interplay between driver intention and traffic context via two independent latent Markov processes. We develop a full Bayesian inference pipeline using MCMC methods, ensuring robust parameter calibration and uncertainty quantification from real-world trajectory data.

    \item \textbf{An interpretable and modular representation:} By disentangling behavioral and contextual components, our model enables interpretable attribution of driving behavior to internal (driver) and external (traffic) factors. This decomposition facilitates regime-aware analysis and enhances the explanatory power of car-following models.
    Empirical results on the HighD dataset show that FHMM-IDM uncovers meaningful regime structures and realistically captures dynamic transitions across driving behaviors and traffic scenarios.
\end{enumerate}

The remainder of this paper is organized as follows: Section~\ref{literature} reviews related work on probabilistic modeling of car-following behavior. Section~\ref{method} introduces the proposed HMM-IDM and FHMM-IDM frameworks, including their mathematical formulation and Bayesian inference algorithms. Section~\ref{experiment} describes the experimental setup, presents the learned interpretable latent states, and provides case studies using the HighD dataset to validate the effectiveness of the models. Finally, Section~\ref{conclusion} concludes with discussions and outlines potential directions for future research.

\section{RELATED WORKS}\label{literature}

\subsection{Probabilistic Models and Behavioral Regimes}

Deterministic car-following models such as the IDM \citep{treiber2000congested} assume a single fixed set of parameters governing driver behavior in all scenarios. This restricts their ability to capture the variability, uncertainty, and abrupt regime changes present in real-world driving \citep{zhang2024bayesian, chen2024combining}. Notably, classic models like Wiedemann and Fritzsche hard-code regime boundaries via perceptual thresholds, yielding a multi-regime structure but requiring extensive manual tuning and lacking adaptability beyond their original calibration context \citep{wiedemann1974simulation, fritzsche1994model}. As a result, these deterministic and threshold-based models tend to underfit behavioral heterogeneity, struggle to model transitions, and suffer from limited interpretability in heterogeneous or context-dependent traffic.

To address these limitations, probabilistic modeling approaches have emerged, treating driving as a stochastic process and enabling the discovery of latent behavioral regimes. HMMs have become foundational in this context \citep{rabiner1986introduction, wang2014modeling}, as they encode both latent driver states and the transitions between them. HMMs enable modeling of short-term regimes, such as aggressive acceleration, steady cruising, or cautious braking, as latent states, naturally accommodating regime shifts and sequential dependencies \citep{vaitkus2014driving, aoude2012driver, gadepally2013framework}. Gaussian Mixture Models (GMMs) have also been adopted, e.g., with delay-embedding \citep{chen2023bayesian} and matrix decomposition \citep{zhang2024learning} to capture multi-modal distributions of driver behavior. It also could be set as emission models for HMMs \citep{wang2018learning}. These probabilistic frameworks, by maintaining distributions over regimes or actions rather than deterministic assignments, increase model flexibility and better reflect the stochastic nature of human driving.

However, most prior work in regime modeling has relied on domain knowledge or heuristic thresholds to define the behavioral regimes themselves, limiting generalizability and transferability \citep{wang2014modeling, vaitkus2014driving}. There remains a need for data-driven methods that can discover and adaptively segment regimes without manual intervention.

\subsection{Advances in Bayesian and Factorial Approaches}

Building on basic HMM and GMM models, Bayesian extensions have been developed to better represent behavioral complexity and uncertainty. One notable extension is the Hidden Semi-Markov Model (HSMM), which explicitly models the dwell time (state duration) in each regime. Standard HMMs assume geometric state durations, which may not reflect how long drivers naturally stay in a given behavior. HSMMs address this by providing a state-specific duration distribution. For example, \citet{taniguchi2014unsupervised} employed an HSMM with a Hierarchical Dirichlet Process (HDP) prior, allowing the model to learn both the duration of maneuvers and the appropriate number of distinct behavioral states from the data. Such a nonparametric Bayesian HMM (using HDP) does not require the researcher to pre-specify the number of driving regimes; instead, the model infers it automatically \citep{fox2009bayesian}. This is especially useful when the set of driving patterns is not known in advance or varies between drivers. \citet{zhang2021spatiotemporal} demonstrated the power of this approach by applying a sticky HDP-HMM to naturalistic driving data, which automatically discovered recurrent interaction patterns (i.e., primitive maneuvers) without any pre-defined labels. This represents a significant advance over earlier HMM studies that assumed a fixed set of driver modes, as the model could flexibly reveal new regime types (and their durations) directly from complex multi-vehicle datasets.

FHMMs \citep{ghahramani1995factorial} further increase modeling expressiveness by combining multiple interacting latent processes, for example, one chain for the driver's intrinsic regime and another for the surrounding traffic scenario. This structure enables the model to disentangle overlapping influences, capturing cases where, for example, a usually relaxed driver becomes aggressive due to external congestion. Though FHMMs remain underutilized in the driving literature, their capability to separate internal and external factors aligns with the motivation for our proposed approach. Bayesian inference methods (e.g., Expectation-Maximization, MCMC) are commonly used to estimate parameters and latent trajectories, providing uncertainty quantification and adaptivity as new data is observed \citep{bishop2006pattern}.

\subsection{Regime-Switching Car-Following Models}

Within car-following modeling, regime-switching has traditionally been implemented through deterministic if-then rules or fixed thresholds, as in multi-regime Wiedemann- or Fritzsche-type models \citep{wiedemann1974simulation, fritzsche1994model}. More recently, data-driven regime-switching has been integrated with car-following models using probabilistic frameworks. For instance, \citet{zaky2015car} proposed a two-stage Markov switching model to classify car-following regimes and estimate regime-specific parameters, allowing for the dynamic detection of abnormal or rare events and more precise behavioral segmentation. Similarly, \citet{zou2022multivariate} applied HMM-based models (including GMM-HMM and HDP-HSMM) to large-scale car-following data, showing that flexible, nonparametric models can automatically identify meaningful regimes (e.g., close following, reactive braking) without manual regime definitions. Recent works, such as \citet{zhang2023interactive}, also integrated IDM with regime-switching frameworks. It proposes distinct action-oriented driving regimes (e.g., interactive/non-interactive driving), with regime transitions governed by an interactive-switching control module. Each regime is characterized by unique IDM parameterizations, allowing the model to dynamically adapt to varying interactive intentions and traffic contexts, significantly improving model fidelity and interpretability. Recent studies have also begun to incorporate regime-switching ideas into deep learning frameworks. Recent advances have also introduced hybrid deep learning frameworks that incorporate discrete regime-switching into car-following prediction. For instance, \citet{zhou2025driving} proposed a regime-embedded architecture that combines Gated Recurrent Units (GRUs) for driving regime classification with Long Short-Term Memory (LSTM) networks for continuous kinematic prediction. Their model targets intra-driver heterogeneity by integrating discrete behavioral modes (e.g., acceleration, cruising, steady-state following) into continuous trajectory forecasting, achieving substantial gains in predictive accuracy. However, such models rely on pre-segmented regime labels and deep architectures that, while powerful, lack the principled probabilistic structure and interpretability.

Despite these advances, most existing approaches still require manual regime boundaries, external calibration, or multi-step procedures. Our work bridges this gap by embedding a Markov switching process directly within the IDM framework, enabling the model to discover, segment, and calibrate regimes in a unified and data-driven manner. This approach is motivated by and extends the probabilistic regime-switching and Bayesian learning literature, aiming to achieve greater realism, interpretability, and context-awareness in microscopic traffic simulation.

\begin{table}[t]
\centering
\caption{{Comparison of related approaches for modeling driver behavior.}}
\resizebox{\textwidth}{!}{
\begin{tabular}{lcccccccc}
\toprule
\textbf{Feature/Model} & \textbf{IDM} & \textbf{Bayesian IDM} & \textbf{GMM} &\textbf{HMM} & \textbf{HMM-GMM} & \textbf{HDP-HMM} & \textbf{NN (LSTM)} & \textbf{FHMM-IDM (Ours)} \\
\midrule
Model Type & Deterministic  & Probabilistic & Probabilistic & Probabilistic & Probabilistic & Probabilistic & Deep Learning & Probabilistic \\
Adaptivity$^1$ & \xmark & \xmark & \xmark & \cmark & \cmark & \cmark & \cmark & \cmark \\
Latent Behavior Type$^2$ & \xmark & \xmark & Discrete & Discrete & Discrete & Discrete & Continuous & Discrete \\
Latent Mode Cardinality$^3$ & - & - & Fixed & Fixed & Fixed & Infinite & Fixed & Factorial Fixed \\
Stochasticity & \xmark & \cmark & \cmark & \cmark & \cmark & \cmark & Implicit & \cmark \\
Parameter Estimation$^4$ & Heuristic & MCMC & EM/MCMC & EM/MCMC & EM/MCMC & EM/MCMC & Gradient descent & MCMC \\
Interpretability$^5$ & High & High & Moderate & Moderate & Moderate & Moderate & Low & High \\
Traffic Context Modeling$^6$ & \xmark & \xmark & \cmark (features) & \xmark & \cmark (features) & \cmark (implicit) & \cmark (learned) & \cmark (explicit) \\
Heterogeneity Handling$^7$ & Poor & Moderate & Moderate & Moderate & Moderate & Excellent & Excellent & Excellent \\
Data-driven Flexibility$^8$ & Low & Moderate & Moderate & Moderate & Moderate & High & High & High \\
Training Complexity$^9$ & Low & Moderate & Low & Moderate & Moderate/High & High & High & High \\
\midrule 
\multicolumn{9}{p{1.5\textwidth}}{\textbf{IDM}: \citet{treiber2000congested, treiber2003memory, treiber2006delays, punzo2021calibration}; \textbf{Bayesian IDM}: \citet{zhang2024bayesian, zhang2024calibrating}; \textbf{GMM}: \citet{chen2023bayesian, zhang2023interactive, zhang2024learning}; \textbf{HMM}: \citet{sathyanarayana2008driver, aoude2012driver, gadepally2013framework, vaitkus2014driving}; \textbf{HMM-GMM}: \citet{wang2018learning, wang2018driving}; \textbf{HDP-HMM}: \citet{taniguchi2014unsupervised, zhang2021spatiotemporal, zou2022multivariate}; \textbf{Neural Networks}: \citet{wang2017capturing, zhu2018human, mo2021physics, yao2025novel, zhou2025driving};}\\
\bottomrule
\end{tabular}}
\raggedright
\footnotesize

$^1$\textit{Can the model dynamically adjust to changing behavior?}\\
$^2$\textit{Type of latent representation: discrete (mode switches) or continuous (trajectory embeddings).}\\
$^3$\textit{Whether the number of latent modes is fixed a priori or inferred.}\\
$^4$\textit{How model parameters are estimated: EM, gradient descent, MCMC, etc.}\\
$^5$\textit{Can latent states or parameters be interpreted as meaningful driving behavior?}\\
$^6$\textit{Whether traffic context (e.g., relative speed, gap) is explicitly used in latent modeling.}\\
$^7$\textit{Ability to capture driver-specific variation (e.g., hierarchical priors, class mixture).}\\
$^8$\textit{Model’s ability to fit and learn from diverse and high-dimensional driving datasets.}\\
$^9$\textit{Overall training/inference complexity: data requirements, convergence cost, parallelism.}\\
\label{tab:model_comparison}
\end{table}

\subsection{Positioning FHMM-IDM among Existing Methods}

To situate our proposed FHMM-IDM framework within the broader spectrum of driver behavior modeling approaches, we summarize and compare representative methods in Table~\ref{tab:model_comparison}. The comparison spans classical deterministic models (e.g., IDM), probabilistic and Bayesian models (e.g., GMM, HMM, HDP-HMM), and more recent learning-based techniques (e.g., LSTM-based deep models), across key modeling characteristics. These include adaptivity, behavioral mode representation, latent state dimensionality, stochasticity, estimation procedures, interpretability, contextual awareness, heterogeneity modeling, and computational complexity.

FHMM-IDM distinguishes itself by explicitly modeling both internal driving regimes and external traffic scenarios through a factorial latent structure. This design allows it to disentangle driver intent from environmental influences---a capability absent in most existing approaches, which either assume a fixed parameterization or rely on indirect context encoding through observed features. Moreover, by adopting a full Bayesian inference framework, FHMM-IDM enables robust parameter estimation and principled uncertainty quantification, which are critical for applications such as behavior prediction, risk assessment, and safety validation.

Compared to existing models, FHMM-IDM strikes a balance between data-driven flexibility and structured interpretability. While deep learning models can learn complex patterns, they often lack transparency and require large-scale training data. In contrast, FHMM-IDM offers interpretable, probabilistically grounded behavioral components that can generalize across scenarios with limited data. This makes it a strong candidate for modeling realistic and context-sensitive driving behaviors in naturalistic traffic environments.

\section{MARKOV REGIME-SWITCHING FRAMEWORK FOR CAR-FOLLOWING}\label{method}
Building on the motivation to model heterogeneous and context-dependent driving behaviors, we develop a probabilistic regime-switching framework that captures the interplay between intrinsic driver actions and external traffic scenarios. Our approach introduces two hybrid models: HMM-IDM and FHMM-IDM, which augment the classical IDM with latent Markovian dynamics. The HMM-IDM captures univariate regime-switching behaviors by associating each latent state with a distinct set of IDM parameters. To further disentangle intrinsic behavioral variability from environmental context, we extend this formulation to a factorial structure, FHMM-IDM, wherein two independent latent Markov chains separately encode driving behaviors and traffic scenarios. This section presents the mathematical formulation, model assumptions, and Bayesian inference procedures used to estimate the latent states and regime-specific parameters.

\subsection{Formulations of HMM-IDM and FHMM-IDM}
\subsubsection{Hidden Markov Model with Intelligent Driver Model (HMM-IDM)}
As discussed in Section~\ref{introduction}, modeling car-following behavior with a fixed parameter set, such as in the deterministic IDM, fails to account for the contextual variability and temporal shifts observed in naturalistic driving. The same input state (e.g., gap, speed, relative speed) can lead to different driver actions depending on latent factors such as intention, caution level, or situational awareness. This ambiguity, or one-to-many mapping, motivates the need for a regime-switching framework that allows behavioral parameters to evolve over time. 

To address this, here we develop a hybrid model that combines the interpretability of IDM with the temporal segmentation power of HMM. The HMM introduces a discrete latent state variable that captures shifts in driving regimes, such as transitions between cruising, closing in, or defensive braking. Each latent state is associated with a distinct set of IDM parameters, enabling the model to account for time-varying human driving behavior while maintaining a physically grounded formulation.

\begin{figure}[!t]
    \centering
    \subfigure[HMM-IDM.]{\centering
    \resizebox{!}{5.2cm}{\usetikzlibrary{positioning,arrows.meta,quotes}
\usetikzlibrary{shapes,snakes}
\usetikzlibrary{bayesnet}
\tikzset{>=latex}
\tikzstyle{plate caption} = [caption, node distance=0, inner sep=0pt,
below left=5pt and 0pt of #1.south]
\tikzset{every picture/.style={line width=0.75pt}} %set default line width to 0.75pt        

% Define layers
\pgfdeclarelayer{background}
\pgfsetlayers{background,main}

\begin{tikzpicture}   

    % Define nodes
    % \node [circle,draw=black,inner sep=0pt,minimum size=0.8cm] (z0) at (0, 3.2) {$z_0^{(d)}$};

    % \node [obs,draw=black,inner sep=0pt,minimum size=0.8cm] (u1) at (0, 3.6) {$u_1^{(d)}$};
    % \node [obs,draw=black,inner sep=0pt,minimum size=0.8cm] (u2) at (1.8, 3.6) {$u_2^{(d)}$};
    % \node [const,inner sep=0pt,minimum size=0.8cm] (u3) at (3, 3.6) {$\cdots$};
    % \node [obs,draw=black,inner sep=0pt,minimum size=0.8cm] (uT) at (4.5, 3.6) {$u_{T_d}^{(d)}$};
    
    \node [circle,draw=black,inner sep=0pt,minimum size=0.8cm, fill=white] (z1) at (0, 1.8) {$z_1^{(d)}$};
    \node [circle,draw=black,inner sep=0pt,minimum size=0.8cm, fill=white] (z2) at (1.8, 1.8) {$z_2^{(d)}$};
    \node [const,inner sep=0pt,minimum size=0.8cm] (z3) at (3, 1.8) {$\cdots$};
    \node [circle,draw=black,inner sep=0pt,minimum size=0.8cm, fill=white] (zT) at (4.5, 1.8) {$z_{T_d}^{(d)}$};
    
    \node [obs,draw=black,inner sep=0pt,minimum size=0.8cm] (y1) at (0,0) {$y_1^{(d)}$};
    \node [obs,draw=black,inner sep=0pt,minimum size=0.8cm] (y2) at (1.8, 0) {$y_2^{(d)}$};
    \node [const,inner sep=0pt,minimum size=0.8cm] (y3) at (3, 0) {$\cdots$};
    \node [obs,draw=black,inner sep=0pt,minimum size=0.8cm] (yT) at (4.5, 0) {$y_{T_d}^{(d)}$};

    \node [obs,draw=black,inner sep=0pt,minimum size=0.8cm] (x1) at (0,-1.8) {$\boldsymbol{x}_1^{(d)}$};
    \node [obs,draw=black,inner sep=0pt,minimum size=0.8cm] (x2) at (1.8, -1.8) {$\boldsymbol{x}_2^{(d)}$};
    \node [const,inner sep=0pt,minimum size=0.8cm] (x3) at (3, -1.8) {$\cdots$};
    \node [obs,draw=black,inner sep=0pt,minimum size=0.8cm] (xT) at (4.5, -1.8) {$\boldsymbol{x}_{T_d}^{(d)}$};
    
    \node [circle,draw=black,inner sep=0pt,minimum size=0.8cm, fill=white] (theta_k) at (-2, 0.6) {$\boldsymbol{\theta}_k$};

    \node [circle,draw=black,inner sep=0pt,minimum size=0.8cm, fill=white] (pi) at (-2, 2.) {$\boldsymbol{\pi}_k$};

    \node [circle,draw=black,inner sep=0pt,minimum size=0.8cm, fill=white] (sigma) at (-2, -.6) {${\sigma}_k$};

    \node [circle,draw=black,inner sep=0pt,minimum size=0.8cm, fill=white] (theta) at (-4.5, 1.5) {$\boldsymbol{\theta}$};

    \node [circle,draw=black,inner sep=0pt,minimum size=0.8cm, fill=white] (Lambda) at (-4.5, -.3) {$\boldsymbol{\Lambda}$};
    
    % Connect the nodes
    % \edge {z0} {z1};
    \edge {z1} {z2};
    \edge {z2} {z3};
    \edge {z3} {zT};
    \edge {z1} {y1};
    \edge {z2} {y2};
    \edge {zT} {yT};
    \edge [color=blue] {x1} {y1};
    \edge [color=blue] {x2} {y2};
    \edge [color=blue] {xT} {yT};
    % \edge {theta} {y1,y2,yT};
    \edge {theta} {theta_k};
    \edge {Lambda} {theta_k};
    \edge {Lambda} {theta};

    % \edge [color=red] {u1} {z1};
    % \edge [color=red] {u2} {z2};
    % \edge [color=red] {uT} {zT};
    
    \path [draw,color=blue,->] (theta_k) edge [bend left] node {} (y1);
    \path [draw,color=blue,->] (theta_k) edge [bend left] node {} (y2);
    \path [draw,color=blue,->] (theta_k) edge [bend left] node {} (yT);

    \path [draw,color=blue,->] (sigma) edge [bend right] node {} (y1);
    \path [draw,color=blue,->] (sigma) edge [bend right] node {} (y2);
    \path [draw,color=blue,->] (sigma) edge [bend right] node {} (yT);

    \path [draw,color=black,->] (pi) edge [bend left] node {} (z1);
    \path [draw,color=black,->] (pi) edge [bend left] node {} (z2);
    \path [draw,color=black,->] (pi) edge [bend left] node {} (zT);

    % Plates
    \node [text width=2.3cm] (k1) at (-2, -1.4) {\small{$k\in\{1,\dots,K\}$}};
    \node [text width=2.3cm] (k2) at (3.9, 2.95) {\small{$d\in\{1,\dots,D\}$}};

    \begin{pgfonlayer}{background} 
        % \plate [color=black,inner sep=0cm,xshift=-0.05cm,yshift=0.2cm] {part1} {(theta_k)(k1)(pi)(sigma)} {};
        \node [fit=(theta_k)(k1)(pi)(sigma), draw,inner sep=0.05cm,xshift=-0.1cm,yshift=0.05cm, fill=Peach!50, fill opacity=0.25, behind path, rounded corners] {};
    
        % \plate [color=black,inner sep=-0.05cm,xshift=-0.15cm,yshift=0.1cm] {part2} {(z1)(xT)(k2)} {}; % (u1)
        \node [fit=(z1)(xT)(k2), draw, inner sep=0.05cm, xshift=-0.15cm, yshift=-0.05cm, fill=gray!30, fill opacity=0.1, behind path, rounded corners] {};
    \end{pgfonlayer}
    
\end{tikzpicture}}
    }
    \subfigure[FHMM-IDM.]{\centering
    \resizebox{!}{6.5cm}{\usetikzlibrary{positioning,arrows.meta,quotes}
\usetikzlibrary{shapes,snakes}
\usetikzlibrary{bayesnet}
\tikzset{>=latex}
\tikzstyle{plate caption} = [caption, node distance=0, inner sep=0pt,
below left=5pt and 0pt of #1.south]
\tikzset{every picture/.style={line width=0.75pt}} %set default line width to 0.75pt       
% Define layers
\pgfdeclarelayer{background}
\pgfsetlayers{background,main}

\begin{tikzpicture}

    % Define nodes
    \node [circle,draw=black,inner sep=0pt,minimum size=0.8cm, fill=white] (z1) at (0, 1.8) {\small$\boldsymbol{z}_1^{(d)}$};
    \node [circle,draw=black,inner sep=0pt,minimum size=0.8cm, fill=white] (z2) at (1.8, 1.8) {\small$\boldsymbol{z}_2^{(d)}$};
    \node [const,inner sep=0pt,minimum size=0.8cm] (z3) at (3, 1.8) {$\cdots$};
    \node [circle,draw=black,inner sep=0pt,minimum size=0.8cm, fill=white] (zT) at (4.5, 1.8) {\small$\boldsymbol{z}_{T_d}^{(d)}$};
    
    \node [obs,draw=black,inner sep=0pt,minimum size=0.8cm] (y1) at (0,0) {$y_1^{(d)}$};
    \node [obs,draw=black,inner sep=0pt,minimum size=0.8cm] (y2) at (1.8, 0) {$y_2^{(d)}$};
    \node [const,inner sep=0pt,minimum size=0.8cm] (y3) at (3.2, 0) {$\cdots$};
    \node [obs,draw=black,inner sep=0pt,minimum size=0.8cm] (yT) at (4.5, 0) {$y_{T_d}^{(d)}$};

    \node [obs,draw=black,inner sep=0pt,minimum size=0.8cm] (x1) at (0,-1.8) {$\boldsymbol{x}_1^{(d)}$};
    \node [obs,draw=black,inner sep=0pt,minimum size=0.8cm] (x2) at (1.8, -1.8) {$\boldsymbol{x}_2^{(d)}$};
    \node [const,inner sep=0pt,minimum size=0.8cm] (x3) at (3.2, -1.8) {$\cdots$};
    \node [obs,draw=black,inner sep=0pt,minimum size=0.8cm] (xT) at (4.5, -1.8) {$\boldsymbol{x}_{T_d}^{(d)}$};

    % \node [circle,draw=black,inner sep=0pt,minimum size=0.8cm, fill=white] (z1S) at (0, -1.8) {\small$\boldsymbol{z}_1^{[\text{S}](d)}$};
    % \node [circle,draw=black,inner sep=0pt,minimum size=0.8cm, fill=white] (z2S) at (1.8, -1.8) {\small$\boldsymbol{z}_2^{[\text{S}](d)}$};
    % \node [const,inner sep=0pt,minimum size=0.8cm] (z3S) at (3, -1.8) {$\cdots$};
    % \node [circle,draw=black,inner sep=0pt,minimum size=0.8cm, fill=white] (zTS) at (4.5, -1.8) {\small$\boldsymbol{z}_{T_d}^{[\text{S}](d)}$};
    
    \node [circle,draw=black,inner sep=0pt,minimum size=0.8cm, scale=1, fill=white] (theta_k) at (-2.55, .6) {$\boldsymbol{\theta}_{{k}^{\text{[B]}}}$};

    \node [circle,draw=black,inner sep=0pt,minimum size=0.8cm,scale=1, fill=white] (sigma) at (-2.55, -.6) {${\sigma}_{{k}^{\text{[B]}}}$};

    \node [circle,draw=black,inner sep=0pt,minimum size=0.8cm, fill=white] (pi) at (-2.55, 2.) {$\boldsymbol{\pi}_{\boldsymbol{k}}$};

    % \node [text width=3cm, color=black] (joint_k) at (-2.3, -1.35) {$\boldsymbol{k}=(k^{[\text{B}]},k^{[\text{S}]})$};

    \node [circle,draw=black,inner sep=0pt,minimum size=0.8cm] (theta) at (-5.1, 1.5) {$\boldsymbol{\theta}$};

    \node [circle,draw=black,inner sep=0pt,minimum size=0.8cm, fill=white] (Lambda) at (-5.1, -.3) {$\boldsymbol{\Lambda}$};

    \node [circle,draw=black,inner sep=0pt,minimum size=0.8cm, fill=white, scale=0.9] (mu_k_S) at (6.9, -1.) {\small$\boldsymbol{\mu}_{\boldsymbol{x}, k^{\text{[S]}}}$};

    \node [circle,draw=black,inner sep=0pt,minimum size=0.8cm, fill=white, scale=0.9] (Sigma_S) at (6.9, -2.5) {\small${\Lambda}_{\boldsymbol{x},k^{\text{[S]}}}$};

    % \node [circle,draw=black,inner sep=0pt,minimum size=0.8cm, fill=white] (pi_S) at (-2.7, -3.9) {$\boldsymbol{\pi}_{k^{\text{[S]}}}^{\text{[S]}}$};

    % Connect the nodes
    % \edge {z0} {z1};
    \edge {z1} {z2};
    \edge {z2} {z3};
    \edge {z3} {zT};
    \edge {z1} {y1};
    \edge {z2} {y2};
    \edge {zT} {yT};

    \edge {theta} {theta_k};
    \edge {Lambda} {theta_k};
    \edge {Lambda} {theta};
    % \edge {z1S} {z2S};
    % \edge {z2S} {z3S};
    % \edge {z3S} {zTS};
    % \edge {z1S} {x1};
    % \edge {z2S} {x2};
    % \edge {zTS} {xT};
    \edge {Sigma_S} {mu_k_S};

    \path [draw,color=blue,->] (theta_k) edge [bend left=20] node {} (y1);
    \path [draw,color=blue,->] (theta_k) edge [bend left=23] node {} (y2);
    \path [draw,color=blue,->] (theta_k) edge [bend left=25] node {} (yT);

    \path [draw,color=blue,->] (sigma) edge [bend right=20] node {} (y1);
    \path [draw,color=blue,->] (sigma) edge [bend right=23] node {} (y2);
    \path [draw,color=blue,->] (sigma) edge [bend right=25] node {} (yT);

    \path [draw,color=black,->] (pi) edge [bend left] node {} (z1);
    \path [draw,color=black,->] (pi) edge [bend left] node {} (z2);
    \path [draw,color=black,->] (pi) edge [bend left] node {} (zT);
    % \path [draw,color=black,->] (pi) edge [bend left=9] node {} (z1S);
    % \path [draw,color=black,->] (pi) edge [bend left=8] node {} (z2S);
    % \path [draw,color=black,->] (pi) edge [bend left=7] node {} (zTS);

    %%%
    \path [draw,color=red,->] (Sigma_S) edge [bend left=20] node {} (x1);
    \path [draw,color=red,->] (Sigma_S) edge [bend left=20] node {} (x2);
    \path [draw,color=red,->] (Sigma_S) edge [bend left=25] node {} (xT);

    \path [draw,color=red,->] (mu_k_S) edge [bend right=20] node {} (x1);
    \path [draw,color=red,->] (mu_k_S) edge [bend right=20] node {} (x2);
    \path [draw,color=red,->] (mu_k_S) edge [bend right=25] node {} (xT);

    \edge [color=blue] {x1} {y1};
    \edge [color=blue] {x2} {y2};
    \edge [color=blue] {xT} {yT};
    % \path [draw,color=blue,->] (x1) edge [bend left] node {} (y1);
    % \path [draw,color=blue,->] (x2) edge [bend left] node {} (y2);
    % \path [draw,color=blue,->] (xT) edge [bend left] node {} (yT);

    \path [draw,color=black,->] (z1) edge [bend left] node {} (x1);
    \path [draw,color=black,->] (z2) edge [bend left] node {} (x2);
    \path [draw,color=black,->] (zT) edge [bend left] node {} (xT);

    % Plates
    \node [text width=3.1cm, color=black] (k1) at (-2.4, -1.35) {\small{$k^{[\text{B}]}\in\{1,\dots,K^{[\text{B}]}\}$}};
    
    \node [text width=2.3cm] (k2) at (3.9, 2.95) {\small{$d\in\{1,\dots,D\}$}};
    
    \node [text width=2.8cm] (k3) at (7.1, -.2) {\small{$k^{[\text{S}]}\in\{1,\dots,K^{[\text{S}]}\}$}};

    \node [text width=3cm, color=black] (k4) at (-2.6, 2.9) {\small{$k^{[\text{S}]}\in\{1,\dots,K^{[\text{S}]}\}$}};

    \begin{pgfonlayer}{background}    
        % \plate [color=black,inner sep=0cm,xshift=-0.1cm,yshift=0cm, fill=Peach!10, fill opacity=0.5] {part1} {(theta_k)(k1)(pi)(sigma)} {};
        \node [fit=(theta_k)(k1)(sigma)(pi), draw, inner sep=0.05cm,xshift=-0.1cm,yshift=0.05cm, fill=Peach!50, fill opacity=0.25, behind path, rounded corners] {};
        
        % \plate [color=black,inner sep=0.cm,xshift=-0.15cm,yshift=0.1cm, fill=green!10, fill opacity=0.5] {part4} {(theta_k)(k4)(sigma)(pi_S)} {};        
        \node [fit=(k4) (pi), draw, inner sep=0.05cm,xshift=-0.25cm,yshift=-0.05cm, fill=RoyalBlue!90, fill opacity=0.1, behind path, rounded corners] {};

        % \plate [color=black,inner sep=0.1cm,xshift=0cm,yshift=0.1cm] {part2} {(z1)(xT)(k2)(zTS)} {}; % (u1)
        \node [fit=(z1)(xT)(k2), draw, inner sep=0.1cm,xshift=-0.05cm,yshift=-0.1cm, fill=gray!30, fill opacity=0.1, behind path, rounded corners] {};
        
        % \plate [color=black,inner sep=-0.03cm,xshift=0cm,yshift=0.1cm] {part3} {(k3)(Sigma_S)} {}; % (u1)
        \node [fit=(k3)(Sigma_S), draw, inner sep=0.cm,xshift=-0.1cm,yshift=-0.05cm, fill=RoyalBlue!90, fill opacity=0.1, behind path, rounded corners] {};
    \end{pgfonlayer}
    
\end{tikzpicture}}
    }
    \caption{Probabilistic graphical model of HMM-IDM and FHMM-IDM.}\label{pgm}
\end{figure}
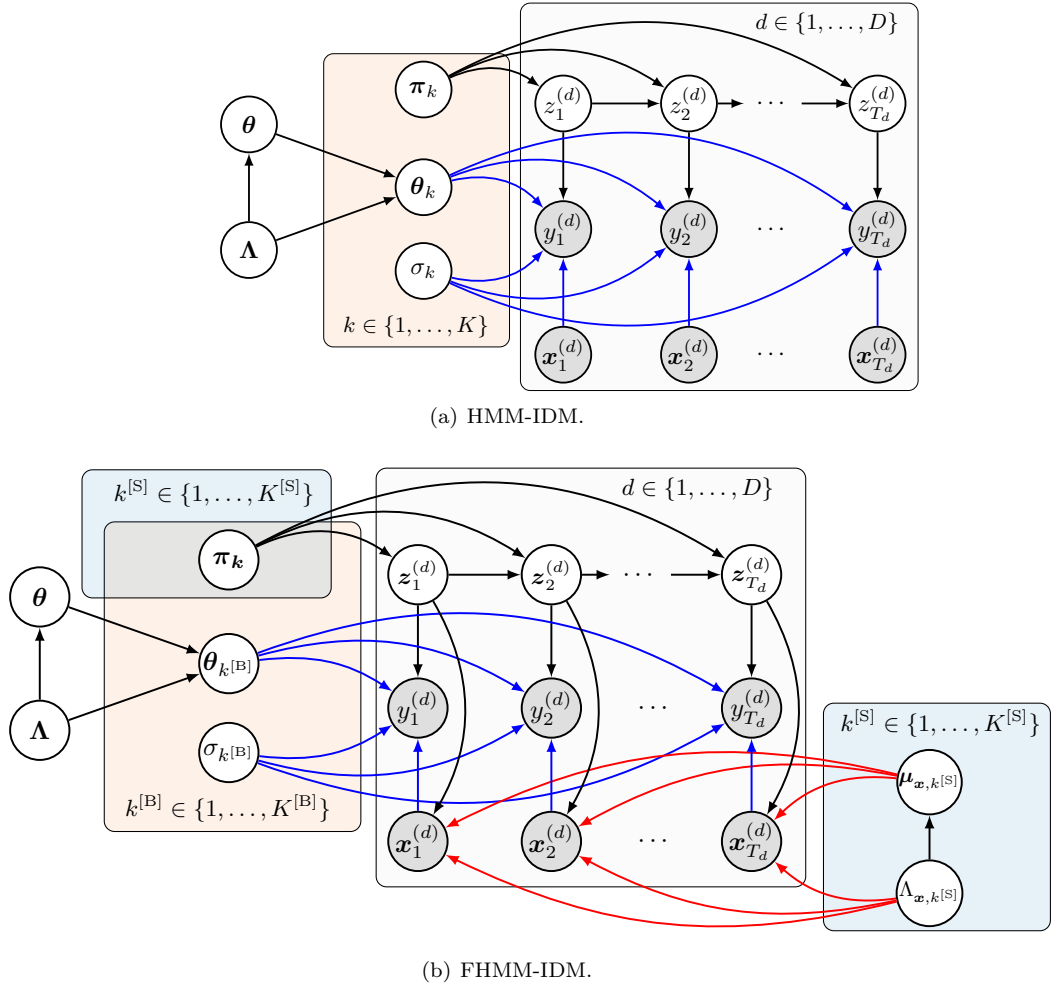

Let $z_{t}^{(d)}$, $\boldsymbol{x}_t^{(d)}$, and $y_t^{(d)}$ denote the latent state, inputs, and outputs for driver $d$ at time $t$, respectively. For simplicity, we omit the superscript $d$ hereafter unless specifically clarified. The two key components of HMM are defined as follows:
\begin{enumerate}
    \item Transition matrix $\boldsymbol{\pi}\in\setR^{K\times K}$: where each entry $\pi_{jk}:=p(z_{t}=k\mid z_{t-1}=j)$ denotes the probability of transitioning from state $j$ to state $k$ according to the Markov property \citep{li2025assessing}. Thus, we express $\boldsymbol{\pi} = [\boldsymbol{\pi}_1,\dots,\boldsymbol{\pi}_K]^\top$.
    \item Local evidence $\boldsymbol{\psi_t}\in\setR^{K}$: the probability of observing $y_{t}$ given the inputs $\boldsymbol{x}_{t}$ and parameters $\boldsymbol{\theta}_{k}$, defined as $\psi_t(k):= p(y_{t}\mid \boldsymbol{x}_{t}, \boldsymbol{\theta}_{k})$.
\end{enumerate}
Formally, the HMM-IDM framework can be summarized by the following equations:
\begin{subequations}
\begin{align}
    z_{t}\mid z_{t-1} &\sim \mathrm{Cat}(\boldsymbol{\pi}_{z_{t-1}}),\\
    y_{t}\mid \boldsymbol{x}_{t}, \boldsymbol{\Theta}, z_{t} &\sim \mathcal{N}(\mathrm{IDM}(\boldsymbol{x}_{t};\boldsymbol{\theta}_{z_{t}}), \sigma^2_{z_{t}}).
\end{align}
\end{subequations}
where $\mathrm{Cat}(\cdot)$ represents the categorical distribution, and $\sigma_{z_{t}}^2$ denotes the variance of the observation noise. Each latent state $z_t$ corresponds uniquely to a driving behavior characterized by specific IDM parameters,  denoted as $\boldsymbol{\theta}_{z_t}$. The complete set of these IDM parameters across all states is indicated by $\boldsymbol{\Theta}=\{\boldsymbol{\theta}_k\}_{k=1}^K$. The overall probabilistic structure of the HMM-IDM model is illustrated in Fig.~\ref{pgm}.

\subsubsection{Factorial Hidden Markov Model with Intelligent Driver Model (FHMM-IDM)}

The FHMM-IDM is an extension of the HMM-IDM framework that incorporates multiple latent processes, referred to as factors. Each factor represents an independent component of driving behaviors, and these factors collectively determine the observed driving behaviors. FHMM-IDM is designed to model the joint effect of these independent latent processes on the observed outputs.

To separate \textit{driving behaviors} (noted by superscript $[\mathrm{B}]$) and \textit{traffic scenarios} (noted by superscript $[\mathrm{S}]$) as two factors, we denote $z_t^{[\mathrm{B}]}$ and $z_t^{[\mathrm{S}]}$ to represent
the latent state of the two factors at time $t$, respectively. The joint latent state vector at time $t$ is represented as $\boldsymbol{z}_t := \left(z_t^{[\mathrm{B}]}, z_t^{[\mathrm{S}]}\right)$. Each factor has a number of $K^{[\mathrm{B}]}$ and $K^{[\mathrm{S}]}$ latent states, respectively. Thus, the joint latent state space $\mathcal{Z}$ is the Cartesian product of the state spaces of both components $\mathcal{Z} = \{1,\dots,K^{[\mathrm{B}]}\}\times \{1,\dots,K^{[\mathrm{S}]}\}$.

In FHMM, the latent states of the two factors evolve jointly over time, defined by a state transition matrix $\boldsymbol{\pi}\in\setR^{|\mathcal{Z}|\times |\mathcal{Z}|}$, where
\begin{subequations}
\begin{align}
{\pi}_{(\boldsymbol{k}', \boldsymbol{k})} :=&~p(\boldsymbol{z}_{t}=\boldsymbol{k}\mid  \boldsymbol{z}_{t-1}=\boldsymbol{k}')\\
=&~ p\left( z_t^{[\mathrm{B}]} = k^{[\mathrm{B}]}, z_t^{[\mathrm{S}]} = k^{[\mathrm{S}]} \mid  z_{t-1}^{[\mathrm{B}]} = k'^{[\mathrm{B}]}, z_{t-1}^{[\mathrm{S}]} = k'^{[\mathrm{S}]} \right),
\end{align}
\end{subequations}
for all $ \boldsymbol{k} = (k^{[\mathrm{B}]}, k^{[\mathrm{S}]}) $ and $ \boldsymbol{k}' = (k'^{[\mathrm{B}]}, k'^{[\mathrm{S}]}) \in \mathcal{Z} $. Also, it should be with
\begin{equation}
    \sum_{{k}\in \mathcal{Z}}\boldsymbol{\pi}_{(\boldsymbol{k}',\boldsymbol{k})}=1,\quad \forall \boldsymbol{k}'\in\mathcal{Z}.
\end{equation}

Then, we define the observation model in FHMM-IDM with separate emission functions for the two factors:
\begin{enumerate}
    \item \textbf{Driving-Behavior Local Evidence} $\boldsymbol{\psi}_t^{[\mathrm{B}]}\in\setR^{k^{[\mathrm{B}]}}$: The observed output $y_t$ is independently influenced by the latent states $z_t^{[\mathrm{B}]}$ and the covariates $\boldsymbol{x}_t$. The emission is modeled as:
\begin{equation}\label{emission:behavior}
y_t \mid \boldsymbol{x}_t, \boldsymbol{\Theta}, z_t^{[\mathrm{B}]} \sim \mathcal{N}\left( \mathrm{IDM}\left(\boldsymbol{x}_t; \boldsymbol{\theta}_{z_t^{[\mathrm{B}]}}\right), \sigma^2_{z_t^{[\mathrm{B}]}} \right),
\end{equation}
where $\mathrm{IDM}\left(\boldsymbol{x}_t; \boldsymbol{\theta}_{z_t^{[\mathrm{B}]}}\right)$ is the predicted output based on the IDM, and $\sigma^2_{z_t^{[\mathrm{B}]}}$ is the variance of the noise for state $z_t^{[\mathrm{B}]}$.

\item \textbf{Traffic-Scenario Local Evidence} $\boldsymbol{\psi}_t^{[\mathrm{S}]}\in\setR^{K^{[\mathrm{S}]}}$: For the traffic scenario, we model the relationship between the covariates $\boldsymbol{x}_t$ and the latent state $z_t^{[\mathrm{S}]}$ as
\begin{equation}\label{emission:scenario}
\boldsymbol{x}_t \mid z_t^{[\mathrm{S}]},\boldsymbol{\mu}_{\boldsymbol{x}}, \boldsymbol{\Lambda}_{\boldsymbol{x}} \sim \mathcal{N}\left( \boldsymbol{\mu}_{\boldsymbol{x}, z_t^{[\mathrm{S}]}} , \boldsymbol{\Lambda}_{\boldsymbol{x},z_t^{[\mathrm{S}]}}^{-1} \right),
\end{equation}
where $\boldsymbol{\mu}_{\boldsymbol{x}, z_t^{[\mathrm{S}]}}$ and $\boldsymbol{\Lambda}_{z_t^{[\mathrm{S}]}}$ are the mean and precision matrix of the scenario-driven input. We represent the collections of these parameters by $\boldsymbol{\mu}_{\boldsymbol{x}}=\{\boldsymbol{\mu}_{\boldsymbol{x}, k^{[\mathrm{S}]}}\}_{k^{[\mathrm{S}]}=1}^{K^{[\mathrm{S}]}}$ and $\boldsymbol{\Lambda}_{\boldsymbol{x}}=\{\boldsymbol{\Lambda}_{\boldsymbol{x}, k^{[\mathrm{S}]}}\}_{k^{[\mathrm{S}]}=1}^{K^{[\mathrm{S}]}}$.
\end{enumerate}
Therefore, the joint local evidence is given as
\begin{subequations}\label{local_evidence}
\begin{align}
p(y_t, \boldsymbol{x}_t \mid \boldsymbol{z}_t, \boldsymbol{\Theta},\boldsymbol{\mu}_{\boldsymbol{x}}, \boldsymbol{\Lambda}_{\boldsymbol{x}}) &= \underbrace{p\left(y_t \mid \boldsymbol{x}_t, \boldsymbol{\Theta}, z_t^{[\mathrm{B}]}\right)}_{:=\psi_t^{[\mathrm{B}]}\left(z_t^{[\mathrm{B}]}\right)} \cdot \underbrace{p\left(\boldsymbol{x}_t \mid z_t^{[\mathrm{S}]},\boldsymbol{\mu}_{\boldsymbol{x}}, \boldsymbol{\Lambda}_{\boldsymbol{x}}\right)}_{:=\psi_t^{[\mathrm{S}]}\left(z_t^{[\mathrm{S}]}\right)}\\
&= \mathcal{N}\left( y_t; \mathrm{IDM}\left(\boldsymbol{x}_t; \boldsymbol{\theta}_{z_t^{[\mathrm{B}]}}\right), \sigma^2_{z_t^{[\mathrm{B}]}} \right) \cdot \mathcal{N}\left( \boldsymbol{x}_t; \boldsymbol{\mu}_{\boldsymbol{x}, z_t^{[\mathrm{S}]}} , \boldsymbol{\Lambda}_{\boldsymbol{x},z_t^{[\mathrm{S}]}}^{-1} \right),
\end{align}
\end{subequations}
Thus, the joint likelihood for the entire sequence of observations $\{y_t, \boldsymbol{x}_t\}_{t=1}^T$ is:
\begin{subequations}\label{Eq:joint_llh}
\begin{align}
p\left(\boldsymbol{y}_{1:T}, \boldsymbol{x}_{1:T} \mid \boldsymbol{z}_{1:T}, \boldsymbol{\Theta},\boldsymbol{\mu}_{\boldsymbol{x}}, \boldsymbol{\Lambda}_{\boldsymbol{x}}\right) &= \prod_{t=1}^T p(y_t, \boldsymbol{x}_t \mid \boldsymbol{z}_t, \boldsymbol{\Theta},\boldsymbol{\mu}_{\boldsymbol{x}}, \boldsymbol{\Lambda}_{\boldsymbol{x}})\\
&= \prod_{t=1}^T \left[\mathcal{N}\left( y_t; \mathrm{IDM}\left(\boldsymbol{x}_t; \boldsymbol{\theta}_{z_t^{[\mathrm{B}]}}\right), \sigma^2_{z_t^{[\mathrm{B}]}} \right) \cdot \mathcal{N}\left( \boldsymbol{x}_t; \boldsymbol{\mu}_{\boldsymbol{x}, z_t^{[\mathrm{S}]}} , \boldsymbol{\Lambda}_{\boldsymbol{x},z_t^{[\mathrm{S}]}}^{-1} \right)\right],
\end{align}
\end{subequations}
where $\boldsymbol{y}_{1:T} = \left\{ y_t\right\}_{t=1}^T$, $\boldsymbol{x}_{1:T} = \left\{ \boldsymbol{x}_t\right\}_{t=1}^T$, and $\boldsymbol{z}_{1:T} = \left\{ \boldsymbol{z}_t \right\}_{t=1}^T$.
To simplify the notation, we define the joint local evidence 
\begin{equation}\label{eq:Psi}
\Psi_t(\boldsymbol{k})=\psi_t^{[\mathrm{B}]}\left(k^{[\mathrm{B}]}\right)\cdot\psi_t^{[\mathrm{S}]}\left(k^{[\mathrm{S}]}\right),\quad \forall \boldsymbol{k}\in\mathcal{Z},
\end{equation}
to be represented by a vector $\boldsymbol{\Psi}_t\in\setR^{|\mathcal{Z}|}$.

\subsection{Prior Distributions}
\subsubsection{Prior for Joint Transition Matrix: $p(\boldsymbol{\pi})$}
A natural prior for $p(\boldsymbol{\pi})$ is the Dirichlet distribution, ensuring each row of the transition matrix sums to 1. For each row $\boldsymbol{k}'$ of $p(\boldsymbol{\pi})$, we set
\begin{equation}
    \boldsymbol{\pi}_{(\boldsymbol{k}',:)}\sim \mathrm{Dir}(\boldsymbol{c}_{\boldsymbol{k}'}),\quad \forall \boldsymbol{k}'\in\mathcal{Z},
\end{equation}
where $\mathrm{Dir}(\cdot)$ denotes a Dirichlet distribution, and $\boldsymbol{c}_{\boldsymbol{k}'}=[c_{\boldsymbol{k}'\rightarrow \boldsymbol{k}}]$ are the concentration parameters for transitions from state $\boldsymbol{k}'$ to all states $\boldsymbol{k}\in\mathcal{Z}$.

\subsubsection{Prior for Latent States: $p(\boldsymbol{z}_{1:T})$}
The prior distribution over the latent states is:
\begin{equation}\label{Eq:state_prior}
p(\boldsymbol{z}_{1:T}) = p(\boldsymbol{z}_1)\prod_{t=2}^T p(\boldsymbol{z}_t \mid \boldsymbol{z}_{t-1}),
\end{equation}
where $p(\boldsymbol{z}_t \mid \boldsymbol{z}_{t-1})={\pi}_{\boldsymbol{z}_{t-1},\boldsymbol{z}_{t}}$, and the prior probabilities $p(\boldsymbol{z}_1)$ is assigned a Dirichlet prior over the joint state space $\mathcal{Z}$:
\begin{equation}
    \boldsymbol{z}_1\sim\mathrm{Dir}(\boldsymbol{c}_{\boldsymbol{z}_1}),
\end{equation}
where $\boldsymbol{c}_{\boldsymbol{z}_1}$ are concentration parameters.

\subsubsection{Prior for IDM Parameters: $p(\boldsymbol{\Theta})$ and $p(\boldsymbol{\mu},\boldsymbol{\Lambda})$}

We assign a log-normal prior on $\boldsymbol{\theta}_{k^{[\mathrm{B}]}}$ and a log-normal-Wishart conjugate prior on its parameters, as follows
\begin{subequations}
\begin{align}
    \ln\left(\boldsymbol{\theta}_{k^{[\mathrm{B}]}}\right)\mid \boldsymbol{\mu},\boldsymbol{\Lambda}^{-1} &\sim \mathcal{N}\left(\ln(\boldsymbol{\mu}),\boldsymbol{\Lambda}^{-1}\right),\quad  {k}^{[\mathrm{B}]}=1,\dots,K^{[\mathrm{B}]},\label{theta_llh}\\
    \ln(\boldsymbol{\mu})\mid \boldsymbol{\Lambda} &\sim \mathcal{N}\left(\ln(\boldsymbol{\mu}_0), (\kappa_0\boldsymbol{\Lambda})^{-1}\right), \\ 
    \boldsymbol{\Lambda} &\sim \mathcal{W}(\nu_0, \boldsymbol{W}_0),
\end{align}
\end{subequations}
where $\mathcal{W}$ denotes a Wishart distribution.

\subsubsection{Prior for Observation Noise Variance: $p(\boldsymbol{\sigma}^2)$}
The variance of the observation noise $\sigma^2_{k^{[\mathrm{B}]}}$ for each joint state $\boldsymbol{k}\in\mathcal{Z}$ is assigned an inverse-Gamma prior:
\begin{equation}
    \sigma^2_{k^{[\mathrm{B}]}}\mid \gamma_a,\gamma_b \sim \mathcal{IG}(\gamma_a,\gamma_b),\quad k^{[\mathrm{B}]}=1,\dots,K^{[\mathrm{B}]}, 
\end{equation}
where ${\mathcal {IG}}$ represents an inverse-Gamma distribution.

\subsubsection{Prior for Traffic Scenario Emission Parameters: $p(\boldsymbol{\mu}_{\boldsymbol{x}},\boldsymbol{\Lambda}_{\boldsymbol{x}})$}

We then put a normal-Wishart conjugate prior on $\boldsymbol{\mu}_{\boldsymbol{x}, k^{[\mathrm{S}]}}$ and $\boldsymbol{\Lambda}_{\boldsymbol{x},k^{[\mathrm{S}]}}$ as
\begin{subequations}
\begin{align}
    \boldsymbol{\mu}_{\boldsymbol{x},k^{[\mathrm{S}]}}\mid \boldsymbol{\Lambda}_{\boldsymbol{x},k^{[\mathrm{S}]}} &\sim \mathcal{N}\left(\boldsymbol{\mu}_{\boldsymbol{x},0}, (\kappa_{\boldsymbol{x},0}\boldsymbol{\Lambda}_{\boldsymbol{x},k^{[\mathrm{S}]}})^{-1}\right), \quad &k^{[\mathrm{S}]}&=1,\dots,K^{[\mathrm{S}]},\\ 
    \boldsymbol{\Lambda}_{\boldsymbol{x},k^{[\mathrm{S}]}} &\sim \mathcal{W}(\nu_{\boldsymbol{x},0}, \boldsymbol{W}_{\boldsymbol{x},0}),\quad &k^{[\mathrm{S}]}&=1,\dots,K^{[\mathrm{S}]}.
\end{align}
\end{subequations}

\subsubsection{Joint Priors for Parameter set: $p(\boldsymbol{\Omega})$}
Here we summarize the prior distribution on the parameters $\boldsymbol{\Omega} := \{\boldsymbol{\pi}, \sigma^2, \boldsymbol{\Theta}, \boldsymbol{\mu}_{\boldsymbol{x}}, \boldsymbol{\Lambda}_{\boldsymbol{x}}, \boldsymbol{\mu}, \boldsymbol{\Lambda}\}$ as:
\begin{align}\label{Eq:theta_priors}
p(\boldsymbol{\Omega}) =&~ p(\boldsymbol{\pi}) \cdot p(\sigma^2) \cdot p(\boldsymbol{\Theta}) \cdot p(\boldsymbol{\mu}_{\boldsymbol{x}},\boldsymbol{\Lambda}_{\boldsymbol{x}}) \cdot p(\boldsymbol{\mu},\boldsymbol{\Lambda})\\
=&~\prod_{\boldsymbol{k}' \in \mathcal{Z}} \mathrm{Dir}(\boldsymbol{\pi}_{(\boldsymbol{k}',:)};\boldsymbol{c}_{\boldsymbol{k}'}) \cdot \prod_{\boldsymbol{k} \in \mathcal{Z}} \left[ \mathcal{N}\left(\ln(\boldsymbol{\theta}_{k^{[\mathrm{B}]}}); \boldsymbol{\mu}, \boldsymbol{\Lambda}^{-1}\right) \cdot \mathcal{IG}(\sigma^2_{k^{[\mathrm{B}]}};\gamma_a, \gamma_b)\right] \cdot \mathcal{N}\left(\boldsymbol{\mu}; \boldsymbol{\mu}_{0}, (\kappa_0 \boldsymbol{\Lambda})^{-1}\right) \nonumber\\
&\cdot \mathcal{W}(\boldsymbol{\Lambda}; \nu_0, \boldsymbol{W}_0) \cdot \prod_{k^{[\mathrm{S}]}=1}^{K^{[\mathrm{S}]}} \left[ \mathcal{N}\left(\boldsymbol{\mu}_{\boldsymbol{x}, k^{[\mathrm{S}]}}; \boldsymbol{\mu}_{\boldsymbol{x}, 0}, (\kappa_{\boldsymbol{x}, 0} \boldsymbol{\Lambda}_{\boldsymbol{x},k^{[\mathrm{S}]}})^{-1}\right) \cdot \mathcal{W}(\boldsymbol{\Lambda}_{\boldsymbol{x},k^{[\mathrm{S}]}}; \nu_{\boldsymbol{x}, 0}, \boldsymbol{W}_{\boldsymbol{x}, 0}) \right].\nonumber
\end{align}

\subsection{Inference with MCMC}
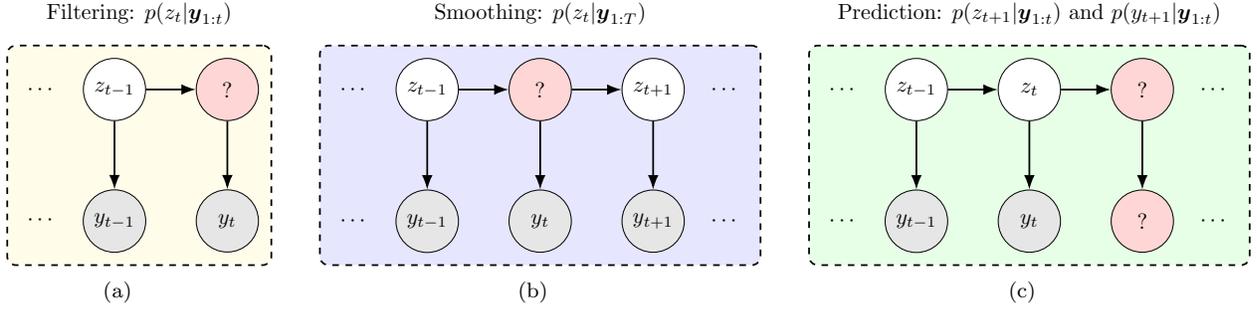
\begin{figure}[t]
    \centering
    \subfigure[]{\centering
    \resizebox{!}{3.6cm}{\usetikzlibrary{arrows.meta, positioning, backgrounds, fit}

\pgfdeclarelayer{background}
\pgfsetlayers{background,main}

\begin{tikzpicture}[node distance=1.1cm and .8cm,
    observed/.style={circle, draw, fill=gray!20, minimum size=1cm},
    hidden/.style={circle, draw, fill=white, minimum size=1cm},
    arrow/.style={-{Latex}, thick},
    target/.style={circle, draw, fill=red!16, minimum size=1cm}]

% Hidden states
\node[ ] (z0) {$\cdots$};
\node[hidden, right=of z0,xshift=-0.5cm] (z1) {$z_{t-1}$};
\node[target, right=of z1] (z2) {$?$};

% Observed states
\node[below=of z0, yshift=-0.6cm] (y0) {$\cdots$};
\node[observed, below=of z1] (y1) {$y_{t-1}$};
\node[observed, below=of z2] (y2) {$y_{t}$};

% Arrows between hidden states
\draw[arrow] (z1) -- (z2);

% Arrows from hidden to observed states
\draw[arrow] (z1) -- (y1);
\draw[arrow] (z2) -- (y2);

\begin{pgfonlayer}{background}
\node[draw, rounded corners, dashed, thick, inner sep=0.2cm, fill=yellow!10,
      fit=(z0)(z1)(z2)(y0)(y1)(y2)] (filtering) {};
\end{pgfonlayer}

\node[above=0.2cm of filtering] {Filtering: $p(z_t|\boldsymbol{y}_{1:t})$};

\end{tikzpicture}}
    }
    \subfigure[]{\centering
    \resizebox{!}{3.6cm}{\usetikzlibrary{arrows.meta, positioning, backgrounds, fit}

\pgfdeclarelayer{background}
\pgfsetlayers{background,main}

\begin{tikzpicture}[node distance=1.1cm and .8cm,
    observed/.style={circle, draw, fill=gray!20, minimum size=1cm},
    hidden/.style={circle, draw, fill=white, minimum size=1cm},
    arrow/.style={-{Latex}, thick},
    target/.style={circle, draw, fill=red!16, minimum size=1cm}]

% Hidden states
\node[ ] (z0) {$\cdots$};
\node[hidden, right=of z0,xshift=-0.5cm] (z1) {$z_{t-1}$};
\node[target, right=of z1] (z2) {$?$};
\node[hidden, right=of z2] (z3) {$z_{t+1}$};
\node[right=of z3, xshift=-0.5cm] (z4) {$\cdots$};

% Observed states
\node[below=of z0, yshift=-0.6cm] (y0) {$\cdots$};
\node[observed, below=of z1] (y1) {$y_{t-1}$};
\node[observed, below=of z2] (y2) {$y_{t}$};
\node[observed, below=of z3] (y3) {$y_{t+1}$};
\node[below=of z4, yshift=-0.6cm] (y4) {$\cdots$};

% Arrows between hidden states
\draw[arrow] (z1) -- (z2);
\draw[arrow] (z2) -- (z3);

% Arrows from hidden to observed states
\draw[arrow] (z1) -- (y1);
\draw[arrow] (z2) -- (y2);
\draw[arrow] (z3) -- (y3);

% \begin{scope}[on background layer]
\begin{pgfonlayer}{background}
\node[draw=black, fill=blue!10, inner sep=0.2cm, rounded corners, thick, dashed,
      fit=(z0)(z2)(y0)(y2)(z4)(y4)] (Smoothing) {};
\end{pgfonlayer}
\node[above=0.2cm of Smoothing] {Smoothing: $p(z_t|\boldsymbol{y}_{1:T})$};
% \end{scope}

\end{tikzpicture}}
    }
    \subfigure[]{\centering
    \resizebox{!}{3.6cm}{\usetikzlibrary{arrows.meta, positioning, backgrounds, fit}

\pgfdeclarelayer{background}
\pgfsetlayers{background,main}

\begin{tikzpicture}[node distance=1.1cm and .8cm,
    observed/.style={circle, draw, fill=gray!20, minimum size=1cm},
    hidden/.style={circle, draw, fill=white, minimum size=1cm},
    arrow/.style={-{Latex}, thick},
    target/.style={circle, draw, fill=red!16, minimum size=1cm}]

% Hidden states
\node[ ] (z0) {$\cdots$};
\node[hidden, right=of z0,xshift=-0.5cm] (z1) {$z_{t-1}$};
\node[hidden, right=of z1] (z2) {$z_{t}$};
\node[target, right=of z2] (z3) {$?$};
\node[right=of z3,xshift=-0.5cm] (z4) {$\cdots$};

% Observed states
\node[below=of z0, yshift=-0.6cm] (y0) {$\cdots$};
\node[observed, below=of z1] (y1) {$y_{t-1}$};
\node[observed, below=of z2] (y2) {$y_{t}$};
\node[target, below=of z3] (y3) {$?$};
\node[below=of z4, yshift=-0.6cm] (y4) {$\cdots$};

% Arrows between hidden states
\draw[arrow] (z1) -- (z2);
\draw[arrow] (z2) -- (z3);

% Arrows from hidden to observed states
\draw[arrow] (z1) -- (y1);
\draw[arrow] (z2) -- (y2);
\draw[arrow] (z3) -- (y3);

\begin{pgfonlayer}{background}
\node[draw=black, fill=green!10, inner sep=0.2cm, rounded corners, thick, dashed,
      fit=(z0)(z2)(z3)(z4)(y0)(y2)(y3)(y4)] (prediction) {};
\end{pgfonlayer}

\node[above=0.2cm of prediction] {Prediction: $p(z_{t+1}|\boldsymbol{y}_{1:t})$ and $p(y_{t+1}|\boldsymbol{y}_{1:t})$};

\end{tikzpicture}}
    }
    \caption{Illustration of the filtering, smoothing, and prediction problem in HMM.}\label{hmm_inference}
\end{figure}

The posterior distribution for the FHMM-IDM model is then proportional to the product of the likelihood, the prior on the latent states, and the prior on the parameters.
% Given the observations $\{ y_t, \boldsymbol{x}_t \}_{t=1}^T$ and the corresponding latent state sequence $\boldsymbol{z}_{1:T}$, we formally derive the maximum a posteriori (MAP) as
% \begin{subequations}
% \begin{align}
% \boldsymbol{z}_{1:T}^\star, \boldsymbol{\Omega}^\star &= \argmax_{\boldsymbol{z}_{1:T}, \boldsymbol{\Omega}}~p(\boldsymbol{z}_{1:T}, \boldsymbol{\Omega} \mid  \boldsymbol{y}_{1:T}, \boldsymbol{x}_{1:T})\\
% &=\argmax_{\boldsymbol{z}_{1:T}, \boldsymbol{\Omega}}~p(\boldsymbol{y}_{1:T}, \boldsymbol{x}_{1:T} \mid \boldsymbol{z}_{1:T}, \boldsymbol{\Omega}) \cdot p(\boldsymbol{z}_{1:T}) \cdot p(\boldsymbol{\Omega})
% \\
% &=\argmax_{\boldsymbol{z}_{1:T}, \boldsymbol{\Omega}}~\underbrace{\prod_{t=1}^T p(y_t, \boldsymbol{x}_t \mid \boldsymbol{z}_t)}_{\text{Eq.~\eqref{Eq:joint_llh}}} \cdot \underbrace{p(\boldsymbol{z}_1)\prod_{t=2}^T p(\boldsymbol{z}_t \mid \boldsymbol{z}_{t-1})}_{\text{Eq.~\eqref{Eq:state_prior}}} \cdot \underbrace{p(\boldsymbol{\pi}) \cdot p(\sigma^2) \cdot p(\boldsymbol{\Theta}) \cdot p(\boldsymbol{\mu}_{\boldsymbol{x}},\boldsymbol{\Lambda}_{\boldsymbol{x}}) \cdot p(\boldsymbol{\mu},\boldsymbol{\Lambda})}_{\text{Eq.~\eqref{Eq:theta_priors}}}.
% \end{align}
% \end{subequations}
It is intractable to find an analytical solution for estimating the posteriors. Therefore, we develop an MCMC sampling algorithm (see Algorithm~\ref{mcmc}) to learn the posteriors of the model parameters and infer the latent states. Note that as shown in Fig.~\ref{hmm_inference}, the three fundamental inference tasks in HMM---filtering, smoothing, and prediction---differ in the set of observations used to estimate the latent state. 

In the filtering task (the left panel), the objective is to estimate the current latent state $z_t$ given observations up to time $t$, i.e., $p(z_t \mid \boldsymbol{y}_{1:t})$. In smoothing (the middle panel), the goal is to retrospectively estimate a past latent state $z_t$ using the entire sequence of observations, $p(z_t \mid \boldsymbol{y}_{1:T})$, thereby incorporating future evidence to improve estimation accuracy. In contrast, prediction (the right panel) aims to estimate future states and observations, such as $p(z_{t+1} \mid \boldsymbol{y}_{1:t})$ or $p(y_{t+1} \mid \boldsymbol{y}_{1:t})$, based on current and past observations.

The figure emphasizes the distinct computational characteristics of these tasks: filtering operates in a causal (forward) manner, smoothing is acausal (utilizing both past and future observations), and prediction is inherently forward-looking. In this work, we focus primarily on the smoothing problem, which enables more accurate inference of the latent states. Nonetheless, our framework can be readily extended to address filtering and prediction tasks depending on the specific application context.

\begin{algorithm}[!t]
    \caption{MCMC Sampling for FHMM-IDM.}\label{mcmc}
    \SetKwInOut{Input}{Input}\SetKwInOut{Output}{Output}\SetKwData{iter}{iteration}
    \Input{Driving behavior observation $\hat{\boldsymbol{y}}_{1:T}$; number of burn-in iterations $m_1$ and number of samples $m_2$ for estimation;  hyperparameters.}
    \Output{Transition matrix ${\boldsymbol{\pi}}$, states assignment $\boldsymbol{z}_{1:T}$, IDM variances ${\boldsymbol{\sigma}}$, IDM parameters ${\boldsymbol{\Theta}}$, mean ${\boldsymbol{\mu}_{\boldsymbol{x}}}$, and precision matrix ${\boldsymbol{\Lambda}_{\boldsymbol{x}}}$.}
    \BlankLine
    Initialize $\boldsymbol{\pi}^{(1)}$, $\boldsymbol{\sigma}^{(1)}$, $\boldsymbol{\Theta}^{(1)}$, $\boldsymbol{z}_{1:T}^{(1)}$, $\boldsymbol{\mu}^{(1)}_{\boldsymbol{x}}$, $\boldsymbol{\Lambda}^{(1)}_{\boldsymbol{x}}$, $\boldsymbol{\mu}^{(1)}$, and $\boldsymbol{\Lambda}^{(1)}$ \;
    \For{\iter $i = 1$ \KwTo $m_1+m_2$}{
        Draw $\{\boldsymbol{\pi}_{(\boldsymbol{k}',:)}^{(i)}\}_{\boldsymbol{k}\in\mathcal{Z}}$ by $\boldsymbol{\pi}_{(\boldsymbol{k}',:)}^{(i)} \sim \mathrm{Dir}(\boldsymbol{c}_{\boldsymbol{k}'}+\boldsymbol{n}_{\boldsymbol{k}'}^{(i)})$ \tcp*[r]{Given $\boldsymbol{z}_{1:T}^{(i)}$ (see Eq.\eqref{post_pi})}
        \For{$\boldsymbol{k}\in \mathcal{Z}$}{
            Compute $\{\boldsymbol{\Psi}_t^{(i)}\}_{t\in\mathcal{T}_{\boldsymbol{k}}}$ \tcp*[r]{Given $\boldsymbol{y}_{1:T}, \boldsymbol{x}_{1:T}, \boldsymbol{z}_{1:T}^{(i)}, \boldsymbol{\Theta}^{(i)}$ (see Eqs.\eqref{eq:Psi} and \eqref{local_evidence})}
        }
        Compute $\{\boldsymbol{\gamma}_t^{(i)}\}_{t=1}^T$ using Algorithm~\ref{fwd_bwd_alg} \tcp*[r]{Given $\boldsymbol{\alpha}_1, \boldsymbol{\beta}_{T},\boldsymbol{\pi}^{(i)},p(\boldsymbol{z}_1),\{\boldsymbol{\Psi}_t^{(i)}\}_{t=1}^T$}
        Draw $\boldsymbol{z}_{1:T}^{(i)}$ by $\boldsymbol{z}_t^{(i)}\sim\mathrm{Cat}(\boldsymbol{\gamma}_t^{(i)})$\tcp*[r]{Given $\{\boldsymbol{\gamma}_t^{(i)}\}_{t=1}^T$ (see Eq.\ref{post_z})}
        Draw $\boldsymbol{\theta}_{k^{[\mathrm{B}]}}^{(i)}$ using Algorithm~\ref{MH} \tcp*[r]{Calibrate IDM given $\boldsymbol{\mu}^{(i)},\boldsymbol{\Lambda}^{(i)},\boldsymbol{\theta}_{k^{[\mathrm{B}]}}^{(i)},\boldsymbol{z}_{1:T}^{(i)}$}
        \For{\iter ${k}^{[\mathrm{B}]}=1$ \KwTo $K^{[\mathrm{B}]}$}{
            Draw $\sigma^{(i)}_{k^{[\mathrm{B}]}}$ by $\sigma^{2(i)}_{k^{[\mathrm{B}]}} \sim \mathcal{IG}(\gamma_a^\star,\gamma_b^\star)$ \tcp*[r]{Given $\boldsymbol{\theta}_{{k}^{[\mathrm{B}]}}^{(i)},\boldsymbol{z}_{1:T}^{(i)}$ (see Eq.\eqref{sigma_post})}
        }
        \For{\iter ${k}^{[\mathrm{S}]}=1$ \KwTo $K^{[\mathrm{S}]}$}{
            Draw $\boldsymbol{\mu}_{\boldsymbol{x},k^{[\mathrm{S}]}}^{(i)},\boldsymbol{\Lambda}_{\boldsymbol{x},k^{[\mathrm{S}]}}^{(i)}$ by $\mathcal{NW}$ \tcp*[r]{Given $\boldsymbol{z}_{1:T}^{(i)}$ (see Eq.\eqref{mux_lambdax_post})}
        }
        Draw $\boldsymbol{\mu}^{(i)}$ and $\boldsymbol{\Lambda}^{(i)}$ by $\mathcal{NW}$ \tcp*[r]{Given $\boldsymbol{\Theta}^{(i)}$ (see Eq.\eqref{Eq:theta_Lambda_post})}
        \If{$i>m_1$}{
            Collect $\boldsymbol{\pi}^{(i)}$, $\boldsymbol{\sigma}^{(i)}$, $\boldsymbol{\Theta}^{(i)}$, $\boldsymbol{z}_{1:T}^{(i)}$, 
            $\boldsymbol{\mu}^{(i)}_{\boldsymbol{x}}$, $\boldsymbol{\Lambda}^{(i)}_{\boldsymbol{x}}$,
            $\boldsymbol{\mu}^{(i)}$ and $\boldsymbol{\Lambda}^{(i)}$\;
        }
    }
    \Return{$\boldsymbol{\pi}$, $\boldsymbol{\sigma}$, $\boldsymbol{\Theta}$, $\boldsymbol{z}_{1:T}$, 
    $\boldsymbol{\mu}_{\boldsymbol{x}}$, $\boldsymbol{\Lambda}_{\boldsymbol{x}}$,
    $\boldsymbol{\mu}$, $\mathrm{and}$ $\boldsymbol{\Lambda}$}.
\end{algorithm}\DecMargin{1em}

\subsubsection{Sample Latent States $\boldsymbol{z}_{1:T}$}\label{sample_z}
In the following, we introduce the Forward-Backward Algorithm (see Algorithm~\ref{fwd_bwd_alg}) to sample $\boldsymbol{z}_{1:T}^{[\mathrm{B}]}$ and $\boldsymbol{z}_{1:T}^{[\mathrm{S}]}$. Firstly, we define
\begin{subequations}
\begin{align}
    \alpha_t\left(\boldsymbol{z}_{t}\right)&:=p\left(\boldsymbol{y}_{1:t}, \boldsymbol{x}_{1:t},\boldsymbol{z}_{t}\right),\\
    \beta_t\left(\boldsymbol{z}_{t}\right)&:=p\left(\boldsymbol{y}_{t+1:T},\boldsymbol{x}_{t+1:T}\mid \boldsymbol{z}_{t}\right),\\    \gamma_t\left(\boldsymbol{z}_{t}\right)&:=p\left(\boldsymbol{z}_{t}\mid \boldsymbol{y}_{1:T},\boldsymbol{x}_{1:T}\right).
\end{align}
\end{subequations}
Then we can obtain
% \begin{subequations}
\begin{align}
    &p\left(\boldsymbol{y}_{1:T},\boldsymbol{x}_{1:T},\boldsymbol{z}_{t}\right)=p\left(\boldsymbol{y}_{1:t},\boldsymbol{x}_{1:t},\boldsymbol{z}_{t}\right)\cdot p\left(\boldsymbol{y}_{t+1:T},\boldsymbol{x}_{t+1:T}\mid \boldsymbol{y}_{1:t},\boldsymbol{x}_{1:t}, \boldsymbol{z}_{t}\right)=\alpha_t\left(\boldsymbol{z}_{t}\right)\cdot \beta_t\left(\boldsymbol{z}_{t}\right),
    % &p\left(\boldsymbol{y}_{1:T}, z_{t}^{[\mathrm{B}]}\right)=p\left(\boldsymbol{y}_{1:t}\mid z_{t}^{[\mathrm{B}]}\right)\cdot p\left(\boldsymbol{y}_{t+1:T}\mid \boldsymbol{y}_{1:t}, z_{t}^{[\mathrm{B}]}\right)=\alpha_t^{[\mathrm{B}]}\left({z}_{t}^{[\mathrm{B}]}\right)\cdot \beta_t^{[\mathrm{B}]}\left({z}_{t}^{[\mathrm{B}]}\right),\\
    % &p\left(\boldsymbol{x}_{1:T}, z_{t}^{[\mathrm{S}]}\right)=p\left(\boldsymbol{x}_{1:t}\mid z_{t}^{[\mathrm{S}]}\right)\cdot p\left(\boldsymbol{x}_{t+1:T}\mid \boldsymbol{x}_{1:t}, z_{t}^{[\mathrm{S}]}\right)=\alpha_t^{[\mathrm{S}]}\left(z_{t}^{[\mathrm{S}]}\right)\cdot \beta_t^{[\mathrm{S}]}\left(z_{t}^{[\mathrm{S}]}\right),
\end{align}
% \end{subequations}
and
% \begin{subequations}
\begin{align}
    &\gamma_t\left(\boldsymbol{z}_{t}\right) = \frac{p\left(\boldsymbol{y}_{1:T},\boldsymbol{x}_{1:T},\boldsymbol{z}_{t}\right)}{p\left(\boldsymbol{y}_{1:T},\boldsymbol{x}_{1:T}\right)} = \frac{\alpha_t\left(\boldsymbol{z}_{t}\right)\cdot\beta_t\left(\boldsymbol{z}_{t}\right)}{p\left(\boldsymbol{y}_{1:T},\boldsymbol{x}_{1:T}\right)} = \frac{\alpha_t\left(\boldsymbol{z}_{t}\right)\cdot\beta_t\left(\boldsymbol{z}_{t}\right)}{\sum_{\boldsymbol{z}_{t}} \alpha_t\left(\boldsymbol{z}_{t}\right)\cdot\beta_t\left(\boldsymbol{z}_{t}\right)}.
    % &\gamma_t^{[\mathrm{B}]}\left({z}_{t}^{[\mathrm{B}]}\right) = \frac{p\left(\boldsymbol{y}_{1:T},z_{t}^{[\mathrm{B}]}\right)}{p\left(\boldsymbol{y}_{1:T}\right)} = \frac{\alpha_t^{[\mathrm{B}]}\left({z}_{t}^{[\mathrm{B}]}\right)\cdot\beta_t^{[\mathrm{B}]}\left({z}_{t}^{[\mathrm{B}]}\right)}{p\left(\boldsymbol{y}_{1:T}\right)} = \frac{\alpha_t^{[\mathrm{B}]}\left({z}_{t}^{[\mathrm{B}]}\right)\cdot\beta_t^{[\mathrm{B}]}\left({z}_{t}^{[\mathrm{B}]}\right)}{\sum_{z_{t}^{[\mathrm{B}]}} \alpha_t^{[\mathrm{B}]}\left({z}_{t}^{[\mathrm{B}]}\right)\cdot \beta_t^{[\mathrm{B}]}\left({z}_{t}^{[\mathrm{B}]}\right)}, \label{post_z_B} \\
    % &\gamma_t^{[\mathrm{S}]}\left({z}_{t}^{[\mathrm{S}]}\right) = \frac{p\left(\boldsymbol{x}_{1:T},z_{t}^{[\mathrm{S}]}\right)}{p\left(\boldsymbol{x}_{1:T}\right)} = \frac{\alpha_t^{[\mathrm{S}]}\left({z}_{t}^{[\mathrm{S}]}\right)\cdot\beta_t^{[\mathrm{S}]}\left({z}_{t}^{[\mathrm{S}]}\right)}{p\left(\boldsymbol{x}_{1:T}\right)} = \frac{\alpha_t^{[\mathrm{S}]}\left({z}_{t}^{[\mathrm{S}]}\right)\cdot\beta_t^{[\mathrm{S}]}\left({z}_{t}^{[\mathrm{S}]}\right)}{\sum_{z_{t}^{[\mathrm{S}]}} \alpha_t^{[\mathrm{S}]}\left({z}_{t}^{[\mathrm{S}]}\right)\cdot \beta_t^{[\mathrm{S}]}\left({z}_{t}^{[\mathrm{S}]}\right)}, \label{post_z_S} \\
\end{align}
% \end{subequations}

In the following, we will derive the iterative form of $\alpha_t(\boldsymbol{z}_{t})$, $\beta_t(\boldsymbol{z}_{t})$, and therefore $\gamma_t(\boldsymbol{z}_{t})$. For the forward passes, $\forall \boldsymbol{z}_t\in\mathcal{Z}$ we have
\begin{subequations}
\begin{align}
    \alpha_1\left(\boldsymbol{z}_1\right) & = \Psi_1(\boldsymbol{z}_1) \cdot p\left(\boldsymbol{z}_1\right), \quad& t&=1,\label{eq:init_fwd}\\
    \alpha_t\left(\boldsymbol{z}_t\right) & = \Psi_t(\boldsymbol{z}_t) \sum_{\boldsymbol{z}_{t-1}} \alpha_{t-1}\left(\boldsymbol{z}_{t-1}\right) \cdot {\pi}_{\boldsymbol{z}_{t-1},\boldsymbol{z}_{t}}, \quad& t&=2,\dots,T. \label{fwd}
\end{align}
\end{subequations}
To simplify the notations, here we organize $\boldsymbol{\alpha}_t\in \setR^{|\mathcal{Z}|}$ as a vector.
Then Eq.~\eqref{fwd} can be expressed in a more efficient form as
\begin{equation}
    \boldsymbol{\alpha}_t = \boldsymbol{\pi}\left( \boldsymbol{\alpha}_{t-1}\odot \boldsymbol{\Psi}_t\right), \label{fwd_mat}
\end{equation}
where $\boldsymbol{a}\odot \boldsymbol{b}$ represents the Hadamard product. 

For the backward passes, $\forall \boldsymbol{z}_t\in\mathcal{Z}$ we can derive
\begin{subequations}
\begin{align}
    \beta_t\left(\boldsymbol{z}_t\right) &= \sum_{\boldsymbol{z}_{t+1}} \beta_{t+1}\left(\boldsymbol{z}_{t+1}\right)\cdot \Psi_{t+1}(\boldsymbol{z}_{t+1}) \cdot {\pi}_{\boldsymbol{z}_{t},\boldsymbol{z}_{t+1}}, \quad& t&=1,\dots,T-1,\label{bwd} \\
    \beta_T\left(\boldsymbol{z}_T\right) &=1, \quad& t&=T. \label{bwd_init}
\end{align}
\end{subequations}

Similarly, we organize $\boldsymbol{\beta}_t\in \setR^{|\mathcal{Z}|}$ as a vector as well, then we can obtain the following form based on Eq.~\eqref{bwd}, written as
\begin{equation}
\boldsymbol{\beta}_{t} =\boldsymbol{\pi}^{\top}\left( \boldsymbol{\beta}_{t+1}\odot \boldsymbol{\Psi}_{t+1}\right). \label{bwd_mat}
\end{equation}
Therefore, we have
\begin{equation}
    \boldsymbol{\gamma}_{t}=\frac{\boldsymbol{\alpha}_{t}\odot \boldsymbol{\beta}_{t}}{\boldsymbol{\alpha}_{t}^{\top} \boldsymbol{\beta}_{t}}\in \setR^{|\mathcal{Z}|}.\label{gamma_mat}
\end{equation}

For each time $t$, we can sample the joint latent state $(z_t^{[\mathrm{B}]}, z_t^{[\mathrm{S}]})$ from the posterior:
\begin{equation}\label{post_z}
    \left(z_t^{[\mathrm{B}]}, z_t^{[\mathrm{S}]}\right) \sim \text{Cat}\left(\boldsymbol{\gamma}_t\right).
\end{equation}
Repeat this process for $t = 1, \dots, T$ to obtain the sequence $\boldsymbol{z}_{1:T}$.

\begin{algorithm}[t]
    % \SetKwData{Left}{left}\SetKwData{This}{this}\SetKwData{Up}{up}
    % \SetKwFunction{Union}{Union}
    \caption{Forward-Backward Algorithm}\label{fwd_bwd_alg}
    \SetKwInOut{Input}{Input}\SetKwInOut{Output}{Output}
    \Input{Initialize $\boldsymbol{\alpha}_1$ and $\boldsymbol{\beta}_{T}$; transition matrix $\boldsymbol{\pi}$; latent state priors $p(\boldsymbol{z}_1)$; local evidence $\{\boldsymbol{\Psi}_t\}_{t=1}^T$. }
    \Output{The latent state posteriors $\{\boldsymbol{\gamma}_t\}_{t=1}^T$.}
    \BlankLine
    % \textit{special treatment of the first line}\;
    Set $\alpha_1\left(\boldsymbol{z}_1\right) = \Psi_1(\boldsymbol{z}_1) \cdot p\left(\boldsymbol{z}_1\right), \forall \boldsymbol{z}_1\in \mathcal{Z}$ \tcp*[r]{(see Eq.\eqref{eq:init_fwd})}
    \For{$t\leftarrow 2$ \KwTo $T$}{
        % \textit{special treatment of the first element of line $i$}\;
        Compute $\boldsymbol{\alpha}_t = \boldsymbol{\pi}\left( \boldsymbol{\alpha}_{t-1}\odot \boldsymbol{\Psi}_t\right)$ \tcp*[r]{Forward passes (see Eq.\eqref{fwd_mat})}
    }
    Set $\boldsymbol{\beta}_T=\mathbf{1}\in\setR^{|\mathcal{Z}|}$\tcp*[r]{(see Eq.\eqref{bwd_init})}
    \For{$t\leftarrow T-1$ \KwTo $1$}{
        % \textit{special treatment of the first element of line $i$}\;
        Compute $\boldsymbol{\beta}_{t} =\boldsymbol{\pi}^{\top}\left( \boldsymbol{\beta}_{t+1}\odot \boldsymbol{\Psi}_{t+1}\right)$ \tcp*[r]{Backward passes (see Eq.\eqref{bwd_mat})}
    }
    
    \For{$t\leftarrow 1$ \KwTo $T$}{
        Compute $\boldsymbol{\gamma}_{t}=({\boldsymbol{\alpha}_{t}\odot \boldsymbol{\beta}_{t}})/({\boldsymbol{\alpha}_{t}^{\top} \boldsymbol{\beta}_{t}})$ \tcp*[r]{Smoothing (see Eq.\eqref{gamma_mat})}
    }
    \Return{$\{\boldsymbol{\gamma}_t\}_{t=1}^T$}.
\end{algorithm}\DecMargin{1em}

\subsubsection{Sample Transition Matrix $\boldsymbol{\pi}$}\label{sample_pi}
For each row $\boldsymbol{\pi}_{(\boldsymbol{k}',:)}$, we define the sufficient statistics as the counts of state transitions from state $\boldsymbol{k}'$ to state $\boldsymbol{k}$ over the entire sequence:
\begin{equation}
    n_{\boldsymbol{k}',\boldsymbol{k}} = \sum_{t=2}^T \mathbb{I}(\boldsymbol{z}_{t-1} = \boldsymbol{k}', \boldsymbol{z}_t = \boldsymbol{k}),
\end{equation}
where $\mathbb{I}$ is the indicator function.

Given the counts $n_{\boldsymbol{k}',\boldsymbol{k}}$, we sample $\boldsymbol{\pi}_{(\boldsymbol{k},:)}$ from the Dirichlet distribution:
\begin{equation}\label{post_pi}
    \boldsymbol{\pi}_{(\boldsymbol{k}',:)} \sim \mathrm{Dir}(\boldsymbol{c}_{\boldsymbol{k}'}+\boldsymbol{n}_{\boldsymbol{k}'}),
\end{equation}
where $\boldsymbol{n}_{\boldsymbol{k}'}\in\setR^{|\mathcal{Z}|}$ collects the transition counts for the $\boldsymbol{k}'$-th row of $\boldsymbol{\pi}$.

\subsubsection{Sample the IDM Parameters $\boldsymbol{\theta}_{k^{[\mathrm{B}]}}$ (Metropolis-Hastings Sampling)}\label{sample_IDM_params}

We define a proposal distribution $q(\boldsymbol{\theta}'_{k^{[\mathrm{B}]}}\mid \boldsymbol{\theta}_{k^{[\mathrm{B}]}}^{(i)})$ as a Gaussian centered at the current state $\boldsymbol{\theta}_{k^{[\mathrm{B}]}}^{(i)}$, such that the proposed parameters $\boldsymbol{\theta}'_{k^{[\mathrm{B}]}}$ is sampled according to
\begin{align}
    \boldsymbol{\theta}'_{k^{[\mathrm{B}]}} \sim \mathcal{N}\left(\boldsymbol{\theta}_{k^{[\mathrm{B}]}}^{(i)}, \boldsymbol{\Sigma}_q\right),\label{propose}
\end{align}
where $\boldsymbol{\Sigma}_q$ is the covariance matrix of the proposal.

According to Eq.~\eqref{emission:behavior}, we have
\begin{subequations}\label{MH_llh}
\begin{align}
    p(\boldsymbol{y}_{1:T} \mid \boldsymbol{x}_{1:T}, \boldsymbol{z}_{1:T},\boldsymbol{\theta}_{k^{[\mathrm{B}]}})&\propto  \prod_{t\in\mathcal{T}_{k^{[\mathrm{B}]}}}\mathcal{N}\left( y_t;\mathrm{IDM}\left(\boldsymbol{x}_t; \boldsymbol{\theta}_{k^{[\mathrm{B}]}}\right), \sigma^2_{k^{[\mathrm{B}]}} \right)\\
    &=\prod_{t\in\mathcal{T}_{k^{[\mathrm{B}]}}} \psi_t^{[\mathrm{B}]}(z_t^{[\mathrm{B}]}), \quad\forall \boldsymbol{k}\in\mathcal{Z},
\end{align}
\end{subequations}
where $\mathcal{T}_{k^{[\mathrm{B}]}}:=\{t\mid \boldsymbol{z}_t^{[\mathrm{B}]}=k^{[\mathrm{B}]}\}$.

The acceptance probability $A$ for the proposed parameters $\boldsymbol{\theta}'_{k^{[\mathrm{B}]}}$ is given by
\begin{align}
    A\left(\boldsymbol{\theta}'_{k^{[\mathrm{B}]}},\boldsymbol{\theta}_{k^{[\mathrm{B}]}}^{(i)}\right) = \min \left( 1, \frac{p(\boldsymbol{\theta}'_{k^{[\mathrm{B}]}})\cdot p(\boldsymbol{y}_{1:T} \mid  \boldsymbol{x}_{1:T}, \boldsymbol{z}_{1:T},\boldsymbol{\theta}'_{k^{[\mathrm{B}]}})}{p\left(\boldsymbol{\theta}_{k^{[\mathrm{B}]}}^{(i)}\right)\cdot p(\boldsymbol{y}_{1:T} \mid  \boldsymbol{x}_{1:T}, \boldsymbol{z}_{1:T},\boldsymbol{\theta}_{k^{[\mathrm{B}]}}^{(i)})}\right),\label{accept_rate}
\end{align}
where according to Eq.~\eqref{theta_llh}, 
\begin{align}\label{MH_priors}
    p\left(\boldsymbol{\theta}_{k^{[\mathrm{B}]}}\right)&= \mathcal{LN}\left(\ln\left(\boldsymbol{\theta}_{k^{[\mathrm{B}]}}\right);\ln(\boldsymbol{\mu}),\boldsymbol{\Lambda}^{-1}\right),
\end{align}
and $\mathcal{LN}(\cdot)$ represents the log-normal distribution. Note that the ratio ${q\left(\boldsymbol{\theta}_{k^{[\mathrm{B}]}}^{(i)}\mid \boldsymbol{\theta}'_{k^{[\mathrm{B}]}}\right)/ q\left(\boldsymbol{\theta}'_{k^{[\mathrm{B}]}}\mid \boldsymbol{\theta}_{k^{[\mathrm{B}]}}^{(i)}\right)}$ of symmetric proposal probabilities simplifies as it cancels out for forward and reverse moves.
The next sample $\boldsymbol{\theta}_{k^{[\mathrm{B}]}}^{(i+1)}$ is then determined by
\begin{align}\label{post_theta}
    \boldsymbol{\theta}_{k^{[\mathrm{B}]}}^{(i+1)}= \begin{cases}
   \boldsymbol{\theta}'_{k^{[\mathrm{B}]}}, &\text{w.p. } A, \\
   \boldsymbol{\theta}_{k^{[\mathrm{B}]}}^{(i)}, &\text{w.p. } 1-A.
\end{cases}
\end{align}

The MH sampling processes are summarized in Algorithm~\ref{MH}.

\begin{algorithm}[!t]
    \caption{Metropolis-Hasting (MH)  Sampling (one step) for IDM Calibration.}\label{MH}
    \SetKwInOut{Input}{Input}\SetKwInOut{Output}{Output}\SetKwData{iter}{iteration}
    \Input{Driving behavior observation $\hat{\boldsymbol{y}}_{1:T}$; latent state assignments $\boldsymbol{z}_{1:T}$; IDM parameter set $\boldsymbol{\Theta}^{(i)}$; proposal covariance matrix $\boldsymbol{\Sigma}_q$; local evidence $\boldsymbol{\psi}_{1:T}$; prior $p(\boldsymbol{\theta})$.}
    \Output{Updated IDM parameter set ${\boldsymbol{\Theta}}^{(i+1)}$.}
    \BlankLine
    \For{\iter $k^{[\mathrm{B}]}=1$ \KwTo $K^{[\mathrm{B}]}$}{
        Draw $\boldsymbol{\theta}'_{k^{[\mathrm{B}]}} \sim \mathcal{N}(\boldsymbol{\theta}_{k^{[\mathrm{B}]}}^{(i)}, \boldsymbol{\Sigma}_q)$ \tcp*[r]{Propose a candidate (see Eq.\eqref{propose})}
        Compute the acceptance rate $A\left(\boldsymbol{\theta}'_{k^{[\mathrm{B}]}},\boldsymbol{\theta}_{k^{[\mathrm{B}]}}^{(i)}\right)$ using Eq.~\eqref{accept_rate}, given Eqs.~\eqref{MH_llh} and \eqref{MH_priors}\;
        Draw a random number $p\sim \mathrm{Uniform}(0,1)$\;
        \If{$p<A(\boldsymbol{\theta}'_{k^{[\mathrm{B}]}},\boldsymbol{\theta}_{k^{[\mathrm{B}]}}^{(i)})$}{
            $\boldsymbol{\theta}_{k^{[\mathrm{B}]}}^{(i+1)}=\boldsymbol{\theta}'_{k^{[\mathrm{B}]}}$\tcp*[r]{Accept candidate (see Eq.\eqref{post_theta})}
        }
        \Else{
            $\boldsymbol{\theta}_{k^{[\mathrm{B}]}}^{(i+1)}=\boldsymbol{\theta}_{k^{[\mathrm{B}]}}^{(i)}$\tcp*[r]{Reject candidate (see Eq.\eqref{post_theta})}
        }
    }
    \Return{$\boldsymbol{\Theta}^{(i+1)}$}.
\end{algorithm}\DecMargin{1em}

\subsubsection{Sample $\boldsymbol{\mu}$ and $\boldsymbol{\Lambda}$}\label{sample_mu_and_Lambda}
Due to the normal-Wishart conjugacy, we derive the posteriors as:
\begin{subequations}\label{Eq:theta_Lambda_post}
\begin{align}
    \ln(\boldsymbol{\mu})\mid \boldsymbol{\Lambda}, \{\boldsymbol{\theta}_{k^{[\mathrm{B}]}}\} &\sim \mathcal{N}\left(\ln(\boldsymbol{\mu}'), (\kappa'\boldsymbol{\Lambda})^{-1}\right), \label{theta_post} \\ 
    \boldsymbol{\Lambda}\mid \{\boldsymbol{\theta}_{k^{[\mathrm{B}]}}\} &\sim \mathcal{W}(\nu', \boldsymbol{W}'), \label{Lambda_post}
\end{align}
\end{subequations}
where
\begin{subequations}
\begin{align}
    \nu' &= \nu_0 +K^{[\mathrm{B}]},\\
    \kappa' &= \kappa_0 + K^{[\mathrm{B}]},\\
    \ln(\bar{\boldsymbol{\mu}}) &= \frac{1}{K^{[\mathrm{B}]}}\sum_{\boldsymbol{k}} \ln(\boldsymbol{\theta}_{k^{[\mathrm{B}]}}),\\
    \boldsymbol{S} &= \sum_{\boldsymbol{k}}\left(\ln(\boldsymbol{\theta}_{k^{[\mathrm{B}]}})-\ln(\bar{\boldsymbol{\mu}})\right)\left(\ln(\boldsymbol{\theta}_{k^{[\mathrm{B}]}})-\ln(\bar{\boldsymbol{\mu}})\right)^\top,\\
    \boldsymbol{W}' &= \boldsymbol{W}_0 + \boldsymbol{S} + \frac{\kappa_0 K^{[\mathrm{B}]}}{\kappa_0 + K^{[\mathrm{B}]}}\left(\ln(\bar{\boldsymbol{\mu}})-\ln(\boldsymbol{\mu}_0)\right) \left(\ln(\bar{\boldsymbol{\mu}})-\ln(\boldsymbol{\mu}_0)\right)^\top,\\
    \ln(\boldsymbol{\mu}') &=\frac{\kappa_0\ln(\boldsymbol{\mu}_0)+K^{[\mathrm{B}]}\ln(\bar{\boldsymbol{\mu}})}{\kappa_0+K^{[\mathrm{B}]}}.
\end{align}
\end{subequations}

\subsubsection{Sample Observation Noise Variance $\sigma^2$}\label{sample_sigma}
Given the normal-inverse-Gamma conjugacy, we have the posterior as
\begin{align}
    \sigma^2_{k^{[\mathrm{B}]}}\mid \{y_t\}_{t\in\mathcal{T}_{k^{[\mathrm{B}]}}},\boldsymbol{\theta}_{k^{[\mathrm{B}]}} \sim \mathcal{IG}(\gamma_a^\star,\gamma_b^\star), \label{sigma_post}
\end{align}
where $\gamma_a^\star = \gamma_a+|\mathcal{T}_{k^{[\mathrm{B}]}}|/2$ and $\gamma_b^\star = \gamma_b+\sum_{t\in\mathcal{T}_{k^{[\mathrm{B}]}}} \left(y_t-\mathrm{IDM}(\boldsymbol{x}_{t};\boldsymbol{\theta}_{k^{[\mathrm{B}]}})\right)^2 /2$.

\subsubsection{Sample $\boldsymbol{\mu}_{\boldsymbol{x}}$ and $\boldsymbol{\Lambda}_{\boldsymbol{x}}$}\label{sample_mux_and_Lambdax}
We define $\mathcal{T}_{k^{[\mathrm{S}]}}:=\{t|\boldsymbol{z}_t^{[\mathrm{S}]}=k^{[\mathrm{S}]}\}$. The posterior distribution of $\boldsymbol{\mu}_{\boldsymbol{x},k^{[\mathrm{S}]}}$ and $\boldsymbol{\Lambda}_{\boldsymbol{x},k^{[\mathrm{S}]}}$ is derived using the normal-Wishart conjugacy
\begin{subequations}\label{mux_lambdax_post}
\begin{align}
    \boldsymbol{\mu}_{\boldsymbol{x},k^{[\mathrm{S}]}}\mid \boldsymbol{\Lambda}_{\boldsymbol{x},k^{[\mathrm{S}]}}, \{x_t\}_{t\in\mathcal{T}_{k^{[\mathrm{S}]}}} &\sim \mathcal{N}\left(\boldsymbol{\mu}'_{\boldsymbol{x},k^{[\mathrm{S}]}}, (\kappa_{\boldsymbol{x},k^{[\mathrm{S}]}}'\boldsymbol{\Lambda}_{\boldsymbol{x},k^{[\mathrm{S}]}})^{-1}\right), \\ 
    \boldsymbol{\Lambda}_{\boldsymbol{x},k^{[\mathrm{S}]}}\mid 
    \{x_t\}_{t\in\mathcal{T}_{k^{[\mathrm{S}]}}} &\sim \mathcal{W}(\nu_{\boldsymbol{x},k^{[\mathrm{S}]}}', \boldsymbol{W}_{\boldsymbol{x},k^{[\mathrm{S}]}}'),
\end{align}
\end{subequations}
where
\begin{subequations}
\begin{align}
    \nu_{\boldsymbol{x},k^{[\mathrm{S}]}}' &= \nu_{\boldsymbol{x},0} + |\mathcal{T}_{k^{[\mathrm{S}]}}|,\\
    \kappa_{\boldsymbol{x},k^{[\mathrm{S}]}}' &= \kappa_{\boldsymbol{x},0} + |\mathcal{T}_{k^{[\mathrm{S}]}}|,\\
    \bar{\boldsymbol{x}}_{{k}^{[\mathrm{S}]}} &= \frac{1}{|\mathcal{T}_{k^{[\mathrm{S}]}}|} \sum_{t\in \mathcal{T}_{k^{[\mathrm{S}]}}} \boldsymbol{x}_t, \\
    \boldsymbol{S}_{\boldsymbol{x},k^{[\mathrm{S}]}} &= \sum_{t\in\mathcal{T}_{k^{[\mathrm{S}]}}}\left({\boldsymbol{x}}_{t}-\bar{\boldsymbol{x}}_{{k}^{[\mathrm{S}]}}\right) \left({\boldsymbol{x}}_{t}-\bar{\boldsymbol{x}}_{{k}^{[\mathrm{S}]}}\right)^\top,\\\boldsymbol{W}_{\boldsymbol{x},k^{[\mathrm{S}]}}' &= \boldsymbol{W}_{\boldsymbol{x},0} + \boldsymbol{S}_{\boldsymbol{x},k^{[\mathrm{S}]}} + \frac{\kappa_{\boldsymbol{x},0} |\mathcal{T}_{k^{[\mathrm{S}]}}|}{\kappa_{\boldsymbol{x},0} + |\mathcal{T}_{k^{[\mathrm{S}]}}|}\left(\bar{\boldsymbol{x}}_{{k}^{[\mathrm{S}]}}-\boldsymbol{\mu}_{\boldsymbol{x},0}\right) \left(\bar{\boldsymbol{x}}_{{k}^{[\mathrm{S}]}}-\boldsymbol{\mu}_{\boldsymbol{x},0}\right)^\top,\\
    \boldsymbol{\mu}_{\boldsymbol{x},k^{[\mathrm{S}]}}' &=\frac{\kappa_{\boldsymbol{x},0}\boldsymbol{\mu}_{\boldsymbol{x},0}+|\mathcal{T}_{k^{[\mathrm{S}]}}|\bar{\boldsymbol{x}}_{{k}^{[\mathrm{S}]}}}{\kappa_{\boldsymbol{x},0}+|\mathcal{T}_{k^{[\mathrm{S}]}}|},
\end{align}
\end{subequations}
and $|\mathcal{T}_{k^{[\mathrm{S}]}}|$, $\bar{\boldsymbol{x}}_{{k}^{[\mathrm{S}]}}$, and $\boldsymbol{S}_{k^{[\mathrm{S}]}}$ are the number, the sample mean, and the sample covariance of the data points assigned to component ${k}^{[\mathrm{S}]}$.

\subsection{Computational Cost Analysis}

The FHMM-IDM model involves a joint latent space of driving regimes ($K^{[\mathrm{B}]}$) and traffic scenarios ($K^{[\mathrm{S}]}$), with total joint states $|\mathcal{Z}| = K^{[\mathrm{B}]} \times K^{[\mathrm{S}]}$. Learning is performed via MCMC, which requires repeated inference on multiple sequences of total length $T$ and over $M = m_1 + m_2$ iterations. And $|\boldsymbol{\theta}|$ is the dimension of the IDM parameters, typically $|\boldsymbol{\theta}| = 5$.

Table~\ref{tab:compute_cost} summarizes the dominant computational costs per MCMC iteration, per trajectory. While the method is computationally intensive, it remains feasible using modern computing resources and can be parallelized over trajectories.

\begin{table}[t]
\centering
\caption{Per-iteration computational cost of the MCMC inference for FHMM-IDM.}
\begin{tabular}{lll}
\toprule
\textbf{Component} & \textbf{Description} & \textbf{Cost} \\
\midrule
Forward-Backward (Section \ref{sample_z}) & Smoothing over $|\mathcal{Z}|$ states & $\mathcal{O}(T \cdot |\mathcal{Z}|^2)$ \\
Transition Counting (Section \ref{sample_pi}) & Compute transitions and Dirichlet update & $\mathcal{O}(T \cdot |\mathcal{Z}| + |\mathcal{Z}|^2)$ \\
IDM Parameter Update (Section \ref{sample_IDM_params}) & Metropolis-Hastings over $T$ points & $\mathcal{O}(T \cdot |\boldsymbol{\theta}| + K^{[\mathrm{B}]} \cdot |\boldsymbol{\theta}|^2)$ \\
IDM Priors Update (Section \ref{sample_mu_and_Lambda}) & Normal-Wishart per $K^{[\mathrm{B}]}$ & $\mathcal{O}(K^{[B]} \cdot |\boldsymbol{\theta}|^2 + |\boldsymbol{\theta}|^3)$ \\
Variance Update (Section \ref{sample_sigma}) & Inverse-Gamma sampling per $K^{[\mathrm{B}]}$ & $\mathcal{O}(K^{[\mathrm{B}]})$ \\
Scenario Parameter Update (Section \ref{sample_mux_and_Lambdax}) & Normal-Wishart per $K^{[\mathrm{S}]}$ & $\mathcal{O}(K^{[\mathrm{S}]})$ \\
\hline
Memory Cost & $\alpha$, $\beta$, and $\gamma$ messages for smoothing & $\mathcal{O}(T \cdot |\mathcal{Z}|)$ \\
\bottomrule
\end{tabular}
\label{tab:compute_cost}
\end{table}

\section{IDENTIFICATION OF INTERPRETABLE DRIVING REGIMES}\label{experiment}
\subsection{Dataset and Preprocessing}
Experiments are performed on the HighD dataset that contains high-resolution naturalistic vehicle trajectories extracted from drone videos of German highways.
Compared to the commonly used NGSIM dataset, the HighD dataset benefits from more reliable data capture methods and advanced computer vision techniques. 
It features 60 recordings captured at different times of the day, ranging from 8:00 to 17:00, and has a resolution of 25 Hz. In our experiment, the original dataset is downsampled to a smaller set with a sampling frequency of $5$ Hz, achieved by uniformly selecting every $5$-th sample. 
The HighD dataset provides detailed information on vehicle trajectories, velocities, and accelerations, which is essential for developing and evaluating car-following models that accurately capture real-world traffic scenarios. 
In this study, we follow the same data processing procedures as in \citet{zhang2024bayesian} to transform the data into a new coordinate system. we selected $100$ leader-follower pairs where the car-following duration lasted for more than 50 seconds. By using pairs with longer car-following duration, we aim to capture more realistic driving behaviors and enable our model to handle complex and dynamic traffic situations better.

\subsection{Modeling Setup and Assumptions}
{
In our FHMM-IDM framework, several standard Bayesian choices---such as Dirichlet priors for the transition matrix, and Normal-Wishart priors for Gaussian emissions---are adopted for analytical tractability and empirical robustness. We now assess whether these assumptions are suitable for the observed driving data and behavior dynamics.

\begin{itemize}
    \item \textbf{Dirichlet Prior on Transition Matrix $\boldsymbol{\pi}$}: Each row of the joint transition matrix $\boldsymbol{\pi}$ is assigned an independent Dirichlet prior with symmetric concentration parameters. This encourages sparse transitions, reflecting empirical observations where drivers remain in the same latent mode over time. While effective for capturing persistence, this assumption does not model structured preferences among state transitions. More flexible priors, such as hierarchical Dirichlet or logistic-normal distributions, could encode such asymmetries, but the standard Dirichlet provides a good balance between simplicity and expressiveness in our setting \citep{zhang2021spatiotemporal}.
    \item \textbf{Normal--Wishart Prior on Scenario Emissions}: For each traffic scenario, the model assumes $\boldsymbol{x}_t = [v_t, \Delta v_t, s_t]$ follows a multivariate Gaussian distribution parameterized by $\boldsymbol{\mu}_x$ and $\boldsymbol{\Lambda}_x^{-1}$, with a Normal--Wishart prior \citep{chen2023bayesian}. Although variables like $s_t$ and $\Delta v_t$ may be skewed in raw form, we apply standardization across the dataset, resulting in approximately symmetric and unimodal distributions within each regime. Therefore, Gaussian emissions are a reasonable assumption. Nonetheless, future work could explore heavy-tailed or skewed distributions to better capture extreme events.
    \item \textbf{Gaussian Noise in Acceleration Residuals $y_t$}: 
    Given the driver behavior state $z_t^{[\mathrm{B}]} = k$, we model the acceleration as $y_t \sim \mathcal{N}(\text{IDM}(\boldsymbol{x}_t; \boldsymbol{\theta}_k), \sigma_k^2)$. This Gaussian residual assumption implies that the model treats the deviations from deterministic IDM responses as temporally uncorrelated noise (i.e., independent and identically distributed). However, prior works (e.g., \citet{zhang2024bayesian, zhang2024calibrating}) have shown that residual acceleration errors in car-following behaviors can exhibit non-negligible temporal autocorrelations, especially under stop-and-go or high-density traffic conditions. While our current formulation assumes independence across time for computational efficiency and clarity, incorporating temporally correlated noise---for example, via a latent residual process or a GP-modulated emission---could enhance realism and improve performance in long-horizon trajectory prediction. This represents a promising direction for extending the FHMM-IDM framework.
    \item \textbf{Log-Normal Prior on IDM Parameters $\boldsymbol{\theta}_k$}: To capture the positive and skewed nature of IDM parameters, we place log-normal priors on $\boldsymbol{\theta}_k$. This choice is supported by empirical distributions from the literature (e.g., \citet{treiber2000congested, zhang2024bayesian}). Posterior samples across different behavior states stay within realistic and interpretable ranges.
\end{itemize}

The modeling assumptions in FHMM-IDM are well-aligned with empirical characteristics of naturalistic driving data. While conjugate priors and Gaussian likelihoods offer tractability and adequate performance, the framework could be extended with more flexible or robust alternatives---such as non-conjugate priors, heavy-tailed noise models, or structured transition dependencies---to better capture rare or extreme behaviors.
}

For the implementation details of the priors, we set $\mathrm{Dir}(1/K^{\mathrm{[b]}},\cdots,1/K^{\mathrm{[b]}})$ for Dirichlet distribution to imply sparse state assignments. The hyperparameters for the prior of IDM is set with $\boldsymbol{\mu}_0=[33, 2, 1.6, 1.5, 1.67]$ (as suggested by \citet{treiber2000congested}), $\kappa_{0}=0.01$, $\nu_{0}=7$, and $\boldsymbol{W}_0=\mathrm{diag}([0.1, 0.1, 0.1, 0.1, 0.1])$ as the covariance matrix. We set $\gamma_a=100$ and $\gamma_b=1$ for the inverse Gamma prior to suppress the noise variances. For the conjugate normal-Wishart priors, we set $\nu_{\boldsymbol{x},0}=5$,  $\kappa_{\boldsymbol{x},0}=0.01$, $\boldsymbol{\mu}_{\boldsymbol{x},0}=[0,0,0]$ (with standardization) and $\boldsymbol{W}_{\boldsymbol{x},0}=\mathrm{diag}([0.1, 0.1, 0.1])$. For each chain, the burn-in iteration is set as $m_1=6000$, and we collect $m_2=2000$ samples to estimate the posteriors. The code will be released at \url{https://github.com/Chengyuan-Zhang/Markov_Switching_IDM} upon acceptance of the paper. It is implemented purely with NumPy, without relying on integrated probabilistic programming frameworks such as PyMC or Stan.

\begin{table}[t]
\centering
\caption{Clarification of key terminologies used in this study. \texttt{Latent States} are latent model variables inferred by the FHMM from trajectory data, and each state jointly characterizes a unique combination of an external \texttt{Traffic Scenario} and a driver's internal \texttt{Driving Regime}. A \texttt{Traffic Scenario} refers to the external contextual conditions (e.g., congestion, free-flow) under which drivers operate, while a \texttt{Driving Regime} represents a short-term behavioral mode or specific driving action (e.g., aggressive acceleration, cautious following). Real-world trajectory \texttt{Cases} are empirical examples selected from the highD dataset to demonstrate representative behaviors and validate the model's capability to capture interactions between scenarios and regimes.}
\begin{tabular}{llp{10cm}}
\toprule
\textbf{Term} & \textbf{Domain} & \textbf{Description} \\
\midrule
\texttt{Latent State} & Latent model variable & Inferred latent states in the HMM that jointly encode traffic scenarios and driving regimes. These are learned from data and do not correspond one-to-one with predefined scenarios or regimes. \\
\texttt{Traffic Scenario} & Traffic context & Inferred external traffic conditions such as free-flow, car-following, or stop-and-go, etc. Used to contextualize the environment in which drivers operate. \\
\texttt{Driving Regime} & Driver behavior & A specific, short-term car-following action or behavioral mode (e.g., aggressive gap-closing, cautious following, free-flow cruising). These are transient states, not fixed driver traits. \\
\texttt{Case} & Empirical examples & Real-world car-following trajectories selected from the highD dataset to illustrate representative behaviors and validate the model. \\
\bottomrule
\end{tabular}
\label{tab:terminology}
\end{table}

\subsection{Interpretable Latent States}
In the following, we demonstrate the experiments results with $(K^{\mathrm{[B]}}=2,K^{\mathrm{[S]}}=2)$ and $(K^{\mathrm{[B]}}=5,K^{\mathrm{[S]}}=5)$, respectively. For each $k^{\mathrm{[B]}}\in[1, \dots,K^{\mathrm{[B]}}]$, we analyze the corresponding driving regime, and for each $k^{\mathrm{[S]}}\in[1, \dots,K^{\mathrm{[S]}}]$, we show the corresponding traffic scenario. A summary of the terminology used to distinguish among latent states, driving regimes, traffic scenarios, and case studies is provided in Table~\ref{tab:terminology}.

\begin{table}[t]
\centering
\caption{Learned IDM parameters $(\boldsymbol{\theta}_k)$ and noise standard deviation $(\sigma_k)$ for each driving regime. }
\begin{tabular}{ccllc} 
\toprule
\textbf{Total states} & \textbf{State No.} & $\boldsymbol{\theta}_k=[v_f,s_0,T,a_{\mathrm{max}}, b]$ & $\boldsymbol{\sigma_k}$ & \textbf{Driving Regime} \\ 
\midrule

$K^{\mathrm{[B]}}=1$ & $\#1$ & [38.57, 2.71, 0.86, 0.14, 1.33] & 0.40 & Averaged Behavior \\
\midrule

\multirow{2}{*}{$K^{\mathrm{[B]}}=2$} & $\#1$ & {[39.26, 1.80, 0.59, 0.30, 1.40]} & 0.47 & High-Speed Seeking \\
& $\#2$ & {[9.84, 5.10, 1.29, 0.08, 0.50]} & 0.15 & Congested Cruising \\
\midrule

\multirow{5}{*}{$K^{\mathrm{[B]}}=5$} & $\#1$ & {[31.51, 4.32, 1.60, 0.13, 1.42]} & 0.11 & Cautious Following \\
& $\#2$ & {[34.10, 1.17, 0.44, 0.90, 4.45]} & 0.33 & Aggressive Following\\
& $\#3$ & {[11.26, 10.20, 3.09, 0.07, 1.51]} & 0.23 & Congested Cruising \\
& $\#4$ & {[33.11, 2.15, 0.90, 0.37, 1.51]} & 0.08 & Steady-State Following \\
& $\#5$ & {[42.11, 1.15, 0.71, 0.62, 1.72]} & 0.11 & High-Speed Seeking \\
\bottomrule
\end{tabular}
\label{tab:idm_params}
\end{table}

Table~\ref{tab:idm_params} outlines the learned IDM parameters along with the corresponding standard deviation $\sigma_k$ for each driving regime $k^{\mathrm{[B]}}$. The standard deviation $\sigma_k$ reflects the uncertainty in the parameter estimates for each driving regime, highlighting the model’s ability to capture the variability in driver behavior across different regimes. These results demonstrate how the FHMM-IDM framework effectively identifies and characterizes multiple driving regimes based on the underlying patterns in car-following behavior. When $K^{\mathrm{[B]}} = 1$, the model reduces to a conventional single-regime IDM (i.e., the pooled Bayesian IDM \citep{zhang2024bayesian}), producing an ``Averaged Behavior'' that aggregates across all driving conditions. While this baseline provides a coarse fit, it fails to account for the diversity and temporal variability present in real-world trajectories.

\begin{table}[t]
\centering
\caption{Learned parameters of each traffic scenario latent state. Each scenario is characterized by the mean speed $\mu_v$, relative speed $\mu_{\Delta v}$, and spacing $\mu_s$, forming the mean vector $\boldsymbol{\mu}_{\boldsymbol{x},k^{\mathrm{[S]}}}$. The interpretation column describes the typical traffic condition reflected by each state, inferred from statistical patterns and their behavioral context.}
\begin{tabular}{cccl} 
\toprule 
\textbf{Total states} & \textbf{Scenario No.} & \boldmath$\boldsymbol{\mu}_{\boldsymbol{x},k^{\mathrm{[S]}}} = [\mu_v, \mu_{\Delta v}, \mu_s]$ & \textbf{Interpretation} \\
\midrule
$K^{\mathrm{[S]}}=1$ 
  & $\#1$ & [5.73, 0.00, 14.32]  & Averaged Traffic \\
\midrule
\multirow{2}{*}{$K^{\mathrm{[S]}}=2$} 
  & \#1 & [3.95, -0.08, 9.12]   & Congested and Dense Traffic \\
  & \#2 & [8.30, 0.10, 21.82]   & High-Speed Cruising \\
\midrule
\multirow{5}{*}{$K^{\mathrm{[S]}}=5$} 
  & \#1 & [5.71, 0.73, 19.04]   & Approaching (Stop-and-Go) \\
  & \#2 & [6.20, -0.34, 38.96]  & Gradual Dissipation \\
  & \#3 & [4.89, 0.02, 12.67]   & Steady-State Following \\
  & \#4 & [3.66, -0.20, 6.90]   & Congested and Dense Traffic \\
  & \#5 & [10.22, -0.17, 16.54] & High-Speed Cruising \\
\bottomrule
\end{tabular}
\label{tab:scenario_params}
\end{table}

It is interesting to observe that when the model is configured with $K^{\mathrm{[B]}} = 2$ and $K^{\mathrm{[S]}} = 2$, the FHMM-IDM yields a binary segmentation of both driving regimes and traffic scenarios (see Table~\ref{tab:idm_params} and Table~\ref{tab:scenario_params}). In this setting, \texttt{Regime~\#1} corresponds to a \textit{High-Speed Seeking} behavior, characterized by a high free-flow speed, short desired time headway, and moderate acceleration and braking capabilities. This regime reflects proactive and assertive driving under relatively unconstrained conditions. In contrast, \texttt{Regime~\#2} reflects a \textit{Congested Cruising} mode, with low speed preference, large desired spacing, long headway, and minimal responsiveness, indicative of passive, slow-paced behavior commonly seen in stop-and-go traffic. Readers interested in similar outcomes may find \citet{zhang2023interactive} to be a useful reference. The two inferred traffic scenarios similarly reflect a coarse partition into high-speed/large-gap and low-speed/small-gap environments, capturing the broad contextual distinctions in which these driving patterns occur.

Although this coarse binary partitioning captures a basic dichotomy between fast, gap-closing behavior and conservative, gap-maintaining behavior, it inevitably oversimplifies the diversity of driving actions observed in naturalistic trajectories. For instance, it fails to distinguish between transitional regimes such as steady-state following, acceleration bursts, or braking responses, which are critical for understanding the dynamics of car-following interactions. To more faithfully represent these variations, we increase the number of latent states to $K^{\mathrm{[B]}} = 5$ and $K^{\mathrm{[S]}} = 5$, which enables the model to uncover a more nuanced and granular structure, revealing five distinct driving regimes and traffic scenarios that better capture the range of human driving behaviors and their contextual dependencies.

To better understand the behavioral distinctions uncovered by the model, we examine the characteristics of each inferred driving regime based on the calibrated IDM parameters listed in Table~\ref{tab:idm_params}. \texttt{Regime \#1} (\textit{Cautious Following}) represents cautious driving with moderate desired speed, relatively large desired gap, long headway, and gentle acceleration capabilities, indicative of defensive and careful gap management. \texttt{Regime \#2} (\textit{Aggressive Following with Abrupt Deceleration}) characterizes assertive driving with short headways, high acceleration capability, and notably large deceleration capacity, indicative of aggressive gap management combined with readiness for abrupt braking events. \texttt{Regime \#3} (\textit{Congested Cruising}) corresponds to cautious driving behaviors in heavy congestion, characterized by very low desired speed, large spacing, long time headway, and minimal acceleration. \texttt{Regime \#4} (\textit{Steady-State Following}) captures balanced and stable tracking behavior, marked by moderate desired speed, moderate headway, and balanced acceleration and deceleration, suitable for stable car-following under moderate conditions. Finally, \texttt{Regime \#5} (\textit{High-Speed Seeking}) represents confident driving aiming for high-speed operation, characterized by very high desired speed, short headway, and high acceleration capability, reflecting proactive, high-speed cruising behavior. Together, these five regimes span a diverse spectrum of driver actions, substantially enhancing the behavioral realism and interpretability of the FHMM-IDM framework.

\begin{figure*}[t]
    \centering
    \subfigure{\centering
    \resizebox{!}{6.cm}{\usetikzlibrary{shapes,snakes,arrows,positioning}
\tikzset{>=latex}

\begin{tikzpicture}

\newcommand{\Kb}{5} % Number of k^{[B]} states (columns)
\newcommand{\Ks}{5} % Number of k^{[S]} states (rows)
\newcommand{\ww}{0.5}
\newcommand{\origin}{0}

\shade[left color=red!80!white, right color=red!10!white]
    (\origin,2.5-0.5) rectangle (\origin+2.5,3-0.5);
\shade[left color=yellow!80!white, right color=yellow!10!white]
    (\origin,2-0.5) rectangle (\origin+2.5,2.5-0.5);
\shade[left color=green!80!white, right color=green!10!white]
    (\origin,1.5-0.5) rectangle (\origin+2.5,2.-0.5);
\shade[left color=cyan!80!white, right color=cyan!10!white]
    (\origin,1.-0.5) rectangle (\origin+2.5,1.5-0.5);
\shade[left color=blue!80!white, right color=blue!10!white]
    (\origin,0.5-0.5) rectangle (\origin+2.5,1.-0.5);
\draw [step=0.5, very thick, color=white] (\origin+0,0.5-0.5) grid (\origin+2.5,3-0.5);
\draw [thick] (\origin,0.5-0.5) rectangle (\origin+2.5,3-0.5);

\node[above] at (1.37,2.7) {\small $k^{\mathrm{[B]}}$};
\node[rotate=90, left] at (-0.55,1.7) {\small $k^{\mathrm{[S]}}$};

% X-ticks Labels (k^{[B]})
\foreach \b in {0,...,4} {
    \pgfmathsetmacro\x{\b * 0.5}
    \pgfmathtruncatemacro\bb{\b + 1}
    \node[below] at (\x+0.25,2.85) {\scriptsize $\bb$};
}

% Y-ticks Labels (k^{[S]})
\foreach \s in {0,...,4} {
    \pgfmathsetmacro\y{2.5 - \s * 0.5}
    \pgfmathtruncatemacro\ss{\s + 1}
    \node[left] at (0,\y-0.25) {\scriptsize $\ss$};
}

%%%%%%%%%%%%%

\foreach \s in {0,...,4} {
    \foreach \b in {0,...,4} {
        \pgfmathsetmacro\x{\b * 0.5}    % Column positions
        \pgfmathsetmacro\y{2.5 - \s * 0.5}  % Row positions
        \pgfmathsetmacro\index{\b + \s * \Kb}  % Compute the 1D index
        
        % Only draw arrows for the first 3 and last 2 indices
        \pgfmathparse{\index<3 ? 1 : (\index>25 ? 1 : 0)}
        \ifnum\pgfmathresult=1
            \draw[->] (\x+0.25,\y-0.25) to[out=-90, in=180] (3.5,-\index*0.5+2.5-0.25);
        \fi

        \pgfmathparse{\index<0 ? 1 : (\index>22 ? 1 : 0)}
        \ifnum\pgfmathresult=1
            \draw[->] (\x+0.25,\y-0.25) to[out=-90, in=180] (3.5,-\index*0.5+1.5-0.25+20*0.5);
        \fi
        
    }
}

\filldraw [fill=red!70!white] (3.5,2.5) rectangle (4.5, 2);
\filldraw [fill=red!50!white] (3.5,2) rectangle (4.5, 1.5);
\filldraw [fill=red!30!white] (3.5,1.5) rectangle (4.5, 1);
\filldraw [fill=blue!20!white] (3.5,0) rectangle (4.5, -0.5);
\filldraw [fill=blue!10!white] (3.5,-0.5) rectangle (4.5, -1);

% --- Draw the vectorized 1D representation ---
% First 3 states
\foreach \index in {0,1,2} {
    \pgfmathsetmacro\yv{-\index * 0.5 + 2.5}  % Compute vertical position
    \pgfmathtruncatemacro\kb{1+floor(\index / \Kb)}  % Compute k^{[S]} using integer division
    \pgfmathtruncatemacro\ks{1+mod(\index, \Kb)}     % Compute k^{[B]} using modulo
    \draw[thick] (3.5,\yv) rectangle (4.5,\yv-0.5);
    \node at (4,\yv-0.25) {\scriptsize $(\kb, \ks)$};
}

% Omitted states with vdots
\node at (4,0.55) {\vdots};

% Last 2 states
\foreach \index in {23,24} {
    \pgfmathsetmacro\yv{-(\index-20) * 0.5 + 1.5}  % Adjusted vertical position
    \pgfmathtruncatemacro\kb{1+floor(\index / \Kb)}  % Compute k^{[S]} using integer division
    \pgfmathtruncatemacro\ks{1+mod(\index, \Kb)}     % Compute k^{[B]} using modulo
    \draw[thick] (3.5,\yv) rectangle (4.5,\yv-0.5);
    \node at (4,\yv-0.25) {\scriptsize $(\kb, \ks)$};
}

% Labels
\node[align=center, anchor=mid] at (4.,-1.37) {\scriptsize vectorized indexing};

\node[text width=2.cm, align=center, anchor=mid] at (1.25,-.95) {\scriptsize bivariate joint latent states};
\node[align=center, anchor=mid] at (1.25,-.5) {\scriptsize $\boldsymbol{k}=(k^{\mathrm{[B]}},k^{\mathrm{[S]}})$};

\node[rotate=90, align=center, anchor=mid] at (5.,1.) {\scriptsize 25 latent states};
\draw[decorate,decoration={brace,amplitude=5pt}, thick]
    (4.6,2.5) -- (4.6,-1.);

\end{tikzpicture}}
    }
    \subfigure{
        \centering
        \includegraphics[width=0.4\textwidth]{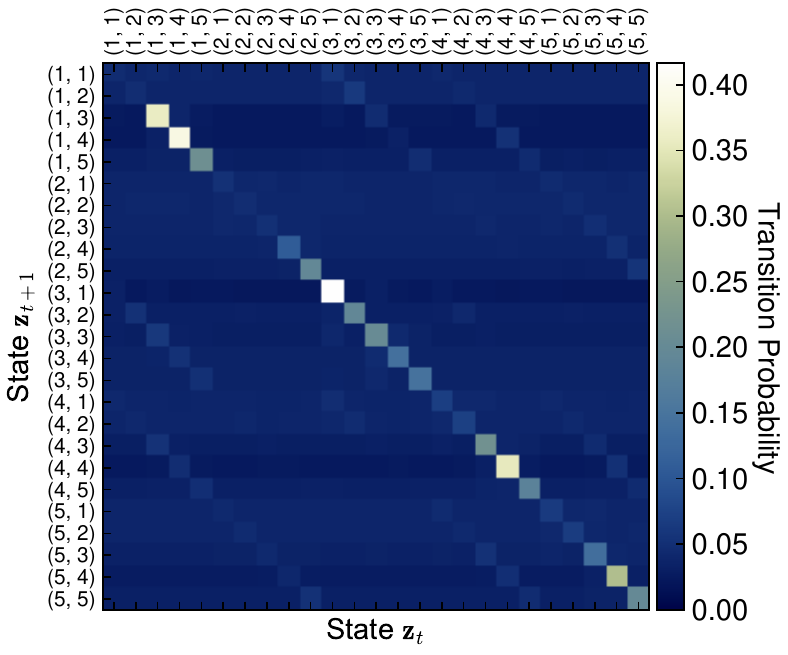}
    }
    \caption{The indexing mechanism and the state transition matrix.}
    \label{fig:trans_mat}
\end{figure*}
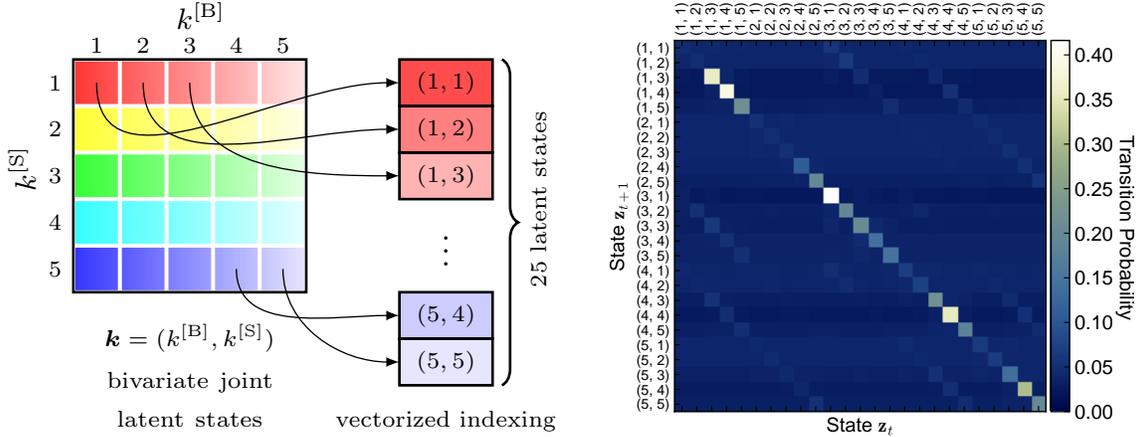

With $K^{\mathrm{[B]}} = 5$ and $K^{\mathrm{[S]}} = 5$, Fig.~\ref{fig:trans_mat} illustrates the indexing mechanism and the state transition matrix of the FHMM-IDM model. In this formulation, the latent state is factorized into two independent components: the driving regime factor, $z_t^{\mathrm{[B]}}$, which reflects intrinsic patterns of driver action (e.g., acceleration, deceleration, cruising), and the traffic scenario factor, $z_t^{\mathrm{[S]}}$, which encodes external conditions such as speed and spacing. Each joint latent state is represented as a pair $\left(k^{\mathrm{[B]}}, k^{\mathrm{[S]}}\right)$ and mapped to a unique index in a vectorized state space, allowing the construction of a unified state transition matrix $\boldsymbol{\pi} \in \mathbb{R}^{|\mathcal{Z}| \times |\mathcal{Z}|}$. Each entry $\boldsymbol{\pi}(\boldsymbol{k}', \boldsymbol{k})$ denotes the probability of transitioning from the joint state indexed by $\boldsymbol{k}'$ to that indexed by $\boldsymbol{k}$, thereby modeling the temporal evolution and interaction between driver action regimes and contextual traffic environments.

\begin{figure}[!t]
    \centering
    \includegraphics[width=.99\linewidth]{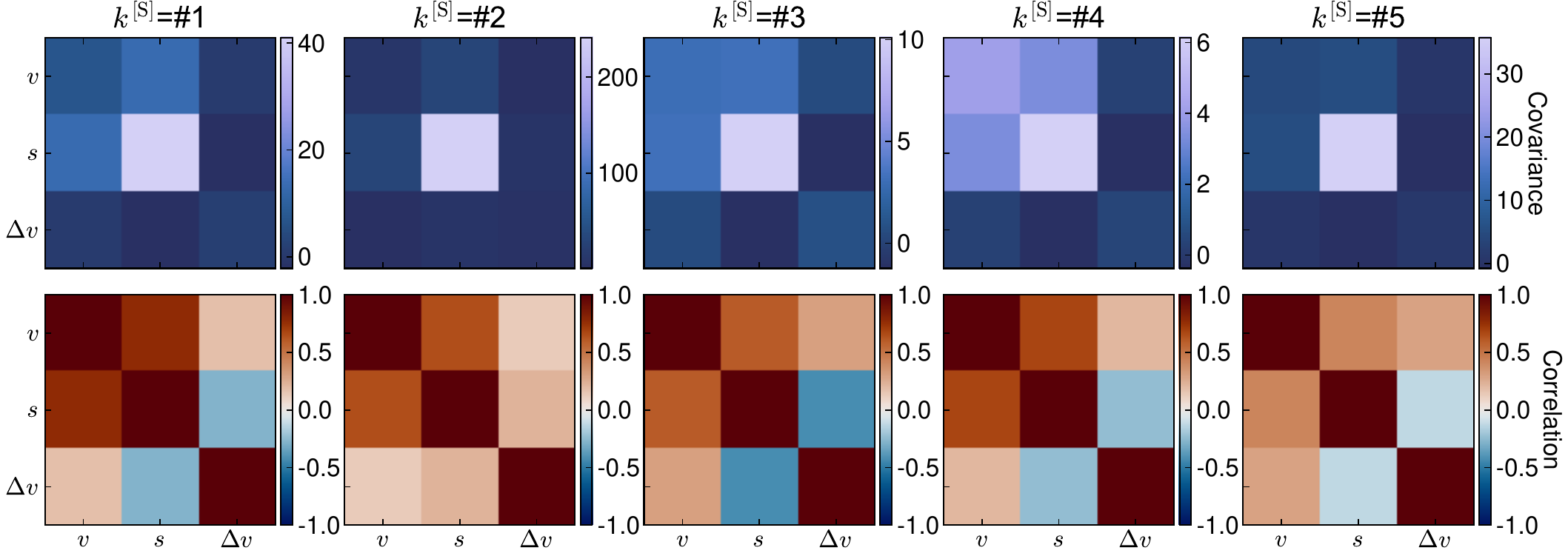}
    \caption{Visualization of the covariance matrices $\boldsymbol{\Lambda}_{\boldsymbol{x},k^{[\mathrm{S}]}}^{-1}$ and the corresponding correlation matrices.}
    \label{fig:cov_and_corr}
\end{figure}

Table~\ref{tab:scenario_params} summarizes the learned mean vectors $\boldsymbol{\mu}_{\boldsymbol{x},k^{\mathrm{[S]}}} = [\mu_v, \mu_{\Delta v}, \mu_s]$ for each latent traffic scenario under varying model complexities. When $K^{\mathrm{[S]}}=1$, the model reduces to a context-agnostic formulation, producing an ``averaged traffic'' scenario that blends behaviors across all regimes and fails to distinguish between qualitatively different traffic conditions. As the number of latent traffic scenarios increases, the model uncovers progressively finer distinctions. With $K^{\mathrm{[S]}}=2$, the model differentiates between \textit{Congested and Dense Traffic} and \textit{High-Speed Cruising}, capturing the broad dichotomy between low-speed/high-density and free-flowing conditions. However, this binary categorization remains too coarse to reflect transient or intermediate states. The five-scenario model ($K^{\mathrm{[S]}}=5$) provides a more expressive segmentation, revealing nuanced traffic contexts such as \textit{approaching} behavior in stop-and-go waves (Scenario~\#1), \textit{gradual dissipation} phases with large spacing and decaying congestion (Scenario~\#2), and \textit{steady-state following} where drivers maintain consistent gaps and speed differentials (Scenario~\#3). In contrast, Scenario~\#4 corresponds to highly congested, short-gap conditions, while Scenario~\#5 captures smooth \textit{high-speed cruising}. These patterns align with observed macroscopic flow phenomena and highlight the model’s ability to extract interpretable latent structure from raw car-following data.

Fig.~\ref{fig:cov_and_corr} provide the learned covariance matrices $\boldsymbol{\Lambda}^{-1}_{\boldsymbol{x},k^{\mathrm{[S]}}}$, with Table~\ref{tab:scenario_params} which jointly show how the FHMM-IDM model distinguishes scenario‐specific relationships among speed $v$, gap $s$, and speed difference $\Delta v$ under each latent driving‐scenario state $k^{\mathrm{[S]}}$. In Fig.~\ref{fig:cov_and_corr},
each column corresponds to a unique latent scenario $k^{\mathrm{[S]}}$. In the top row, the color scales capture the covariance between these variables, while in the bottom row, the normalized correlation matrices highlight the same relationships bounded within $[-1,1]$. Notably, different states exhibit distinct off‐diagonal elements, revealing that each latent scenario reflects a characteristic pattern of co‐movements and variability among $\{v, s, \Delta v\}$. As a result, the FHMM-IDM framework effectively uncovers traffic regimes in which drivers may exhibit strong correlations in speed and gap in one scenario, but a contrasting pattern in another.

\begin{figure}[t]
    \centering
    \includegraphics[width=0.7\linewidth]{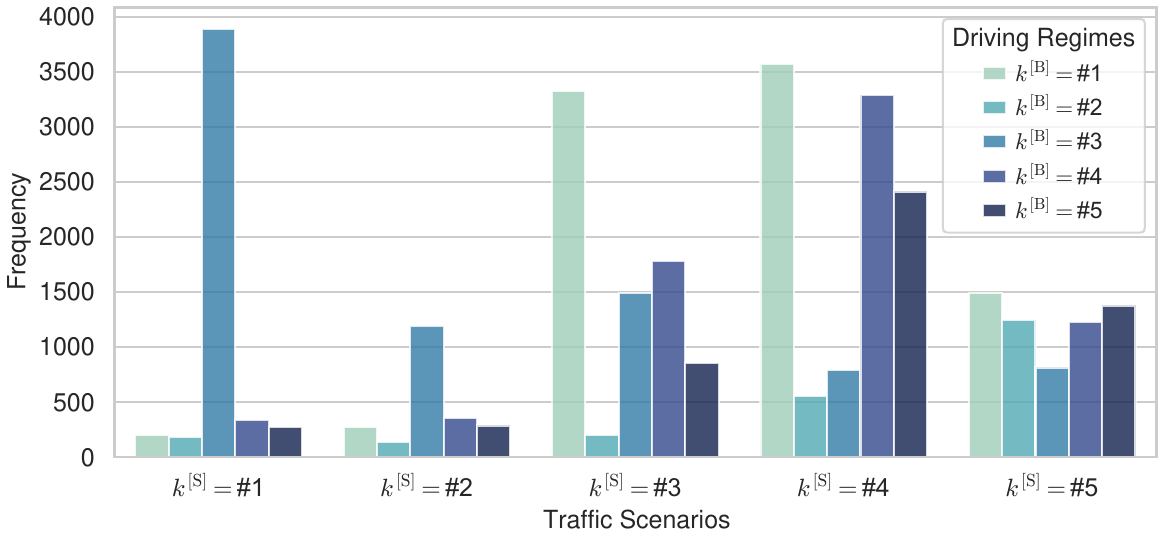}
    \caption{The histogram of driving regimes for each scenario.}
    \label{fig:histogram}
\end{figure}

\begin{itemize} 
    \item \textbf{\texttt{Scenario \#1} ($k^{\mathrm{[S]}}=1$).} This state distinctly represents an \textit{approaching scenario} commonly observed in congested or stop-and-go traffic conditions. As shown in Table~\ref{tab:scenario_params}, this scenario is characterized by a moderate vehicle speed ($\mu_v = 5.71\,\text{m/s}$), a substantial positive relative speed ($\mu_{\Delta v} = 0.73\,\text{m/s}$), and a relatively large headway ($\mu_s = 19.04\,\text{m}$). These statistics describe a traffic context in which the lead vehicle is nearly stationary or moving slowly, while the following vehicle continues to approach at a significantly higher speed—resulting in a rapidly closing gap despite active deceleration.

    The correlation matrices in Fig.~\ref{fig:cov_and_corr} further support this interpretation. A strong positive correlation between speed and gap suggests that vehicles traveling at higher speeds initially maintain longer headways. Meanwhile, a moderate negative correlation between gap and relative speed implies that drivers experience increasing closing rates as the gap shrinks, consistent with anticipatory deceleration in response to a slow or stopped lead vehicle.
    
    Moreover, the histogram in Fig.~\ref{fig:histogram} indicates that \texttt{Scenario \#1} frequently co-occurs with \texttt{Regime \#3} (congested cruising, $k^{\mathrm{[B]}}=3$), as summarized in Table~\ref{tab:idm_params}. This regime is marked by low desired speed, large spacing preferences, and cautious acceleration and braking parameters. The predominance of this pairing suggests that drivers tend to adopt conservative and anticipatory behaviors when approaching slower traffic—gradually reducing speed to maintain safety margins and avoid abrupt maneuvers. This finding highlights the model’s ability to capture meaningful interactions between contextual traffic scenarios and immediate driver action patterns.
    
    \item \textbf{\texttt{Scenario \#2} ($k^{\mathrm{[S]}}=2$).} This scenario represents a distinctive car-following context characterized by relatively low mean speed ($\mu_v = 6.20\,\text{m/s}$), a notably large gap ($\mu_s = 38.96\,\text{m}$), and a slightly negative relative speed ($\mu_{\Delta v} = -0.34\,\text{m/s}$), as reported in Table~\ref{tab:scenario_params}. The combination of low speed and generous spacing suggests a transitional state in which traffic is beginning to recover from congestion. Drivers in this scenario likely engage in \textit{gradual congestion dissipation} or exhibit cautious behavior in a \textit{dense yet slowly improving} traffic environment.

    Fig.~\ref{fig:cov_and_corr} shows a moderate positive correlation between gap ($s$) and relative speed ($\Delta v$), indicating that as vehicles maintain or slightly increase their headways, the magnitude of negative relative speed decreases. This pattern reflects drivers’ gentle adjustments in speed to preserve spacing and mitigate abrupt maneuvers, consistent with defensive behavior in transitional flow.
    
    From the histogram in Fig.~\ref{fig:histogram}, \texttt{Scenario \#2} also frequently co-occurs with \texttt{Regime \#3} (congested cruising, $k^{\mathrm{[B]}}=3$). This pairing further reinforces the interpretation that drivers in this scenario tend to adopt conservative, safety-oriented strategies by maintaining comfortably long gaps and adjusting their speeds cautiously in response to recovering traffic dynamics.

    \item \textbf{\texttt{Scenario \#3} ($k^{\mathrm{[S]}}=3$).}  
    This scenario corresponds to moderate traffic conditions, characterized by a relatively low average speed ($\mu_v = 4.89\,\text{m/s}$), a moderate headway ($\mu_s = 12.67\,\text{m}$), and an almost neutral relative speed ($\mu_{\Delta v} = 0.02\,\text{m/s}$), as shown in Table~\ref{tab:scenario_params}. The near-zero relative speed indicates that the following vehicle maintains a speed closely matched to that of the leader, resulting in a stable gap. This pattern is indicative of \textit{steady-state car-following} behavior, where drivers operate under moderately dense but stable traffic flow.
    
    The histogram in Fig.~\ref{fig:histogram} reveals that \texttt{Scenario \#3} frequently co-occurs with \texttt{Regime \#1} (cautious following, $k^{\mathrm{[B]}}=1$), \texttt{Regime \#3} (congested cruising, $k^{\mathrm{[B]}}=3$), and \texttt{Regime \#4} (steady-state following, $k^{\mathrm{[B]}}=4$). These regimes, according to Table~\ref{tab:idm_params}, span a range of cautious to responsive behaviors, including large spacing preferences, low acceleration capacity, and moderate headway maintenance. This distribution suggests that drivers in this scenario engage in smooth, adaptive tracking of the lead vehicle, with limited acceleration or deceleration, depending on their behavioral disposition. The convergence of multiple regimes further highlights the versatility of steady-state following, as drivers with varying levels of conservatism or responsiveness consistently stabilize their speed in relation to the leader under moderate traffic conditions.

    \item \textbf{\texttt{Scenario \#4} ($k^{\mathrm{[S]}}=4$).}  
    This scenario reflects a \textit{congested or dense traffic} condition, with low average speed ($\mu_v = 3.66\,\text{m/s}$), small headway ($\mu_s = 6.90\,\text{m}$), and a slightly negative relative speed ($\mu_{\Delta v} = -0.20\,\text{m/s}$). The tight spacing and low speed indicate that the follower is closely trailing the leader and making frequent small adjustments to maintain a safe distance, a hallmark of car-following in congested flow.
    
    Figure~\ref{fig:histogram} reveals that \texttt{Scenario \#4} co-occurs with multiple driving regimes, including \texttt{Regime \#1} (cautious following), \texttt{Regime \#4} (steady-state tracking), and \texttt{Regime \#5} (high-speed cruising with quick adaptation). This wide range of co-occurring regimes suggests that drivers with different behavioral tendencies all converge on similarly dense following behaviors under constrained traffic conditions. Whether adopting which regime, drivers consistently manage close headways and subtle adjustments in speed to ensure safe following during congestion.
    
    \item \textbf{\texttt{Scenario \#5} ($k^{\mathrm{[S]}}=5$).}  
    This scenario captures a \textit{higher-speed car-following} state, characterized by the highest mean speed across all scenarios ($\mu_v = 10.22\,\text{m/s}$), a moderate gap ($\mu_s = 16.54\,\text{m}$), and a slightly negative relative speed ($\mu_{\Delta v} = -0.17\,\text{m/s}$). The slight deceleration trend in the relative speed indicates that the follower maintains a stable cruising distance behind the lead vehicle, with gentle corrections to preserve spacing.

    According to Fig.~\ref{fig:histogram}, \texttt{Scenario \#5} co-occurs with all five identified driving regimes, highlighting its prevalence across diverse driver behaviors. This widespread occurrence suggests that high-speed conditions with moderate spacing are a common operational state encountered by both assertive and cautious drivers alike. The consistency in this scenario’s co-occurrence pattern underscores its role as a baseline driving context in which individuals with varying action tendencies regulate headway and speed in a similarly stable manner.
\end{itemize}

\begin{figure}[!t]
    \centering
    \includegraphics[width=\linewidth]{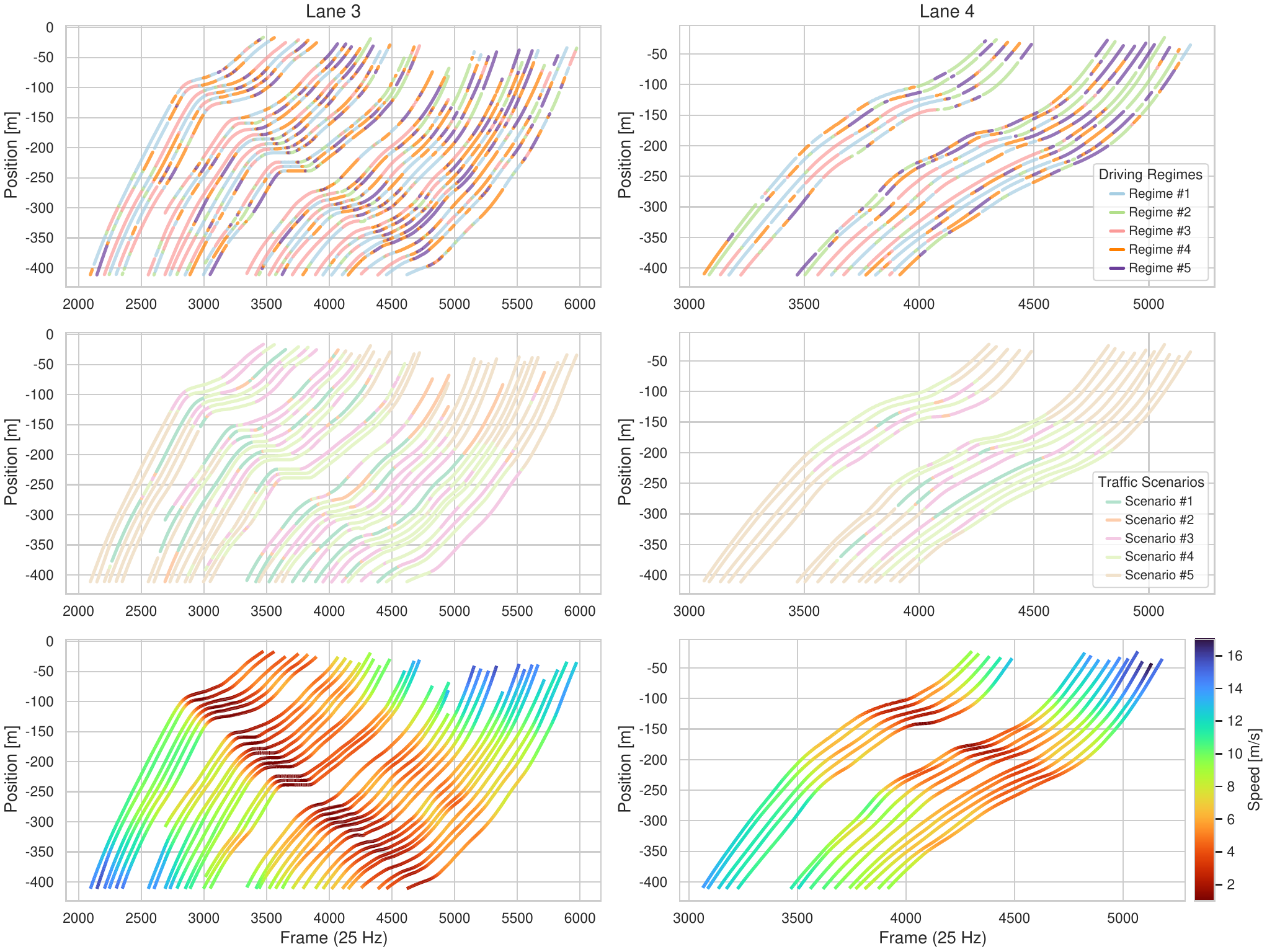}
    \caption{Samples of time–space trajectories for vehicles in Lane 3 (left) and Lane 4 (right) from the HighD dataset. \textit{First row (Driving Regime Coloring)}: Each vehicle’s longitudinal position $x_t$ (in meters) is plotted against the frame index (25 Hz), with pastel colors indicating the rounded posterior mean of the driving‐regime state $z_t^{[\mathrm{B}]}$. Whenever the inferred regime changes, a new line segment is drawn with the corresponding color from the \texttt{Paired} colormap (see legend in the Lane 4 panel). This view illustrates how drivers switch among discrete behavioral modes (e.g., aggressive, defensive, relaxed) as they travel. \textit{Second row (Scenario Coloring):} The same trajectories are recolored according to the rounded posterior mean of the traffic‐scenario state $z_t^{[\mathrm{S}]}$. Again, changes in the inferred scenario trigger a new line segment, with pastel colors representing different regimes (\texttt{Scenario \#1}–\texttt{\#5}). This highlights how vehicles transition among traffic contexts (e.g., free‐flow vs.\ congested) over time. \textit{Third row (Speed Coloring):} The trajectories are colored by actual speed $v_t$ (in m/s) using a continuous colormap (blue $\approx$ high‐speed, red $\approx$ low‐speed). The vertical colorbar at right indicates the speed scale. Comparing all three rows reveals how driving regimes and traffic‐scenario assignments correspond to underlying speed patterns---e.g., \texttt{Scenario \#4} often aligns with lower‐speed segments, while \texttt{Scenario \#5} consistently aligns with higher‐speed segments; \texttt{Scenario \#1} often co-occurs with \texttt{Regime \#3} in the approaching situations.
    }
    \label{fig:time_space_diagram}
\end{figure}

Figure~\ref{fig:time_space_diagram} demonstrates how the FHMM-IDM framework disentangles short-term driving regimes, traffic scenarios, and actual speed profiles from naturalistic vehicle trajectories on Lanes 3 and 4 of the HighD dataset. In the top row, each trajectory segment is colored by its inferred driving regime $z_t^{[\mathrm{B}]}$, revealing frequent transitions among driving regimes. The middle row displays the corresponding traffic scenario assignments $z_t^{[\mathrm{S}]}$, which capture contextual states ranging from \textit{Approaching (Stop-and-Go)} through \textit{Gradual Dissipation} and \textit{Dense Traffic}, to \textit{High-Speed Cruising}. Finally, the bottom row shows the raw vehicle speed $v_t$ over time, allowing a direct visual comparison between the latent assignments and the true kinematic behavior.

A closer inspection of the layered panels reveals consistent co-occurrence patterns that validate the interpretability of the model. For example, \texttt{Regime \#3} (Congested Cruising) often coincides with \texttt{Scenario \#1} (Approaching) during deceleration phases; \texttt{Regime \#4} (Steady-State Following) aligns with dense traffic in low-speed, tightly spaced flow; and \texttt{Scenario \#5} (High-Speed Cruising) consistently matches the segments where the speed trace is highest. These results illustrate that FHMM-IDM successfully separates internal driver intent from external traffic context, and that the latent scenario labels correspond closely to observed speed dynamics.

\subsection{Case Study}
\begin{figure}[!th]
    \centering
    \includegraphics[width=.7\linewidth]{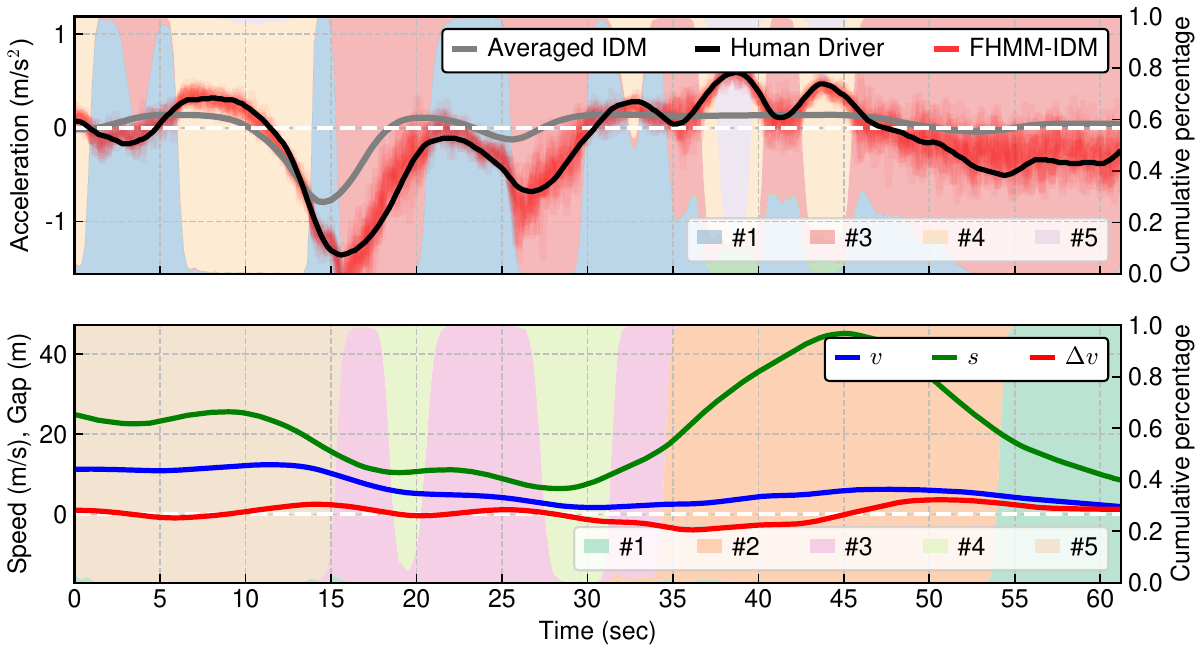}
    \caption{Case I: Comparison of human‐driver acceleration (solid black) and FHMM‐IDM predicted acceleration (solid red) with inferred driving regimes (background shading, top), and corresponding speed ($v$, blue), gap ($s$, green), and relative speed ($\Delta v$, red) with inferred traffic scenarios (background shading, bottom).}
    \label{fig:case2}
\end{figure}

\begin{figure}[!th]
    \centering
    \includegraphics[width=.7\linewidth]{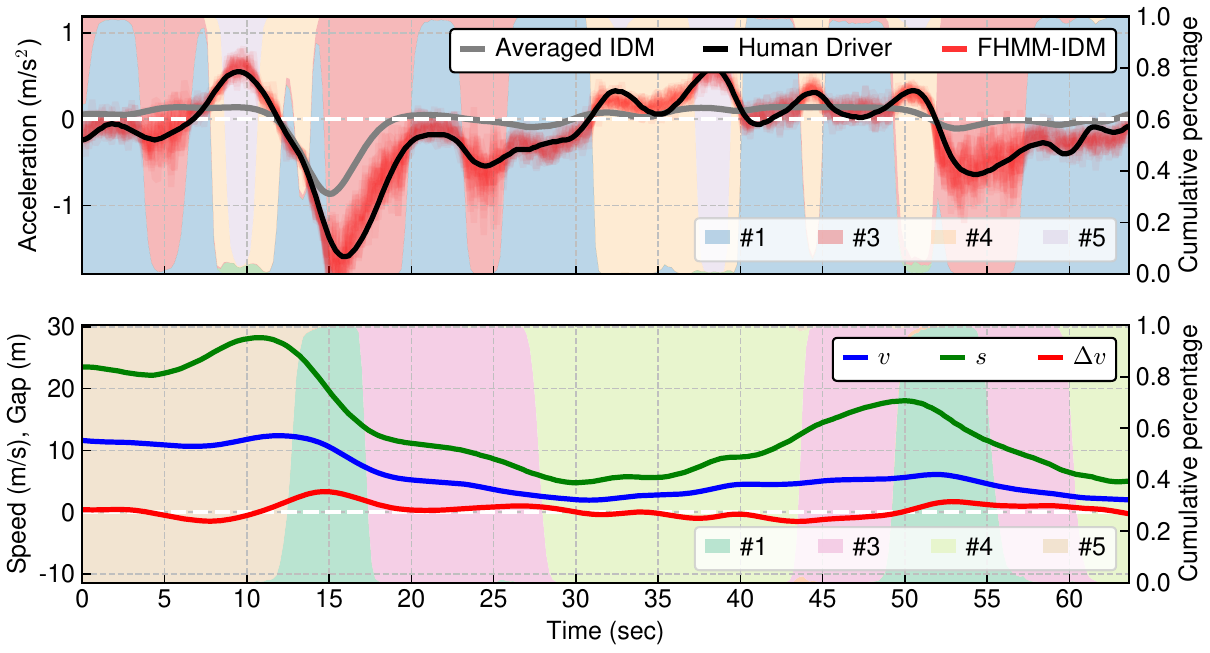}
    \caption{Case II: Comparison of human‐driver acceleration (solid black) and FHMM‐IDM predicted acceleration (solid red) with inferred driving regimes (background shading, top), and corresponding speed ($v$, blue), gap ($s$, green), and relative speed ($\Delta v$, red) with inferred traffic scenarios (background shading, bottom).} 
    \label{fig:case3}
\end{figure}

\begin{figure}[!th]
    \centering
    \includegraphics[width=.7\linewidth]{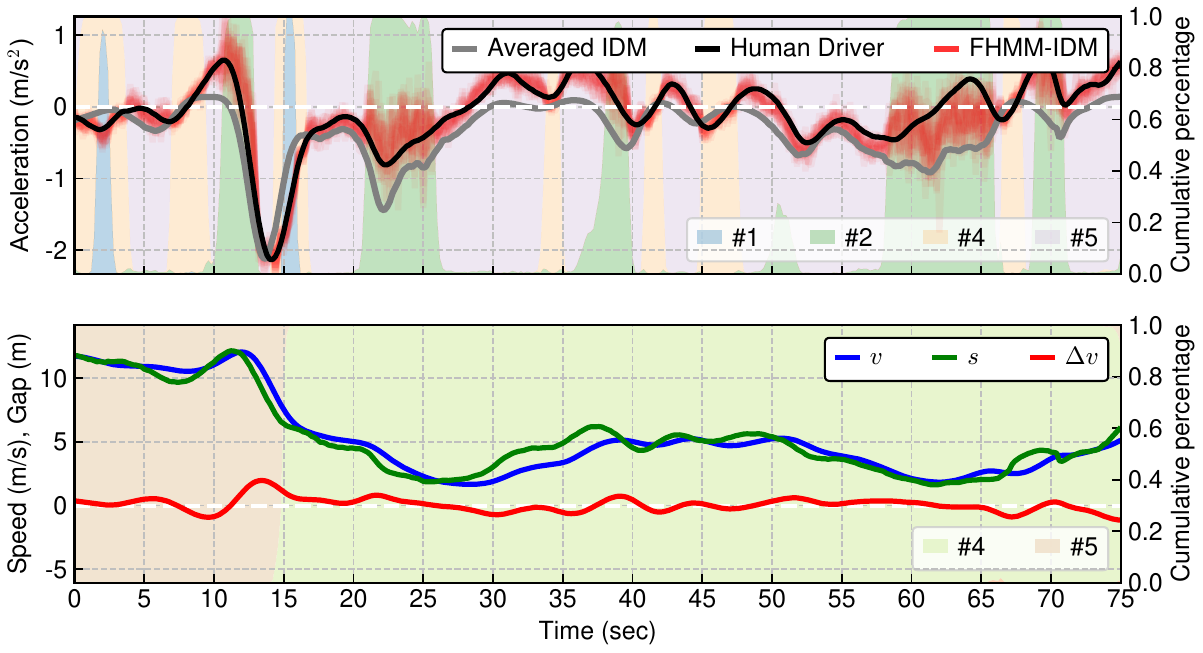}
    \caption{Case III: Comparison of human‐driver acceleration (solid black) and FHMM‐IDM predicted acceleration (solid red) with inferred driving regimes (background shading, top), and corresponding speed ($v$, blue), gap ($s$, green), and relative speed ($\Delta v$, red) with inferred traffic scenarios (background shading, bottom).} 
    \label{fig:case6}
\end{figure}

\begin{figure}[!th]
    \centering
    \includegraphics[width=.7\linewidth]{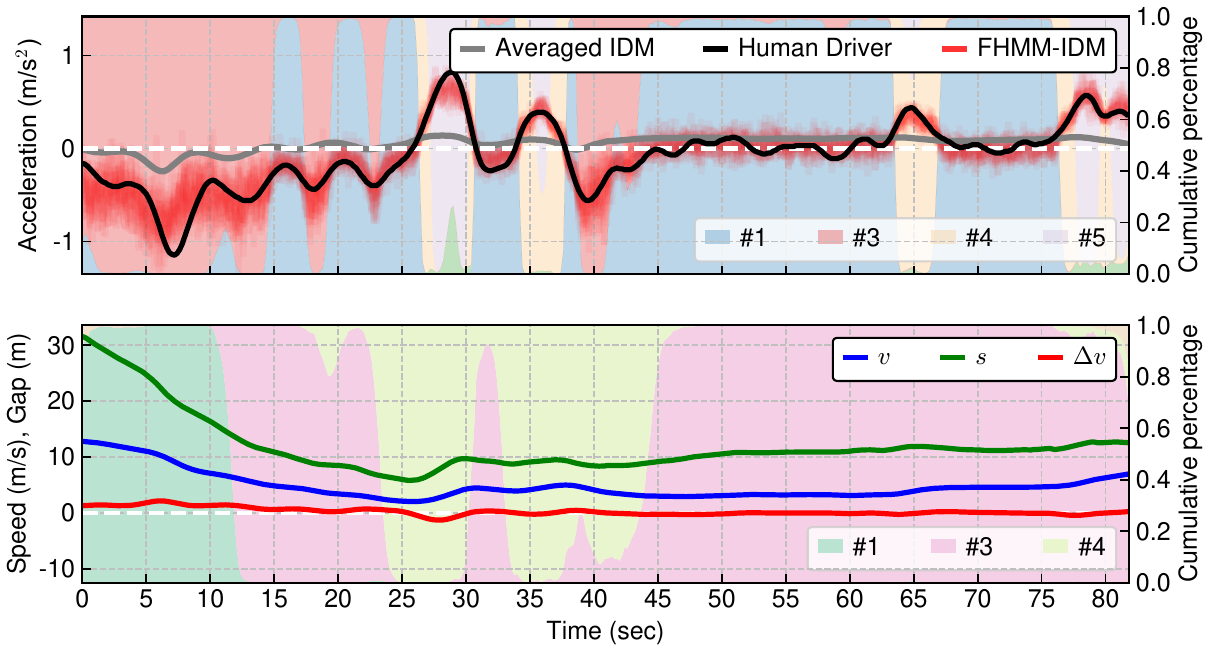}
    \caption{Case IV: Comparison of human‐driver acceleration (solid black) and FHMM‐IDM predicted acceleration (solid red) with inferred driving regimes (background shading, top), and corresponding speed ($v$, blue), gap ($s$, green), and relative speed ($\Delta v$, red) with inferred traffic scenarios (background shading, bottom).} 
    \label{fig:case9}
\end{figure}

Figures~\ref{fig:case2}, \ref{fig:case3}, \ref{fig:case6}, and \ref{fig:case9} illustrate four representative trajectories in which the FHMM-IDM framework jointly infers short-term driving regime states ($k_t^{[\mathrm{B}]}$) in the upper panels and traffic scenario states ($k_t^{[\mathrm{S}]}$) in the lower panels. In each upper plot, the human driver’s measured acceleration (black line) is overlaid with the model’s regime-specific prediction (red line, with $K^{[\mathrm{B}]}=5$), while the colored background denotes the inferred driving regime—ranging from \texttt{Regime~\#1}: \textit{Cautious Following} to \texttt{Regime~\#5}: \textit{High-Speed Seeking}. A grey line is included to represent the prediction from a single averaged IDM model (with $K^{[\mathrm{B}]}=1$), highlighting the discrepancy caused by the one-to-one mapping assumption and underscoring the benefit of regime switching. In the corresponding lower plot, the vehicle’s speed ($v$, blue), gap ($s$, green), and relative speed ($\Delta v$, red) are shown, with the shaded background indicating the inferred traffic scenario states. Together, these visualizations demonstrate how the FHMM-IDM dynamically adapts to changing contexts, assigning interpretable regime and scenario labels while closely tracking the driver’s control inputs.

In Case I (Fig.~\ref{fig:case2}), frequent transitions among \texttt{Regime \#1}, \texttt{Regime \#3}, and \texttt{Regime \#4} occur over successive segments of the trajectory, roughly covering the intervals 0--15 seconds, 15--35 seconds, and 35--60 seconds. These regime switches correspond closely to the sequence of scenario transitions: \texttt{Scenario \#5} $\rightarrow$ \texttt{\#3} $\rightarrow$ \texttt{\#4} $\rightarrow$ \texttt{\#3} $\rightarrow$ \texttt{\#2} $\rightarrow$ \texttt{\#1}. \texttt{Scenario \#5} represents high-speed conditions with moderate gaps, indicative of smooth, free-flowing traffic. Transitioning to \texttt{Scenario \#3} reflects moderate-speed conditions with relatively stable gaps, typical of steady-state traffic. \texttt{Scenario \#4} represents dense traffic characterized by reduced speeds and tighter spacing, requiring greater interaction and adaptation. A brief return to \texttt{Scenario \#3} indicates temporary relief from congestion. \texttt{Scenario \#2} introduces low-speed conditions with large gaps, reflecting cautious, transitional driving behavior during gradual flow dissipation. Finally, \texttt{Scenario \#1} captures highly congested, stop-and-go traffic marked by minimal speeds and short spacing. This chain of scenario transitions illustrates how the driver progressively adapts from free-flow to congested environments, with the model accurately capturing both behavioral responses and contextual changes.

Case II (Fig.~\ref{fig:case3}) predominantly highlights \texttt{Regime \#1} (Cautious Following), \texttt{Regime \#3} (Congested Cruising), and \texttt{Regime \#4} (Steady-State Following), corresponding to the scenario sequence: \texttt{Scenario \#5} $\rightarrow$ \texttt{\#1} $\rightarrow$ \texttt{\#3} $\rightarrow$ \texttt{\#4} $\rightarrow$ \texttt{\#3} $\rightarrow$ \texttt{\#1} $\rightarrow$ \texttt{\#3} $\rightarrow$ \texttt{\#4}. The sequence begins with \texttt{Scenario \#5}, reflecting high-speed cruising with moderate spacing, before shifting abruptly to \texttt{Scenario \#1}, which denotes approaching behavior in highly congested conditions, marked by minimal gaps and low speeds. The subsequent transitions between \texttt{Scenarios \#3} and \texttt{\#4} indicate alternation between steady-state car-following and dense traffic, while brief returns to \texttt{Scenario \#1} capture intermittent stop-and-go phases. These dynamic scenario changes are mirrored by shifts among cautious, responsive, and congestion-aware driving regimes. The frequent reappearance of \texttt{Regime \#1} during congested intervals suggests that drivers adopt defensive behaviors to maintain safety under uncertain and variable traffic conditions.

In Case III (Fig.~\ref{fig:case6}), aggressive driving regimes---\texttt{Regime \#2} (Aggressive Braking) and \texttt{Regime \#5} (High-Speed Cruising)---dominate throughout the observed period. These regimes occur primarily under alternating conditions between \texttt{Scenario \#5} (high-speed cruising) and \texttt{Scenario \#4} (congested and dense traffic). This relatively simple transition pattern (\texttt{Scenario \#5} $\rightarrow$ \texttt{\#4}) reflects a driver who persistently adopts assertive control strategies to manage speed and spacing. The consistent preference for aggressive regimes under both free-flow and denser traffic conditions suggests a strong intent to maintain efficiency and assertiveness in car-following behavior.

Lastly, Case IV (Fig.~\ref{fig:case9}) illustrates a more intricate interplay among \texttt{Regime \#1} (Cautious Following), \texttt{Regime \#3} (Congested Cruising), and \texttt{Regime \#4} (Steady-State Following). The associated traffic scenarios transition through the sequence \texttt{Scenario \#1} $\rightarrow$ \texttt{\#3} $\rightarrow$ \texttt{\#4} $\rightarrow$ \texttt{\#3}. Initially, \texttt{Scenario \#1} captures a congested, stop-and-go context, followed by a transition into \texttt{Scenario \#3}, denoting moderate speed with stable spacing indicative of steady-state following. As traffic becomes denser and spacing narrows, \texttt{Scenario \#4} arises, prompting a shift to more responsive driving adjustments. The final return to \texttt{Scenario \#3} suggests relief from congestion and restoration of steady-state conditions. These transitions underscore the driver’s adaptive strategy, modulating between cautious, stable, and responsive regimes in response to the evolving traffic environment.

Taken together, these two layers of latent states, driving regime ($k^{\mathrm{[B]}}$) and traffic scenario ($k^{\mathrm{[S]}}$), demonstrate how the FHMM-IDM framework captures the nuanced interplay between intrinsic driver behavior and the surrounding traffic environment. The detailed case analyses underscore the interpretability of the proposed model, showing how drivers dynamically transition between distinct behavioral modes in response to evolving traffic contexts. By disentangling internal decision patterns from external conditions, the FHMM-IDM enhances the realism and behavioral richness of microscopic traffic simulations.

\section{CONCLUSION AND DISCUSSIONS}\label{conclusion}

This paper presents FHMM-IDM, a novel probabilistic framework that formalizes car-following behavior as a regime-switching process governed by interacting latent dynamics. The key contribution lies in the explicit separation of intrinsic driving regimes (e.g., aggressive acceleration, steady-state following) from external traffic scenarios (e.g., free-flow, congestion), achieved through a factorial hidden Markov structure. This decomposition enables interpretable, context-dependent behavioral modeling, resolving the ambiguity and non-identifiability issues inherent in conventional one-to-one parameter mappings. Compared to prior models, FHMM-IDM offers a principled and scalable solution that combines interpretable physics-based dynamics with probabilistic inference, supporting adaptive prediction and behavioral reasoning under uncertainty.

Through extensive experiments conducted on the HighD naturalistic driving dataset, we demonstrated that FHMM-IDM successfully identifies multiple distinct driving regimes (e.g., aggressive acceleration, steady-state following, cautious cruising) and differentiates among a range of traffic scenarios, from high-speed free-flow to congested stop-and-go conditions. The use of Bayesian inference via MCMC ensures robust parameter estimation and allows for principled uncertainty quantification. Case studies further illustrate the model’s ability to realistically capture regime-switching behavior, offering insight into how drivers adapt their actions in response to evolving traffic conditions.

Future research may extend the FHMM-IDM framework to incorporate lane-changing maneuvers and broader multi-lane traffic interactions. This extension is feasible by augmenting the observation vector to include lateral displacement and a discrete lane index, although modeling such maneuvers presents challenges due to their shorter time scales and the involvement of additional control variables such as steering and yaw rate. Moreover, integrating latent factors representing environmental influences or driver internal states (e.g., emotion, attention) could further enhance the model's representational capacity. Overall, FHMM-IDM offers a foundational step toward developing interpretable and adaptive probabilistic models of driver behavior for applications in intelligent transportation systems and autonomous driving technologies.

\section*{Acknowledgment}
This research is supported by the Natural Sciences and Engineering Research Council of Canada (NSERC). C. Zhang would like to thank the McGill Engineering Doctoral Awards (MEDA), the Interuniversity Research Centre on Enterprise Networks, Logistics and Transportation (CIRRELT), Fonds de recherche du Québec -- Nature et technologies (FRQNT) for providing scholarships to support this study.

%%%%%%%%%%%%%%%%%%%%%%%%%%%%%%%%%%%%%%%%%%%%%%%%%%%%%%%%%%%%%
%% BIBLIOGRAPHY
%%%%%%%%%%%%%%%%%%%%%%%%%%%%%%%%%%%%%%%%%%%%%%%%%%%%%%%%%%%%%

\bibliographystyle{elsarticle-harv} 
\bibliography{references}

\begin{thebibliography}{35}
\expandafter\ifx\csname natexlab\endcsname\relax\def\natexlab#1{#1}\fi
\providecommand{\url}[1]{\texttt{#1}}
\providecommand{\href}[2]{#2}
\providecommand{\path}[1]{#1}
\providecommand{\DOIprefix}{doi:}
\providecommand{\ArXivprefix}{arXiv:}
\providecommand{\URLprefix}{URL: }
\providecommand{\Pubmedprefix}{pmid:}
\providecommand{\doi}[1]{\href{http://dx.doi.org/#1}{\path{#1}}}
\providecommand{\Pubmed}[1]{\href{pmid:#1}{\path{#1}}}
\providecommand{\bibinfo}[2]{#2}
\ifx\xfnm\relax \def\xfnm[#1]{\unskip,\space#1}\fi
%Type = Article
\bibitem[{Aoude et~al.(2012)Aoude, Desaraju, Stephens and How}]{aoude2012driver}
\bibinfo{author}{Aoude, G.S.}, \bibinfo{author}{Desaraju, V.R.}, \bibinfo{author}{Stephens, L.H.}, \bibinfo{author}{How, J.P.}, \bibinfo{year}{2012}.
\newblock \bibinfo{title}{Driver behavior classification at intersections and validation on large naturalistic data set}.
\newblock \bibinfo{journal}{IEEE Transactions on Intelligent Transportation Systems} \bibinfo{volume}{13}, \bibinfo{pages}{724--736}.
%Type = Book
\bibitem[{Bishop and Nasrabadi(2006)}]{bishop2006pattern}
\bibinfo{author}{Bishop, C.M.}, \bibinfo{author}{Nasrabadi, N.M.}, \bibinfo{year}{2006}.
\newblock \bibinfo{title}{Pattern recognition and machine learning}. volume~\bibinfo{volume}{4}.
\newblock \bibinfo{publisher}{Springer}.
%Type = Article
\bibitem[{Chen et~al.(2024)Chen, Zhu, Sun and Yang}]{chen2024combining}
\bibinfo{author}{Chen, K.}, \bibinfo{author}{Zhu, M.}, \bibinfo{author}{Sun, L.}, \bibinfo{author}{Yang, H.}, \bibinfo{year}{2024}.
\newblock \bibinfo{title}{Combining time dependency and behavioral game: A deep markov cognitive hierarchy model for human-like discretionary lane changing modeling}.
\newblock \bibinfo{journal}{Transportation Research Part B: Methodological} \bibinfo{volume}{189}, \bibinfo{pages}{102980}.
%Type = Article
\bibitem[{Chen et~al.(2023)Chen, Zhang, Cheng, Hou and Sun}]{chen2023bayesian}
\bibinfo{author}{Chen, X.}, \bibinfo{author}{Zhang, C.}, \bibinfo{author}{Cheng, Z.}, \bibinfo{author}{Hou, Y.}, \bibinfo{author}{Sun, L.}, \bibinfo{year}{2023}.
\newblock \bibinfo{title}{A bayesian gaussian mixture model for probabilistic modeling of car-following behaviors}.
\newblock \bibinfo{journal}{IEEE Transactions on Intelligent Transportation Systems} \bibinfo{volume}{25}, \bibinfo{pages}{5880--5891}.
%Type = Phdthesis
\bibitem[{Fox(2009)}]{fox2009bayesian}
\bibinfo{author}{Fox, E.B.}, \bibinfo{year}{2009}.
\newblock \bibinfo{title}{Bayesian nonparametric learning of complex dynamical phenomena}.
\newblock Ph.D. thesis. Massachusetts Institute of Technology.
%Type = Article
\bibitem[{Fritzsche and Ag(1994)}]{fritzsche1994model}
\bibinfo{author}{Fritzsche, H.T.}, \bibinfo{author}{Ag, D.b.}, \bibinfo{year}{1994}.
\newblock \bibinfo{title}{A model for traffic simulation}.
\newblock \bibinfo{journal}{Traffic Engineering+ Control} \bibinfo{volume}{35}, \bibinfo{pages}{317--21}.
%Type = Article
\bibitem[{Gadepally et~al.(2013)Gadepally, Krishnamurthy and Ozguner}]{gadepally2013framework}
\bibinfo{author}{Gadepally, V.}, \bibinfo{author}{Krishnamurthy, A.}, \bibinfo{author}{Ozguner, U.}, \bibinfo{year}{2013}.
\newblock \bibinfo{title}{A framework for estimating driver decisions near intersections}.
\newblock \bibinfo{journal}{IEEE Transactions on Intelligent Transportation Systems} \bibinfo{volume}{15}, \bibinfo{pages}{637--646}.
%Type = Article
\bibitem[{Ghahramani and Jordan(1995)}]{ghahramani1995factorial}
\bibinfo{author}{Ghahramani, Z.}, \bibinfo{author}{Jordan, M.}, \bibinfo{year}{1995}.
\newblock \bibinfo{title}{Factorial hidden markov models}.
\newblock \bibinfo{journal}{Advances in neural information processing systems} \bibinfo{volume}{8}.
%Type = Article
\bibitem[{Li et~al.(2021)Li, Chen, Zhou, Laval and Xie}]{li2021car}
\bibinfo{author}{Li, T.}, \bibinfo{author}{Chen, D.}, \bibinfo{author}{Zhou, H.}, \bibinfo{author}{Laval, J.}, \bibinfo{author}{Xie, Y.}, \bibinfo{year}{2021}.
\newblock \bibinfo{title}{Car-following behavior characteristics of adaptive cruise control vehicles based on empirical experiments}.
\newblock \bibinfo{journal}{Transportation research part B: methodological} \bibinfo{volume}{147}, \bibinfo{pages}{67--91}.
%Type = Article
\bibitem[{Li et~al.(2025)Li, Meng, Ma, Ma and Li}]{li2025assessing}
\bibinfo{author}{Li, Z.}, \bibinfo{author}{Meng, H.}, \bibinfo{author}{Ma, C.}, \bibinfo{author}{Ma, K.}, \bibinfo{author}{Li, X.}, \bibinfo{year}{2025}.
\newblock \bibinfo{title}{Assessing markov property in driving behaviors: Insights from statistical tests}.
\newblock \bibinfo{journal}{arXiv preprint arXiv:2501.10625} .
%Type = Article
\bibitem[{Mo et~al.(2021)Mo, Shi and Di}]{mo2021physics}
\bibinfo{author}{Mo, Z.}, \bibinfo{author}{Shi, R.}, \bibinfo{author}{Di, X.}, \bibinfo{year}{2021}.
\newblock \bibinfo{title}{A physics-informed deep learning paradigm for car-following models}.
\newblock \bibinfo{journal}{Transportation research part C: emerging technologies} \bibinfo{volume}{130}, \bibinfo{pages}{103240}.
%Type = Article
\bibitem[{Punzo et~al.(2021)Punzo, Zheng and Montanino}]{punzo2021calibration}
\bibinfo{author}{Punzo, V.}, \bibinfo{author}{Zheng, Z.}, \bibinfo{author}{Montanino, M.}, \bibinfo{year}{2021}.
\newblock \bibinfo{title}{About calibration of car-following dynamics of automated and human-driven vehicles: Methodology, guidelines and codes}.
\newblock \bibinfo{journal}{Transportation Research Part C: Emerging Technologies} \bibinfo{volume}{128}, \bibinfo{pages}{103165}.
%Type = Article
\bibitem[{Rabiner and Juang(1986)}]{rabiner1986introduction}
\bibinfo{author}{Rabiner, L.}, \bibinfo{author}{Juang, B.}, \bibinfo{year}{1986}.
\newblock \bibinfo{title}{An introduction to hidden markov models}.
\newblock \bibinfo{journal}{ieee assp magazine} \bibinfo{volume}{3}, \bibinfo{pages}{4--16}.
%Type = Inproceedings
\bibitem[{Sathyanarayana et~al.(2008)Sathyanarayana, Boyraz and Hansen}]{sathyanarayana2008driver}
\bibinfo{author}{Sathyanarayana, A.}, \bibinfo{author}{Boyraz, P.}, \bibinfo{author}{Hansen, J.H.}, \bibinfo{year}{2008}.
\newblock \bibinfo{title}{Driver behavior analysis and route recognition by hidden markov models}, in: \bibinfo{booktitle}{2008 IEEE International Conference on Vehicular Electronics and Safety}, \bibinfo{organization}{IEEE}. pp. \bibinfo{pages}{276--281}.
%Type = Article
\bibitem[{Taniguchi et~al.(2014)Taniguchi, Nagasaka, Hitomi, Takenaka and Bando}]{taniguchi2014unsupervised}
\bibinfo{author}{Taniguchi, T.}, \bibinfo{author}{Nagasaka, S.}, \bibinfo{author}{Hitomi, K.}, \bibinfo{author}{Takenaka, K.}, \bibinfo{author}{Bando, T.}, \bibinfo{year}{2014}.
\newblock \bibinfo{title}{Unsupervised hierarchical modeling of driving behavior and prediction of contextual changing points}.
\newblock \bibinfo{journal}{IEEE Transactions on Intelligent Transportation Systems} \bibinfo{volume}{16}, \bibinfo{pages}{1746--1760}.
%Type = Article
\bibitem[{Treiber and Helbing(2003)}]{treiber2003memory}
\bibinfo{author}{Treiber, M.}, \bibinfo{author}{Helbing, D.}, \bibinfo{year}{2003}.
\newblock \bibinfo{title}{Memory effects in microscopic traffic models and wide scattering in flow-density data}.
\newblock \bibinfo{journal}{Physical Review E} \bibinfo{volume}{68}, \bibinfo{pages}{046119}.
%Type = Article
\bibitem[{Treiber et~al.(2000)Treiber, Hennecke and Helbing}]{treiber2000congested}
\bibinfo{author}{Treiber, M.}, \bibinfo{author}{Hennecke, A.}, \bibinfo{author}{Helbing, D.}, \bibinfo{year}{2000}.
\newblock \bibinfo{title}{Congested traffic states in empirical observations and microscopic simulations}.
\newblock \bibinfo{journal}{Physical review E} \bibinfo{volume}{62}, \bibinfo{pages}{1805}.
%Type = Article
\bibitem[{Treiber et~al.(2006)Treiber, Kesting and Helbing}]{treiber2006delays}
\bibinfo{author}{Treiber, M.}, \bibinfo{author}{Kesting, A.}, \bibinfo{author}{Helbing, D.}, \bibinfo{year}{2006}.
\newblock \bibinfo{title}{Delays, inaccuracies and anticipation in microscopic traffic models}.
\newblock \bibinfo{journal}{Physica A: Statistical Mechanics and its Applications} \bibinfo{volume}{360}, \bibinfo{pages}{71--88}.
%Type = Inproceedings
\bibitem[{Vaitkus et~al.(2014)Vaitkus, Lengvenis and {\v{Z}}ylius}]{vaitkus2014driving}
\bibinfo{author}{Vaitkus, V.}, \bibinfo{author}{Lengvenis, P.}, \bibinfo{author}{{\v{Z}}ylius, G.}, \bibinfo{year}{2014}.
\newblock \bibinfo{title}{Driving style classification using long-term accelerometer information}, in: \bibinfo{booktitle}{2014 19th international conference on methods and models in automation and robotics (MMAR)}, \bibinfo{organization}{IEEE}. pp. \bibinfo{pages}{641--644}.
%Type = Article
\bibitem[{Wang et~al.(2022)Wang, Wang, Zhang, Liu, Sun et~al.}]{wang2022social}
\bibinfo{author}{Wang, W.}, \bibinfo{author}{Wang, L.}, \bibinfo{author}{Zhang, C.}, \bibinfo{author}{Liu, C.}, \bibinfo{author}{Sun, L.}, et~al., \bibinfo{year}{2022}.
\newblock \bibinfo{title}{Social interactions for autonomous driving: A review and perspectives}.
\newblock \bibinfo{journal}{Foundations and Trends{\textregistered} in Robotics} \bibinfo{volume}{10}, \bibinfo{pages}{198--376}.
%Type = Article
\bibitem[{Wang et~al.(2014)Wang, Xi and Chen}]{wang2014modeling}
\bibinfo{author}{Wang, W.}, \bibinfo{author}{Xi, J.}, \bibinfo{author}{Chen, H.}, \bibinfo{year}{2014}.
\newblock \bibinfo{title}{Modeling and recognizing driver behavior based on driving data: A survey}.
\newblock \bibinfo{journal}{Mathematical Problems in Engineering} \bibinfo{volume}{2014}, \bibinfo{pages}{245641}.
%Type = Article
\bibitem[{Wang et~al.(2018a)Wang, Xi and Zhao}]{wang2018driving}
\bibinfo{author}{Wang, W.}, \bibinfo{author}{Xi, J.}, \bibinfo{author}{Zhao, D.}, \bibinfo{year}{2018}a.
\newblock \bibinfo{title}{Driving style analysis using primitive driving patterns with bayesian nonparametric approaches}.
\newblock \bibinfo{journal}{IEEE Transactions on Intelligent Transportation Systems} \bibinfo{volume}{20}, \bibinfo{pages}{2986--2998}.
%Type = Article
\bibitem[{Wang et~al.(2018b)Wang, Xi and Zhao}]{wang2018learning}
\bibinfo{author}{Wang, W.}, \bibinfo{author}{Xi, J.}, \bibinfo{author}{Zhao, D.}, \bibinfo{year}{2018}b.
\newblock \bibinfo{title}{Learning and inferring a driver's braking action in car-following scenarios}.
\newblock \bibinfo{journal}{IEEE Transactions on Vehicular Technology} \bibinfo{volume}{67}, \bibinfo{pages}{3887--3899}.
%Type = Article
\bibitem[{Wang et~al.(2017)Wang, Jiang, Li, Lin, Zheng and Wang}]{wang2017capturing}
\bibinfo{author}{Wang, X.}, \bibinfo{author}{Jiang, R.}, \bibinfo{author}{Li, L.}, \bibinfo{author}{Lin, Y.}, \bibinfo{author}{Zheng, X.}, \bibinfo{author}{Wang, F.Y.}, \bibinfo{year}{2017}.
\newblock \bibinfo{title}{Capturing car-following behaviors by deep learning}.
\newblock \bibinfo{journal}{IEEE Transactions on Intelligent Transportation Systems} \bibinfo{volume}{19}, \bibinfo{pages}{910--920}.
%Type = Article
\bibitem[{Wiedemann(1974)}]{wiedemann1974simulation}
\bibinfo{author}{Wiedemann, R.}, \bibinfo{year}{1974}.
\newblock \bibinfo{title}{Simulation des strassenverkehrsflusses.} .
%Type = Article
\bibitem[{Yao et~al.(2025)Yao, Calvert and Hoogendoorn}]{yao2025novel}
\bibinfo{author}{Yao, X.}, \bibinfo{author}{Calvert, S.C.}, \bibinfo{author}{Hoogendoorn, S.P.}, \bibinfo{year}{2025}.
\newblock \bibinfo{title}{A novel framework for identifying driving heterogeneity through action patterns}.
\newblock \bibinfo{journal}{IEEE Transactions on Intelligent Transportation Systems} .
%Type = Inproceedings
\bibitem[{Zaky et~al.(2015)Zaky, Gomaa and Khamis}]{zaky2015car}
\bibinfo{author}{Zaky, A.B.}, \bibinfo{author}{Gomaa, W.}, \bibinfo{author}{Khamis, M.A.}, \bibinfo{year}{2015}.
\newblock \bibinfo{title}{Car following markov regime classification and calibration}, in: \bibinfo{booktitle}{2015 IEEE 14th International Conference on Machine Learning and Applications (ICMLA)}, \bibinfo{organization}{IEEE}. pp. \bibinfo{pages}{1013--1018}.
%Type = Inproceedings
\bibitem[{Zhang et~al.(2024a)Zhang, Chen, Zhu, Yang and Sun}]{zhang2024learning}
\bibinfo{author}{Zhang, C.}, \bibinfo{author}{Chen, K.}, \bibinfo{author}{Zhu, M.}, \bibinfo{author}{Yang, H.}, \bibinfo{author}{Sun, L.}, \bibinfo{year}{2024}a.
\newblock \bibinfo{title}{Learning car-following behaviors using bayesian matrix normal mixture regression}, in: \bibinfo{booktitle}{2024 IEEE Intelligent Vehicles Symposium (IV)}, \bibinfo{organization}{IEEE}. pp. \bibinfo{pages}{608--613}.
%Type = Inproceedings
\bibitem[{Zhang et~al.(2023)Zhang, Chen, Zhu, Wang, Liu and Sun}]{zhang2023interactive}
\bibinfo{author}{Zhang, C.}, \bibinfo{author}{Chen, R.}, \bibinfo{author}{Zhu, J.}, \bibinfo{author}{Wang, W.}, \bibinfo{author}{Liu, C.}, \bibinfo{author}{Sun, L.}, \bibinfo{year}{2023}.
\newblock \bibinfo{title}{Interactive car-following: Matters but not always}, in: \bibinfo{booktitle}{2023 IEEE 26th International Conference on Intelligent Transportation Systems (ITSC)}, \bibinfo{organization}{IEEE}. pp. \bibinfo{pages}{5120--5125}.
%Type = Article
\bibitem[{Zhang and Sun(2024)}]{zhang2024bayesian}
\bibinfo{author}{Zhang, C.}, \bibinfo{author}{Sun, L.}, \bibinfo{year}{2024}.
\newblock \bibinfo{title}{Bayesian calibration of the intelligent driver model}.
\newblock \bibinfo{journal}{IEEE Transactions on Intelligent Transportation Systems} .
%Type = Article
\bibitem[{Zhang et~al.(2024b)Zhang, Wang and Sun}]{zhang2024calibrating}
\bibinfo{author}{Zhang, C.}, \bibinfo{author}{Wang, W.}, \bibinfo{author}{Sun, L.}, \bibinfo{year}{2024}b.
\newblock \bibinfo{title}{Calibrating car-following models via bayesian dynamic regression}.
\newblock \bibinfo{journal}{Transportation Research Part C: Emerging Technologies} \bibinfo{volume}{168}, \bibinfo{pages}{104719}.
%Type = Article
\bibitem[{Zhang et~al.(2021)Zhang, Zhu, Wang and Xi}]{zhang2021spatiotemporal}
\bibinfo{author}{Zhang, C.}, \bibinfo{author}{Zhu, J.}, \bibinfo{author}{Wang, W.}, \bibinfo{author}{Xi, J.}, \bibinfo{year}{2021}.
\newblock \bibinfo{title}{Spatiotemporal learning of multivehicle interaction patterns in lane-change scenarios}.
\newblock \bibinfo{journal}{IEEE Transactions on Intelligent Transportation Systems} \bibinfo{volume}{23}, \bibinfo{pages}{6446--6459}.
%Type = Article
\bibitem[{Zhou et~al.(2025)Zhou, Yan, Tian, Wang, Li and Zhong}]{zhou2025driving}
\bibinfo{author}{Zhou, S.}, \bibinfo{author}{Yan, J.}, \bibinfo{author}{Tian, J.}, \bibinfo{author}{Wang, T.}, \bibinfo{author}{Li, Y.}, \bibinfo{author}{Zhong, S.}, \bibinfo{year}{2025}.
\newblock \bibinfo{title}{A driving regime-embedded deep learning framework for modeling intra-driver heterogeneity in multi-scale car-following dynamics}.
\newblock \bibinfo{journal}{arXiv preprint arXiv:2506.05902} .
%Type = Article
\bibitem[{Zhu et~al.(2018)Zhu, Wang and Wang}]{zhu2018human}
\bibinfo{author}{Zhu, M.}, \bibinfo{author}{Wang, X.}, \bibinfo{author}{Wang, Y.}, \bibinfo{year}{2018}.
\newblock \bibinfo{title}{Human-like autonomous car-following model with deep reinforcement learning}.
\newblock \bibinfo{journal}{Transportation research part C: emerging technologies} \bibinfo{volume}{97}, \bibinfo{pages}{348--368}.
%Type = Article
\bibitem[{Zou et~al.(2022)Zou, Zhu, Xie, Zhang and Zhang}]{zou2022multivariate}
\bibinfo{author}{Zou, Y.}, \bibinfo{author}{Zhu, T.}, \bibinfo{author}{Xie, Y.}, \bibinfo{author}{Zhang, Y.}, \bibinfo{author}{Zhang, Y.}, \bibinfo{year}{2022}.
\newblock \bibinfo{title}{Multivariate analysis of car-following behavior data using a coupled hidden markov model}.
\newblock \bibinfo{journal}{Transportation research part C: emerging technologies} \bibinfo{volume}{144}, \bibinfo{pages}{103914}.

\end{thebibliography}

\end{document}